\documentclass[a4paper,12pt,openany]{book}
\usepackage[english,frenchb]{babel}
\usepackage{graphicx}
\usepackage{fancyhdr}
\usepackage[latin1]{inputenc}
\usepackage{mathrsfs}
\usepackage{array,multirow}
\usepackage{latexsym}
\usepackage{slashbox}
\usepackage{amsfonts}
\usepackage{amsmath}
\usepackage{amssymb}
\usepackage[T1]{fontenc}
\usepackage{ae,aecompl}
\usepackage{smallcap}
\usepackage{url}
\usepackage{scrextend}
\usepackage{enumitem}
\usepackage{amsthm}
\usepackage{longtable}
\usepackage{amsmath,algorithm,algpseudocode}
\usepackage{subfigure}

\makeatletter
\renewcommand{\thesection}{\@arabic\c@section}
\makeatother

\newtheorem{definition}{Definition}

\fancypagestyle{plain}{ \fancyhead{}
}
\providecommand{\keywords}[1]{\textbf{\textit{Keywords:}} #1}
\begin{document}
\selectlanguage{english}
\frontmatter
\thispagestyle{empty}
\vspace*{-1.5\baselineskip}
\hspace{-0.05\linewidth}
\renewcommand{\fboxrule}{1pt}
\begin{minipage}[b][\textheight][b]{1.05\linewidth}
\centerline{\textbf{TUNISIAN REPUBLIC}}
\centerline{\textbf{MINISTRY OF HIGHER EDUCATION AND SCIENTIFIC RESEARCH}}
\centerline{\textbf{UNIVERSITY OF TUNIS EL MANAR}}
\vspace{\baselineskip}
\centerline{\resizebox{0.3\textwidth}{!}{\includegraphics[scale=0.5]{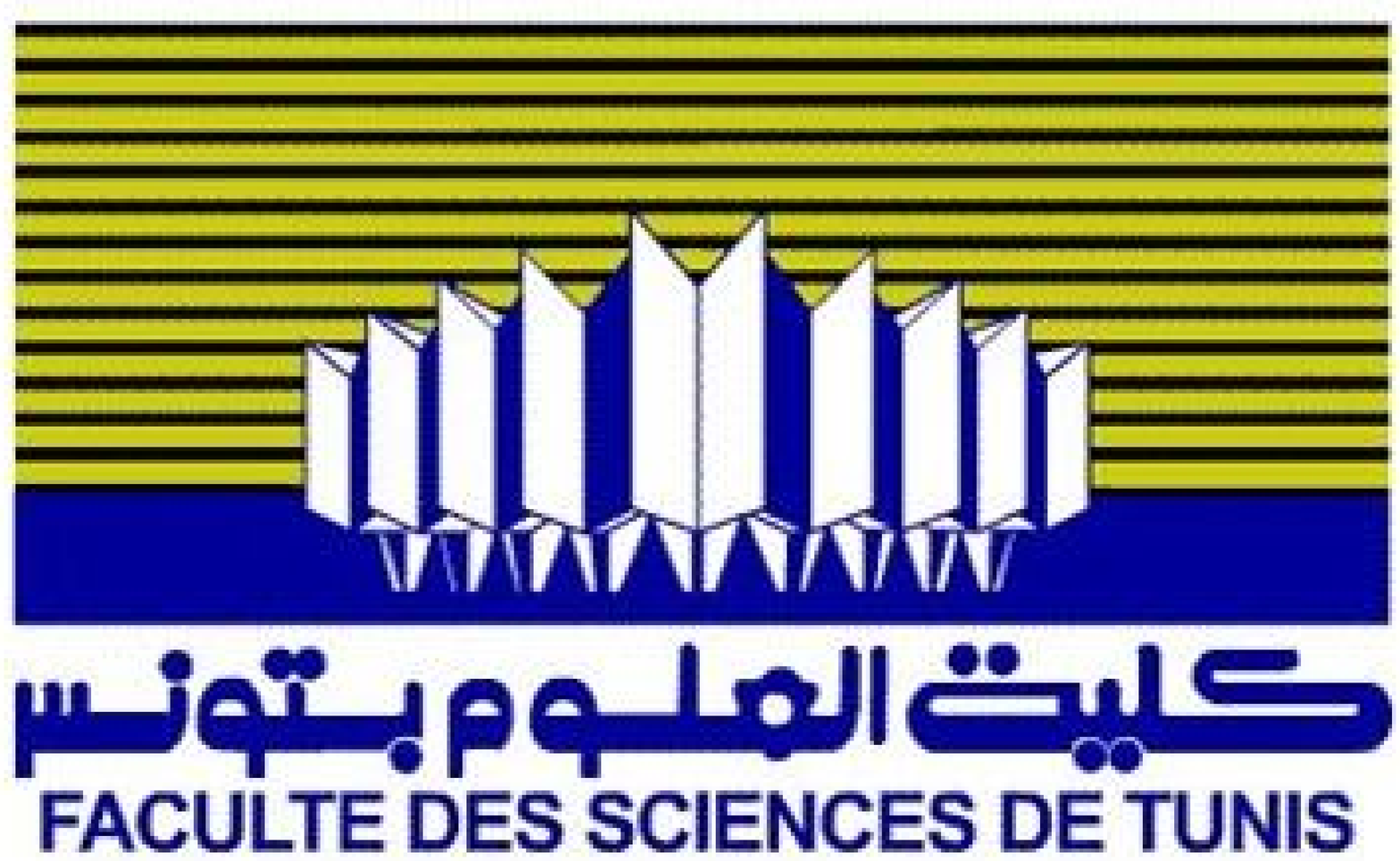}}}
 \centerline{ \textbf{FACULTY OF SCIENCES OF TUNIS}}
 \centerline{ \textbf{DEPARTMENT OF COMPUTER SCIENCE  }}
\vfill \centerline{\LARGE \textbf{MASTER THESIS}}
 \vfill \centerline{{Presented to obtain the}}
 \centerline{\textbf{Master's degree in Computer Science}}
\vspace{\baselineskip}
\vfill \centerline{\Large \textbf{By~:}}
\vfill \centerline{\Large \textbf{Sawsen BEN NASR}}
\vfill \renewcommand{\fboxrule}{1.5pt}
\centerline{\fbox{\parbox{0.95\textwidth}{\bf
    \LARGE\center \vspace{0.3\baselineskip} New approach for a stable multi-criteria ridesharing system \\[0.3\baselineskip]}}}
\vspace{\baselineskip}
\vfill  \centerline{\textbf{To be defended on December 3, 2018}}
\vfill  \textbf {\underline{Graduation committee~:}}\\
  \textbf{Chairperson : Pr. Faouzi Moussa}\\
  \textbf{Examiner : Dr. Narjes Doggaz}\\
  \textbf{Master director : Pr. Sadok BEN YAHIA}\\
\vfill \centerline{\textbf{2017/2018}}
\end{minipage}

\newpage
\vspace*{\stretch{1}}
\begin{flushright}
\begin{em}
To my dear parents Mohamed and Amel
\\
For the love they gave me
\paragraph{}
To my exceptional husband Tarek
\\
For his support and patience
\paragraph{}
To my loving daughters Yasmine \& Nermine
\\
For the happiness they bring to my life

\end{em}
\end{flushright}
\vspace*{\stretch{2}}

\newpage
\thispagestyle{empty}
\centerline{\Large \textbf{Acknowledgments}}
\paragraph{} 
\paragraph{} 
While I am convinced that the result of this work is the fruit of the collaboration of several individuals, I am proud to note, on the occasion of the completion of the report of my master thesis, my thanks for those who in one way or another contributed and extended their valuable assistance in the preparation and completion of this study.
\\
\paragraph{} 
It is a pleasant duty for me to express my deep gratitude to my supervisor, Pr. Sadok Ben Yahia, for his great human qualities. Professor at the Faculty of Sciences of Tunis, he knew how to make me benefit from his confirmed experience as well as his wide scientific knowledge. He guided me throughout this work, bringing encouragement and understanding in addition to his knowledge and expertise.
\\
\paragraph{}
I would also like to thank Dr. Taoufik Yaferney, assistant professor at the Higher institute of Multimedia Art of Manouba, for supporting me continuously and tirelessly. He has demonstrated an impressive scientific presence to see this work takes shape gradually and finally succeed.
\\
\paragraph{}
Finally, I would like to extend my sincere gratitude to the members of the jury for serving on my defense committee and for the time and the effort given to examine this dissertation.
\\

\newpage
\thispagestyle{empty}

\centerline{\Large \textbf{Abstract}}
\vspace{\baselineskip}
The witnessed  boom in mobility results in many problems such as urbanization, costly construction of many highways and air pollution. In an attempt to address these problems, in this master, we are interested in the implementation of a ridesharing system.\\
Ridesharing is recognized as a highly effective means of transport to solve energy consumption, environmental pollution and traffic congestion issues. Indeed, ridesharing can reduce  the number of vehicles on the roads to avoid traffic jams and thus it contributes to a reduction in greenhouse gas emissions. Its main thrust resides in sharing transport expenses, meeting different people and making traveling more enjoyable. 
\\
In this respect, we introduce in this dissertation an effective ridesharing system, called the Stable Multi-Criteria Rideshare Matching (SMRM) system, that (i) considers users' personal preferences when sharing a private space with others and (ii) enables a stable matching between driver and passenger sets. The performed experiments show that the introduced system outperforms its competitors in terms of stability quality and cost. 



\tableofcontents

\newpage
\addcontentsline{toc}{chapter}{\listfigurename}
\listoffigures

\newpage
\addcontentsline{toc}{chapter}{\listtablename}
\listoftables

\chapter{Introduction}

Since the transport revolution of the twentieth century, the car has rapidly emerged as the main means of transportation in and around large cities. In 2006, nearly 900 million cars roamed the planet. In 2015, the billion was exceeded \cite{r35}. Under the shadow of this mobility boom, there are major misdeeds and pejorative impacts affecting the quality of life as well as the biological and environmental balance necessary for the survival of human beings. Indeed, public authorities hope to find solutions to reduce the existing problems and to make transport systems more durable, i.e., economically viable, socially equitable and environmentally sustainable \cite{rr8}.
\\
The focus is on alternatives to individual cars and on the radical change in human behavior. Thus, similar to public transport, walking or cycling, ridesharing has been recognized as a highly effective way of transport to solve energy consumption, environmental pollution and traffic congestion issues \cite{rr9}. Indeed, it reduces the number of vehicles on the roads in order to avoid traffic jams and thus helps decrease greenhouse gas emissions. Moreover, it allows sharing transportation expenses between several individuals. In addition to the economic and ecological benefits, ridesharing permits, through the grouping of people who know each other or not, restoring a certain communication and to creating and fortifying social bonds. 
Ridesharing has thus made its entry into the field of research and several systems have been proposed. The purpose of such systems is to enable drivers and passengers to submit respectively their ridesharing offers and requests by specifying their constraints such as their origin, destination and detour tolerance, etc. Once the request is submitted, the system provides a list of pairs (driver, passenger) that satisfies these constraints and maximizes its global objective.
\\
Although there are many attempts to provide ridesharing systems, they are not meeting the expected success, given the lack of motivation of individuals to turn to such systems.

This lack of enthusiasm for this practice is explained in particular by the boredom of traveling with an unknown person. Indeed, sharing a private space as the vehicle is not an interesting idea for many persons, especially if their tastes and habits are different. Passengers and drivers often aspire to have a pleasant ridesharing time and this is not always the case if they have different social preferences. Everyone has their own personality and their moods. However, classical ridesharing systems mainly focus on the improvement of their potential and performance in ridesharing to fulfill spatio-temporal constraints and to maximize certain objectives. In addition, a rideshare matching solution aiming to maximize the total system objective or the total number of matches may not necessarily maximize the objective of each individual participant.
Launched on the path of improvement, we propose in our work to set up an optimized dynamic ridesharing system. Aspiring to the development of a functioning, efficient and competitive system on a large scale, we focused on the notion of satisfaction of personal constraints. Our study was then oriented in this direction and resulted in a new ridesharing system that we call Stable Multi-Criteria Rideshare Matching. The main thrust of our proposal is the selection of matches that promise the satisfaction of drivers and passengers' social preferences. Furthermore , we consider stability for rideshare matches.
\\
Indeed, the ridesharing problem is treated as a matching problem where drivers are assigned to passengers. Our objective is to create a perfect matching of drivers and passengers such that there does not exists any pair of a driver and a passenger who prefers each other to their current partners. This notion of stability is similar to that of the stable marriage problem. To achieve this, we rely on the multi-criteria decision-making method (MCDM), which is considered one of the most important branches of operational research and decision theory.
\\
This report is divided into four chapters: the first introduces the ridesharing that is recognized as a highly effective means of transport to solve energy consumption, environmental pollution and traffic congestion issues. In the second chapter, we highlight a state of the art of existing rideshare systems with a focus on their limitations and gaps and we summarize the methodological tools underlying our approach. In the third chapter, we present the architecture of our system by detailing its main components. The methodological concepts and algorithms leading to such modeling are also described in this chapter. Evaluation and experimentation will be the subject of the forth chapter. The latter emphasizes the realization of the approach, the choice of the evaluation metrics and the experimental environment.

\mainmatter
\chapter{Ridesharing: An alternative, ecological and economical means of transport}

\section{Introduction}
With the lengthening of distances and travel times, causing an explosion of mobility that is not without harmful consequences, the private car has become a source of trouble after the dazzling success of which it was crowned. Indeed, in addition to noise pollution, the pressure it exerts on individuals and huge expenses, the massive use of the personal car causes the extermination of the own ecological concept. 
\\
Meeting the challenge of reducing the excessive use of the private car is all the more difficult as it exerts an immeasurable pressure on individuals creating strong links of dependence. One solution lies in the reasonable non-abusive use of the private car. Group access, which in addition makes use of private cars, helps to considerably reduce the number of moving cars. This concept is the definition of ridesharing.

\section{Motivation: Effects of the car on society}
The European Union has a fairly dense transport network, including road, rail, metropolitan, maritime, etc. The use of personal vehicles is popular, indeed, 80\% of urban travel is done with this means of transport \cite{rr2}. The environmental impact and congestion of the road network have become a major concern for the authorities, who are trying to reduce or to see better use of this means of transport.
\subsection{Private car: Access and convenience}
\begin{figure}[h!]
\centering
\fbox{
\includegraphics[scale=0.7]{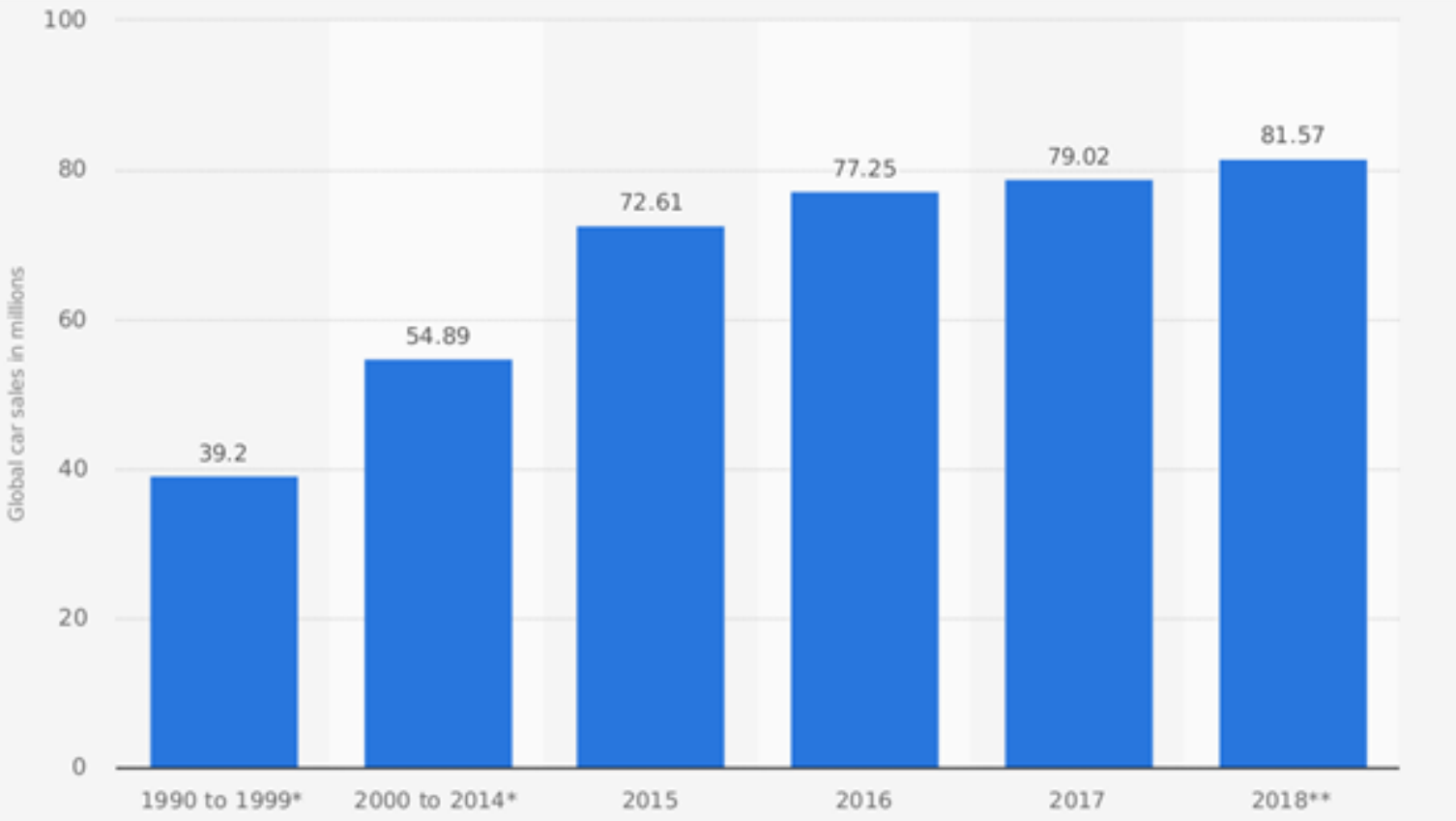}}
\caption{Number of cars sold worldwide from 1990 to 2018 (in million units)} \label{f1}
\end{figure}

The road transport has been one of the most important means of transport and has been indispensable to the development of commerce and industry. All the movement of people, freight and information has begun and ultimately has ended by making use of roads. These movements have always been fundamental components of the economic and social life of societies. 
\\
Contemporary economic processes have been accompanied by a significant increase in mobility and higher number of individual cars. Although this trend can be traced back to the industrial revolution, it significantly accelerated in the second half of the twentieth century and the demand of individual car has rapidly increased as the main means of transportation in and around large cities. 
\\
The statistic in Fig.\ref{f1}\footnote{\url{https://www.statista.com/statistics/200002/international-car-sales-since-1990/}} represents the number of cars sold worldwide from 1990 through 2018. 81.6 million automobiles are expected to be sold by the end 2018. This figure equals double of all ten-year car sales from 1990 to 2000.

Indeed, due to its many advantages: its performance in terms of speed, handling, comfort, safety and reliability and the great adaptability it represents (short or long distance travel, urban, peri-urban or extra-urban area ...); the individual car has become a necessity in people's lives. Therefore, joining convenience to ease of access and spatio-temporal flexibility, individuals are increasingly moving towards this mode of transportation. The latter has become, over time, necessary for their psychological balance and peace of mind \cite{rr3}. Despite its decisive benefits, even changing the behavior and lifestyles of individuals \cite{rr4}, the private car prevents the prospects of advanced and sustainable mobility to go to good advantage. It is therefore behind many obstacles that run counter to the implementation of a clean and sustainable development process. 
\subsection{Disadvantages of the private car use}
The car threatens the centrality of the urban organization. Also, it affects the relationships between people because it destroys urban sociability thus generating economic distortions.
\subsubsection{Environmental impact}
\begin{figure}[h!]
\centering
\fbox{
\includegraphics[scale=0.4]{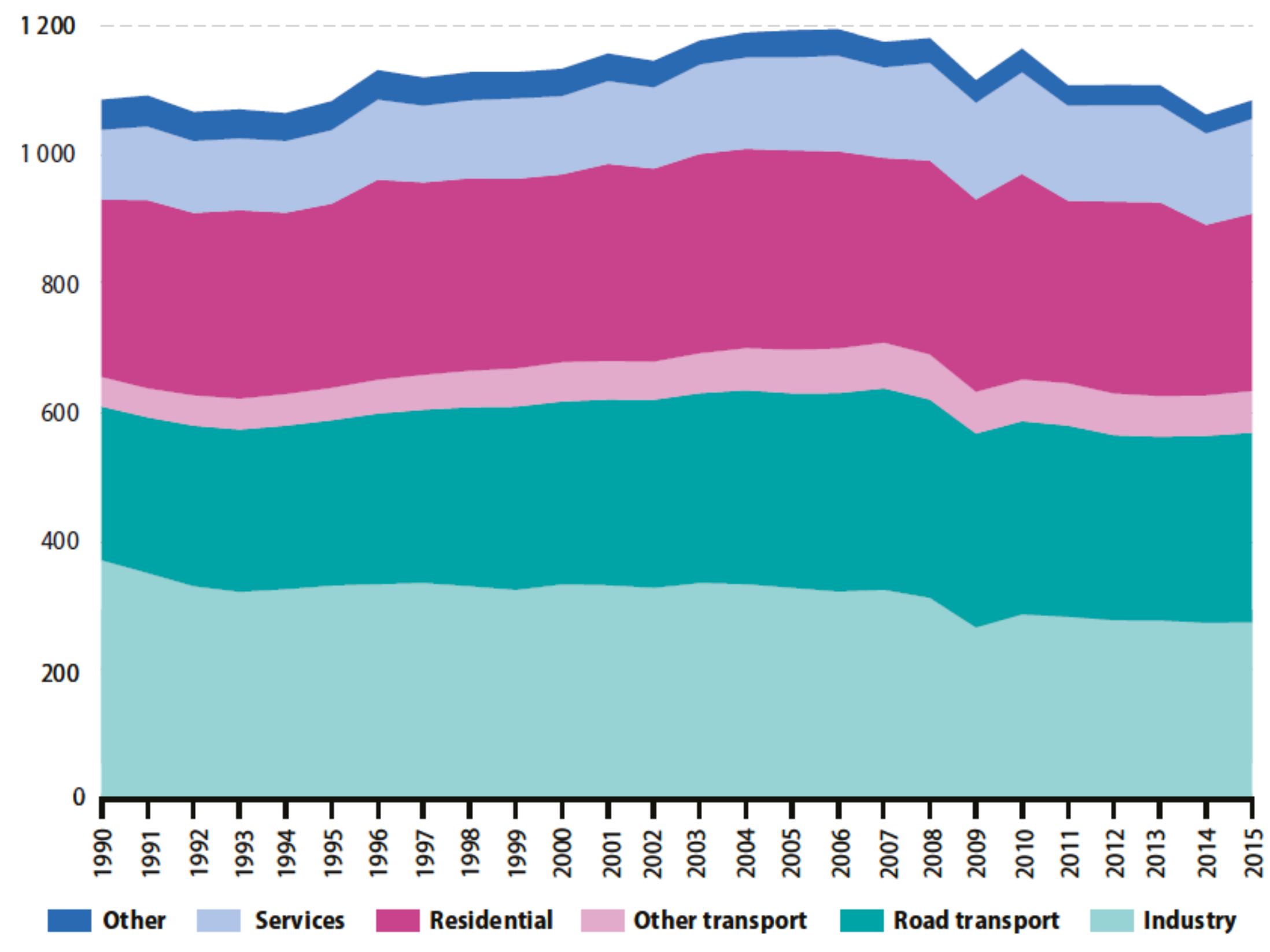}}
\caption{Final energy consumption by sector, EU-28, 1990-2015} \label{f2}
\end{figure}

\begin{figure}[h!]
\centering
\fbox{
\includegraphics[scale=0.7]{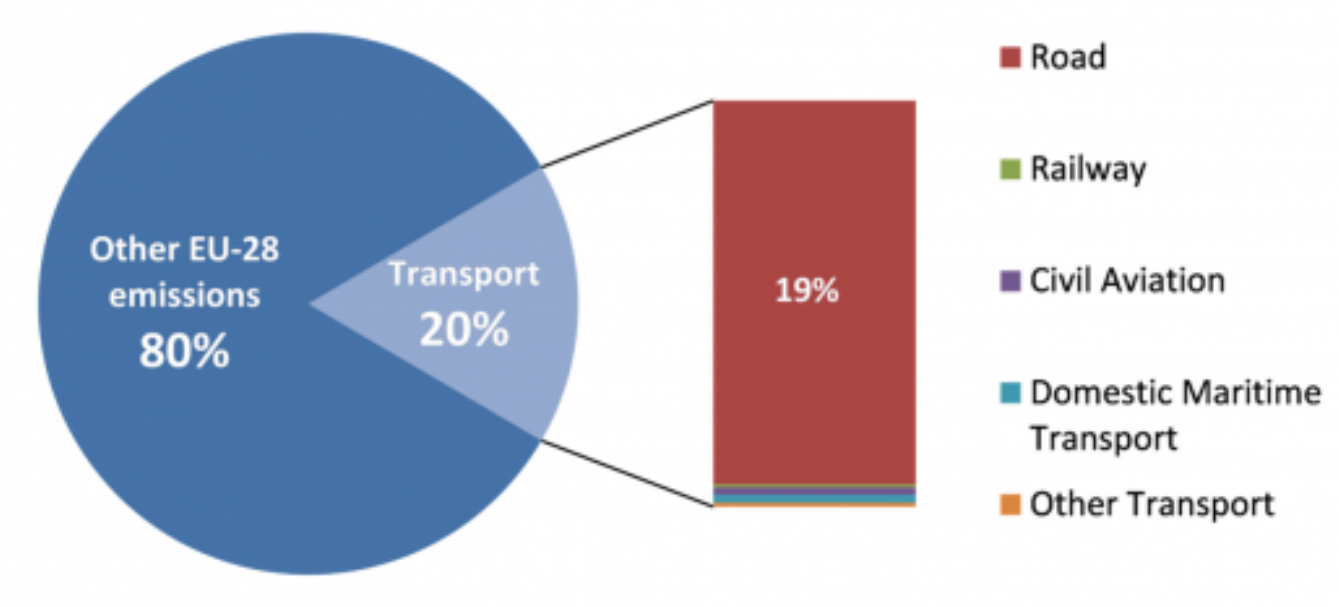}}
\caption{EU-28 GHG emissions by mode of transport in 2015} \label{f3}
\end{figure}

~~\\
To move, the car needs gasoline, diesel or natural gas, etc. It contains a battery, lubricating oil, and a catalyst composed of precious-metal such as platinum and rhodium. These few examples are enough to understand how difficult it is to measure the environmental impact of a car.
Road transportation is linked with a wide range of environmental considerations. The nature of these environmental impacts is related to the infrastructures over which they operate, their energy supply systems and their emissions. While consuming large quantities of energy vehicles also emits numerous pollutants such as carbon dioxide and nitrogen oxide which damages many ecological systems. The structure of final energy consumption in the European Union in 2015 by sector shows that road transport with 30\% accounted for the biggest share of energy consumption (Fig.\ref{f2}\footnote{Source: Energy, transport and environment indicators, 2015 Edition.\label{refnote}}).
A breakdown by sector presented in Fig.\ref{f3}\footref{refnote} is tangible evidence of this, based on figures extracted on CO2 emissions from different modes of transport. Statistics at the origin of these histograms reveal that road transport in the European Union remains the most polluting mode with 94\% of the CO2 emissions. 
\subsubsection{Safety issues}
~~\\
The large number of cars on the road generates negative impacts not only on the environment but also on the fluidity of the traffic thus creating traffic jams and a congestion which can be at the origin of accidents or any other type of incident. Indeed, according to data from the World Health Organisation\footnote{\url{http://www.who.int/violence_injury_prevention/road_traffic/en/}}, over three thousand people die every day on the world's roads and tens of millions of people are disabled or injured every year. In parallel, public transport helps to make road transport more fluid by transporting large numbers of people on specific routes, thus reducing the number of cars on the roads. Indeed, for 60 people transported by bus, the occupied surface does not exceed that intended to support two cars. While a car only carries an average of 1.5 passengers \cite{rr6}. In addition, reserved lanes intended for transport far from road traffic considerably help to smooth traffic and thus reduce accidents. 
\subsubsection{Societal issues}
~~\\
Since the 1970s, the proliferation of ring roads, highways and suburban road networks have encouraged many city dwellers to leave city centers with rents whose prices are too high, to join the suburban housing estates. The car contributes to the dissolution of the limits of the city (peri-urbanization) and it is the birth of "rurbains", who depend on the car to go to work or do their shopping.
\\
Thus the emergence of this automobile civilization has caused a brutal transformation of the classical city towards a city where the car progresses irresistibly thus causing an increasing separation of the zones of activity of the zones of habitat which increases the distance to be traveled and the road traffic.
\\
Various societal effects remain after the growth of car traffic: noise, stress, pollution, lack of space for pedestrians, etc. All of these negative effects contribute to the dispersal and remoteness of habitat, businesses, services, workplaces and leisure facilities that in turn cause increased travel needs and increased car traffic. This is a vicious circle, the phenomenon of which is further reinforced by the fact that the dispersion and distance of the various functions leads to a reduction in accessibility for non-motorized users and for those in common.
\subsection{Towards sustainable mobility}
Public authorities hope to find solutions to reduce existing problems and encourage innovative initiatives that aim to make transport systems more durable, i.e., economically viable, socially equitable and environmentally sustainable \cite{rr8}.
\\
There are several innovative projects to develop decision support software platforms that offer cost-effective and flexible solutions. These platforms therefore offer sustainable mobility development solutions in the sense that they propose easy solutions with the required knowledge. The latter are, for example, real-time information on traffic conditions, measurements of harmful gas emissions, and so on. Also in the context of innovative advanced mobility projects, some organizations offer to help travelers in their travels based on calculations of travel times for example. Their impact in this context may concern more efficient management of travel, control of the environmental impact, an incentive for the use of public transport or even the minimization of costs, etc. 
\\
Efforts in this direction have shifted towards green technologies and clean modes of transport while trying to remain in the cozy environment of on-demand transport services or the personal cars. The aim is to reduce the number of cars on the road, thereby reducing its negative impact on traffic flow and the environment while maintaining its flexibility. A key solution lies in the concept of a shared car that heals the image of the particular automobile, dislodging it from the prevailing context of which it has always appeared while preserving the advantages it presents. Researchers and industrialists have collaborated to immerse innovation projects in this context and put into practice the ideas expressed by these projects, thus combining pragmatism with theoretical ideologies. In doing so, many practices are born and systems offering more or less advanced services compete with originality and performance. 
\\
Transport problems, which are now partially solved, remain far-reaching. Advanced mobility is thus placed at a probative stage in the fields of industry as well as research. In our work, we address the problem of the shared car and more specifically the concept of ridesharing.

\section{Ridesharing}
With ridesharing, the private car gave birth to a means of transport that was in nature individual but became collective. Several ridesharing systems have been emerged as a result of the development of several works to improve the quality of life, in particular by considering the development of the transport field in an improving context.
\subsection{A mode with high potential}
Ridesharing has been recognized as a highly effective way of transport to solve energy consumption, environmental pollution and traffic congestion issues \cite{rr9}. Indeed, it reduces the number of vehicles on the roads in order to avoid traffic jams and thus helps decrease greenhouse gas emissions. Moreover, it allows sharing transportation expenses between several individuals. In addition to the economic and ecological benefits, ridesharing permits, through the grouping of people who know each other or not, restoring a certain communication and to creating and fortifying social bonds.
\subsection{Principle and definitions}
Ridesharing refers to the shared use of a vehicle by a non-professional driver and one or more passengers for the purpose of performing all or part of a common travel \cite{rr10}. Ridesharing is then the matching of travelers doing all or part of a trip they would otherwise have done alone. 
\\
To better present the concept of ridesharing, we present some definitions:
\paragraph{Passengers:} 
At the base, they are pedestrians looking for a possible rideshare offer that can bring them to a specific place, these people may or may not own their private cars. They are thus defined as service providers initiating requests to be driven between two given points in the context of a desired trip.
\paragraph{Request:} 
A rideshare trip can only be done if a request exists. This is a request from a passenger who wishes to travel by car from one place to another. A request relates to a specific travel requirement according to which the passenger determines the date and time of travel, where he wants to go, etc.
\paragraph{Drivers:} 
This refers to the driver of a car in the context of ridesharing with one or more other passengers. These users of the service use in the general case their own vehicle. In addition, the driver has his own travel needs which he defines through the submission of a ridesharing offer. 
\paragraph{Offer:} 
It refers to the driver's travel parameters (usually the owner of the car used for ridesharing). These parameters define the specificities of the trip to be traveled: origin, destination, date, departure time, number of places available, etc.
\subsection{Ridesharing problem classification}	 
\begin{figure}[h!]
\centering
\fbox{
\includegraphics[scale=0.6]{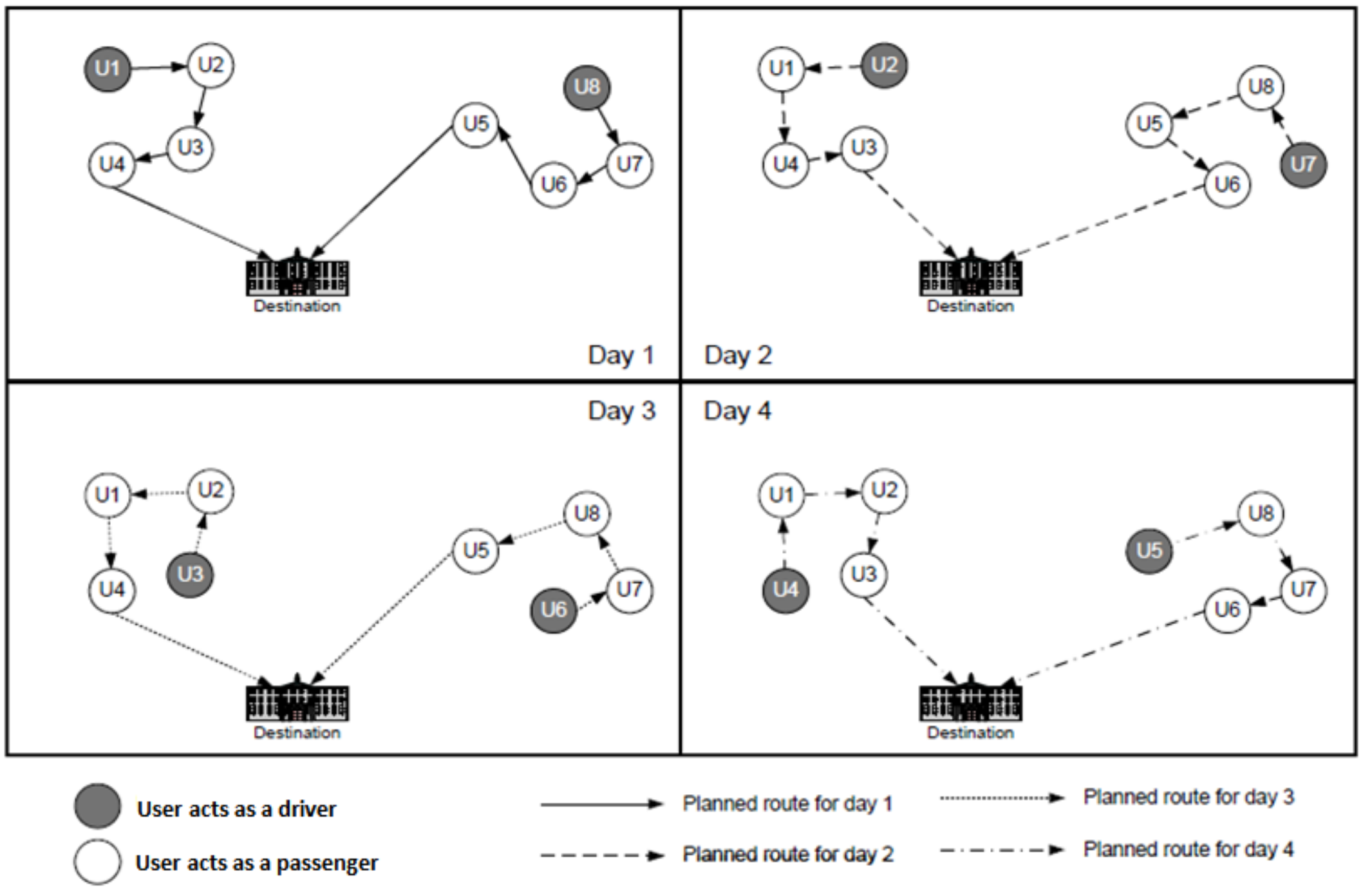}}
\caption{An Example of the LTRSP} \label{f5}
\end{figure}

\begin{figure}[h!]
\centering
\fbox{
\includegraphics[scale=0.6]{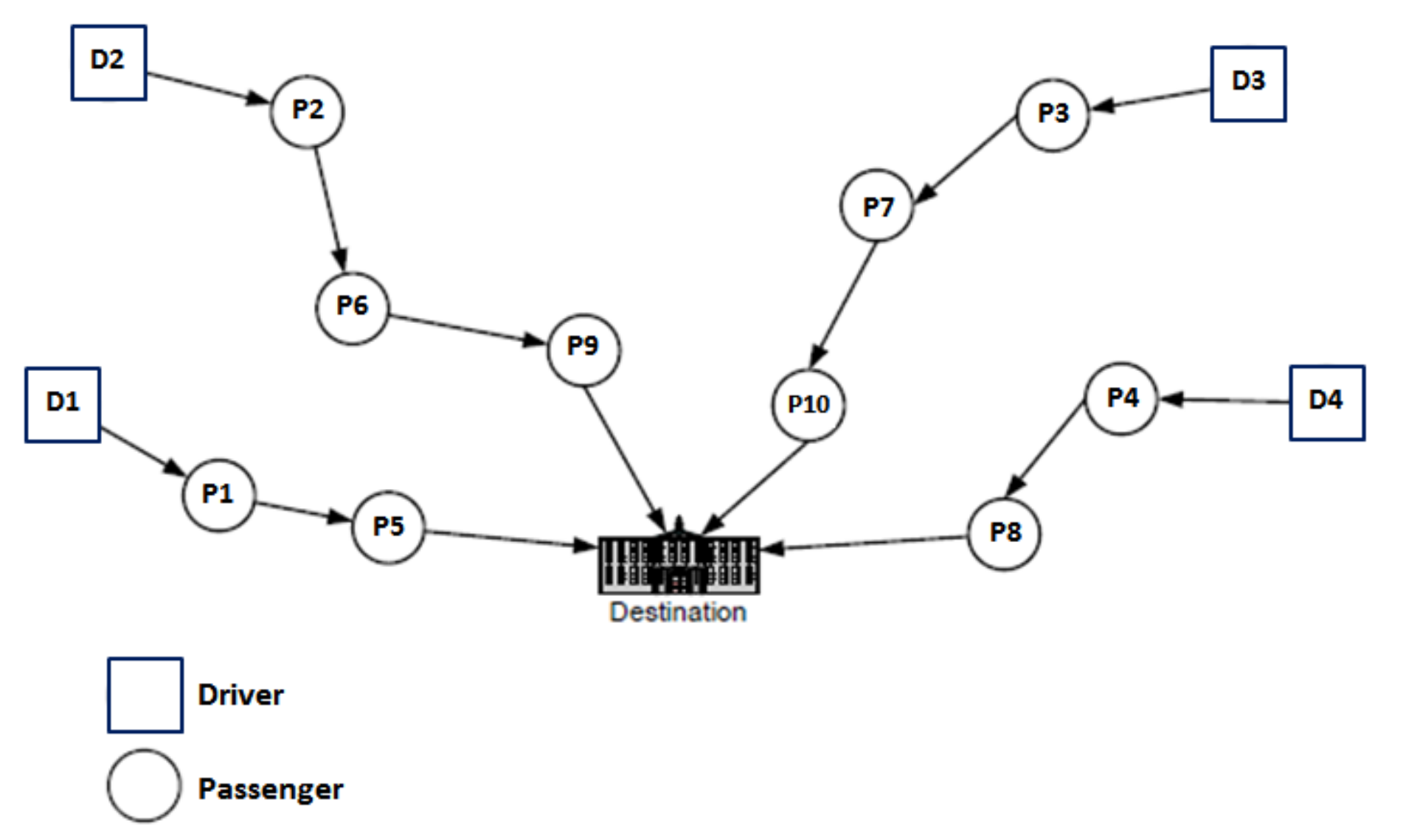}}
\caption{An Example of the DRSP} \label{f4}
\end{figure}

ccording to the procedure for using the ridesharing service, we split ridesharing problem into two different types: Long-term RideSharing Problem (LTRSP) and Daily RideSharing Problem (DRSP). 
\\
In the LTRSP, each user has to act as both a driver and a passenger and a solution is to define rideshares where each user will in turn, on different days, pick up the remaining rideshare members. The objective is to minimize the amount of vehicles used and the total distance traveled by all users, subject to car capacity and time window constraints. The LTRSP can be considered as a combination of a clustering problem and a routing problem. It requires finding the rideshare members relatively close to each other and identifying the route and schedule for each member in the rideshare. Fig.\ref{f5} presents an example of LTRSP.
On the contrary, in the DRSP, a number of users declare their availability for picking up or bringing back other users on one particular day. Hence, these users are considered as drivers, and the other users being picked up or bringing back are considered as passengers. Then the problem becomes to assign passengers to drivers and to identify the routes to be driven by the drivers. Since in the DRSP, the drivers and the passengers are known in advance, the objective is to construct path starting from each driver and going through as many passengers as possible with respect to the car capacity and time window constraints, and to minimize the total travel cost. 
\\
The DRSP model is based on daily schedule, so the participants change every day. It is a model normally used by the commercial website which organizes daily rideshare service among different members. Fig.\ref{f4} shows an example of the DRSP. 
The LTRSP is a more stable ridesharing model; the users in LTRSP will not change frequently in a relatively long period of time. This model is usually used by large companies, organizations and universities which provide long-term rideshare service for their employees or students. In the academic point of view, solving the DRSP can simply be done by clustering users into rideshares based on a long-term schedule, and each rideshare member has to act as a driver on a different day. On the other hand, we believe that the DRSP is a more valuable topic for research, since it has its own characteristics and its optimization challenges. Therefore, the DRSP is chosen to be the focused ridesharing.
\section{Ridesharing system}	 	
\subsection{System principle}	 
The ridesharing system connects drivers and passengers wishing to share a trip and the associated costs. Drivers publish their available seats and passengers buy them online, on trips like home-work. It enables drivers and passengers to submit respectively their ridesharing offers and requests by specifying their constraints such as their origin, destination and detour tolerance etc. Once the request is submitted, it provides automatic ride-matching between participants. This rideshare system may use technology such as Global Positioning Software (GPS), continuous internet connection and smart phone software to remove the requirement to pre-arrange trip schedules well in advance as matches are made based on current proximity. Payment can be made in cash in the car or also through PayPal on the ridesharing's mobile phone application. Registration is typically on a handset with fewer requirements for details as the current location will already be known through the phone's GPS function.
\\
This system can also include a post-ride rating for the driver or passenger which would be available to other program users to help them decide whether they want to share a ride with that person.
\subsection{Matching constraints}
Ridesharing does not make sense and has no place to be unless certain conditions are fulfilled. Classical ridesharing systems mainly rely on the matching of spatio-temporal constraints between a driver offer and a passenger request. These constraints essentially refer to the correspondence between offer and demand, thus making it possible to verify:
\begin{itemize}
\item Timing of trips offered and required; it is probably the most important consideration since time tends to be a most constraining factor. Both riders and drivers must provide information on their time schedule preferences. These must come nearer, or even better, confused.
\item Coincidences between trips; the passenger trip fits within to the driver itinerary. This does not exclude the fact that a small detour is tolerated to pick up or to drop off one or more passengers, depending on the motivation of the driver and the other passenger.
\end{itemize}

\subsection{Kinds of ridesharing systems}
There are several distinct forms of ridesharing systems, they each address different trip needs and have different characteristics. 
\paragraph{Commuter}
~~\\
Commuter ridesharing system caters for individuals and organizations looking for regular rideshare matches for work or other regular trips. These are the typical commercial rideshares that provide free membership for individuals who register but if an organization wishes to create a closed network available only to their employees or students it will typically be required to pay a fee for the service. 
\paragraph{Long distance / once off}
~~\\	
Long distance ridesharing system provides matches for people travelling long distances, typically as once-off trips. This mode is targeted towards travelers who have planned schedules well in advance but who may have some flexibility in terms of departure timings.
As a form of pre-organized hitchhiking, this rideshare system typically has less registration and security requirements than commuter ridesharing as the users only utilize the service occasionally. Travel to events and festivals would also be considered part of this category. 

\paragraph{Casual/ flexible}
~~\\
Casual ridesharing system, also described as 'slugging' system, is rideshare network that operates without pre-organization and contact. This rideshare system involves drivers and passengers simply turning up to the same departure location and matches are made on-the-spot for trips to a similar destination. It has typically developed as a means to use high-occupancy vehicle (HOV) lanes or to share toll payments. This ridesharing system is not supported by public or employer organizations.
Advantage of this ridesharing is the reduced commitment for participants; if they don't turn up on the day they won't be letting anyone down. The associated disadvantages are a lack of security and reduced ability to be replicated in other locations as this ridesharing arises organically to address a specific need in locations with a large number of participants making similar trips.
\paragraph{Dynamic}
~~\\
Also known as real-time ridesharing system, it provides automatic ride-matching between participants at short notice or while the driver is already travelling. This rideshare system uses technology such as Global Positioning Software (GPS), continuous internet connection and smart phone software to remove the requirement to pre-arrange trip schedules well in advance as matches are made based on current proximity. Because of the spontaneous nature of this system, the latter will typically suggest a cost per kilometer or a ridesharing fare. Payment can be made in cash in the car or also through PayPal on the ridesharing's mobile phone application. Registration is typically on a handset with fewer requirements for details as the current location will already be known through the phone's GPS function.
This system can also include a post-ride rating for the driver or passenger which would be available to other program users to help them decide whether they want to share a ride with that person.
\vspace{0.5cm}

Being interested in the issue of dynamic ridesharing, we focus essentially on this concept and we propose an optimized approach for ridesharing system.

\section{Conclusion}
In this chapter we discussed how ridesharing systems contribute to improve the environmental conditions in which people evolve. The concept of ridesharing has many advantages, including reducing the number of cars in circulation per kilometer. Indeed, the personal car has become, due to this concept, a means of collective travel accessible to the general public. It is within this context that we directed our work to propose an innovative approach whose foundations were built on the basis of a study of the principles and concepts lacking the systems erected by the development of this phenomenon.
\\
A thorough study of the history of ridesharing as well as a description of the methodological tools used in our proposed approach will be highlighted in the next chapter.
 
\chapter{Related work}
\section{Introduction}
In recent years, ridesharing has become a very popular research topic for the industry and the researchers. In this chapter, we first summarize the pioneering approaches of ridesharing and we define our main motivations justifying the great interest that we carry to it. In section~\ref{sec:1} and section~\ref{sec:2} we present an overview of the methodological tools that form the basis of our approach, which are Multi-Criteria Decision-Making and stable matching.
\section{Survey on ridesharing }
\subsection{Industrial work on ridesharing}
\subsubsection{Static ridesharing}
Numerous Internet sites allow the proposition and the demand of rideshares, whether regular or occasional, of proximity or long distance. In the latter case, some sites offer online search engines for ridesharing, which calculates the routes and the best possibilities for the driver and the passenger. These rideshare bulletin board services are often free and easy to use.
In Tunisia, ridesharing is not yet well recognized. Only a few static internet sites are being hosted like tawsila.tn \footnote{\url{https://www.tawsila.tn/}} and partagi.tn \footnote{\url{https://www.partagi.tn}}.
In France, after the launch of BlaBlaCar \footnote{\url{https://www.blablacar.fr/}} in 2006, ridesharing is booming. By 2010 this site had more than 600,000 registered members and was attracting more than 10 million page views per month. BlaBlaCar has since expanded significantly, with 25 million users across 22 countries \cite{c2r2}.
\\
In 2014, SNCF just played a double blow by opening a new ridesharing platform called IDvroom \footnote{\url{https://www.idvroom.com/}}. It is an open portal; i.e., it is not reserved for SNCF customers. The goal was to attract new revenue and not be overtaken by the already ubiquitous BlaBlaCar. In addition, this project is part of its "door-to-door" travel strategy, which began with personal services such as car rental on arrival of the train, chauffeur service that takes passengers or bring back from the station or take charge of luggage at home. The principle of iDVROOM is about the same as that of its competitor. The user has the choice between a single trip and regular commute. The driver fixes to passengers the cost per kilometer. Passengers can pay the amount of their trip directly online. On arrival, they send the driver their passenger code sent by the site when booking. Afterwards, the money is transferred to the iDVROOM wallet, which will then pay the driver after pocketing a commission \cite{c2r3}.
Static systems require booking in advance based on static reasoning, ignoring instant events. Thus, these systems do not allow a real-time allocation of vehicles, which does not provide an instant response to the user.
In order to cope with this deficit, dynamic ridesharing systems are developed with real-time management of the service. This type of ridesharing has a strong potential for development due to its main principles: real time, optimization of trips and the guarantee of a reliable service.
\subsubsection{Dynamic ridesharing}
Dynamic ridesharing consists of providing users in real time with an opportunity to rideshare at short notice. Conversely to static ridesharing, it allows more flexibility, credibility and less interdependence between participants. 
The principle is based on an instantaneous exchange of data, between drivers and potential passengers, via at least a Smartphone equipped with a GPS tool, allowing real-time geolocation of passengers and drivers, and allowing them to be connected. With a sufficient number of members, the operation is flexible and offers a good quality of service: the probability of finding the right correspondent is considerable. In recent years, dynamic rideshare experiments are emerging. Among the systems made, we can mention:
\begin{itemize}
\item Piggyback \footnote{\url{http://www.piggybackmobile.com/}}: developed by a French research team. Via Piggyback, the driver can enter his destination. If passengers are interested, they send a request, via mobile phone equipped with GPS, which can be accepted or not. After each rideshare trip, the driver has the opportunity to report the passengers transported as favorites or not.
Users have the option to cancel their queries, but if they wait until the last moment penalties can be applied. These penalties can be financial or in terms of "rating".
\item GreenMonkeys \footnote{\url{http://clem.mobi/covoiturage}}: supported by the Clem' company, it offers ridesharing services for businesses. Drivers and passengers specify the origin and destination of their trips on the platform, and Greenmonkeys guarantees that transport will be available to make the trip. If the rideshare solution is impossible, the company agrees to pay the taxi. The payment of the trip is done automatically by an electronic wallet system.
\item Carma Carpooling \footnote{\url{https://www.gocarma.com/}}: set up by the transportation technology company Carma, available from an Iphone equipped with GPS. It allows for dynamic carpooling. Carma is also developing solutions for shuttle services and Transportation On Demand.
The driver enters his destination and the empty seats are offered to potential passengers. If a passenger wants a trip at a certain time, the system selects the most suitable driver and offers the driver the detour to make. If the driver accepts, a voice can guide the driver to the appropriate stop where driver and passenger can meet. On the Iphone, the driver can evaluate between 1 and 5 his experience with the passenger in question. Carma automatically manages the sharing of fees between carpoolers.
\end{itemize}
Nevertheless, despite technological advances and efforts in this domain, most of the existing systems implementing a real-time ridesharing service have remained at an embryonic stage due mainly to lack of security and the automation gap. In addition, compared to static systems, the platforms dedicated to dynamic ridesharing have the advantage of consulting in real time the list of offers of vehicles in circulation. On the other hand, the optimization aspect is completely ignored. Indeed, these systems do not integrate optimization algorithms to generate Driver / Passenger matching.

\subsection{Literature review on ridesharing}
With regard to the dynamic rideshare systems, we can say that they have three components. A rideshare system is composed of 1) an algorithm that matches the participants with each other, 2) according to their constraints, 3) to finally optimize a certain objective.
\\
In the following, we outline how these three notions have been treated in the literature.
\subsubsection{Matching algorithms}
There are several types of matchmaking algorithms for ridesharing. A state of the art has been realized by Agatz et al. \cite{c2r4}. According to the latters, dynamic rideshare systems can be classified according to the number of drivers and passengers considered. Systems that allow the driver to take a single passenger differ from systems that can take more than one on the same trip. On the passenger side, the systems are classified according to whether the passenger can only ride with one driver during his trip, or can connect portions of the trip with different drivers. According to Agatz et al., we can classify ridesharing systems into four categories: Single Driver - Single Passenger, Single Driver - Multiple Passengers, Multiple Drivers - Single Passenger and Multiple Drivers - Multiple Passengers. 
\\
Table \ref{tab1} summarizes for each variant the problem associated with it.

\begin{table}[h!]
\centering
\begin{tabular}{| l | m{5cm} | m{5cm} | }
\hline
    & Single Passenger & Multiple Passengers \\
\hline
   Single Driver & Matching of pairs of drivers and passengers & Routing of drivers to pick up and drop off passengers \\
\hline
   Multiple Drivers & Routing of passengers to transfer between drivers & Routing of drivers and passengers \\
\hline
 \end{tabular}
 \caption{Ridesharing variants}\label{tab1}
\end{table}
\paragraph{Single Driver - Single Passenger Systems}
~~\\
These are the simplest and are the basis of the rideshare study. Research on this subject aim to match a passenger and a driver while fulfilling certain constraints. In this case, we find ourselves in the context of matching problems, that is to say that we have agents (drivers) who must be matched to tasks (passengers) the costs of each match (detours for example). A survey of the most useful of the variations of the matching problem was presented in \cite{c2r5}.
\paragraph{Single Driver - Multiple Passengers Systems}
~~\\
By adding a capacity to the vehicles used, we fall into this variant. This adds a dimension to the problem and we go completely out of the conventional matching problem, since the constraints of the passengers are multiple and different. In this class of problems, there are systems for picking up employees to go to the workplace. Baldacci et al. \cite{c2r6} offer a method to respond to a company that wants its employees to come to work and return home with as few vehicles as possible, i.e. by ridesharing. The method must therefore determine the vehicles to be used and the paths that must be traveled to pick up all employees. This is what is commonly referred to as "dial-a-ride" problems and can be found in the literature review presented by Cordeau et al. \cite{c2r7}. Clearly, a problem "dial-a-ride" is to find the routes and times of pick up and drop off of n users by m vehicles knowing the origins and destinations of n users while minimizing the costs of travel and respecting a set of constraints, such as user preferences. The main difference between this type of problem and the rideshare problems is that in ridesharing, drivers do not come from one or more sources and are independent entities, not attached to a company or service \cite{c2r4}.
\paragraph{Multiple Drivers - Single Passenger}
~~\\ 
It is a question of finding a sequence of drivers that would bring the passenger to destination. In practice, however, it is difficult to take three or more different cars in succession to make a single trip, except perhaps for very long trips. It is for this reason that this category of problem is much less studied than the preceding categories. Drews et al. \cite{c2r8} were interested in the subject and proposed a "multi-hop" ridesharing method, that is to say, with vehicle changes, based on network modeling and graphical timetables. Similarly, using a graph whose network is only composed of drivers' geographic and temporal offerings, Herbawi et al. \cite{c2r9} stated a method where passengers had to find an acceptable path to go from their origin to their destination on this graph.
\subsubsection{Constraints of rideshare matching}
\paragraph{Spatio-temporal constraints}
~~\\ 	 
Classical ridesharing systems mainly rely on the matching of spatio-temporal constraints between a driver supply and a passenger request. Indeed, the travel date and time of both the passenger and the driver should be closely matched. In addition, it is mandatory that the passenger trip fits within to the driver itinerary. This does not exclude the fact that a small detour is tolerated to pick up or to drop off one or more passengers, depending on the motivation of the driver and the other passengers.
\\
Wen et al. \cite{r1} presented an approach that mined regular routes from the historical GPS trajectories of a user for ridesharing recommendations. In this work, only the origin and destination regions as well as the time property of each travel were taken into account to match driver /passenger candidates.  In this respect, an optimization model based on mixed continuous-integer linear programming was proposed in \cite{r2} to maximize the performance of dynamic ridesharing systems. This approach looked for the best path in the considered transportation network to minimize the difference between the desired departure and arrival times. 
\\
It is worth mentioning that other approaches have been interested in checking the tolerance in detour constraints for drivers. For instance, a routing optimization model for ridesharing a taxi was suggested in \cite{r3}. The objective of the proposed model was to minimize the travel cost and to maximize the passenger satisfaction. The latter was defined by the direct travel time, the extra riding time caused by ridesharing and the waiting time. Additionally, in \cite{r4} a matching algorithm for dynamic ridesharing based on network partitioning was presented. A route was expressed as a sequence of tiles which was referred to as a corridor. Only passengers, whose origin and destination were inside the corridor of an existing trip, were matched in possible ridesharing. To be matched, the additional time set by both the driver and the passenger associated to ridesharing had to be below a specific limit. Moreover, the approach in \cite{r5} dynamically matched trip requests to vehicles while satisfying two constraints: a waiting time defining the maximal time allowed between making the request and receiving the service and a service constraint, defining the acceptable extra detour time from the shortest possible trip duration. Interestingly enough, Schreieck et al. \cite{r6} put forward a matching algorithm for ride requests and offers that would check whether a driver could ride with a passenger without violating the maximum detour constraint they had set. In the same vein, Cici et al. \cite{r7} took as an input some information about the desired trajectories and spatio-temporal constraints of drivers and passengers and returned a matching that not only met individual user constraints but also maximized the total revenue for the system. A match would be feasible as far as the driver's tolerance in detour and the passenger's timeline were fulfilled.
\\
On the other hand, an approach that took into account both the maximal price the passenger was willing to pay for the service and the maximal waiting time before being picked up was defined in SHAREK \cite{r8}. Among the set of drivers that could provide such a service, SHAREK reported only the skyline element, i.e. maximal vector of these drivers according to the price and the waiting time.
\paragraph{Personal preferences}
~~\\ 
The aforementioned systems mainly focused on the improvement of the potential and performance of ridesharing to satisfy spatio-temporal constraints. However, a few approaches in literature have paid attention to social constraints to provide an optimal matching that would satisfy users' preferences.In this respect, the iCAP system \cite{r9} proposed a probablistic method that provided an optimal matching  of drivers and passengers' preferences. The system considered several parameters related to personal profiles such as smoking, gender, social behavior, as well as service related parameters like punctuality and itinerary cost, which increase the degree of reliability of the decisions reached. Despite the high-level and formal description of the iCAP functionality, the approach lacked detailled descriptions on the matching algorithm to find the best passenger.
\\
Furthermore, a topic based publish-Subscribe model, where the publisher was the vehicle rider and the subscriber stood for the ride seeker, was introduced in \cite{r10}. The system provided gender, smoking, age and marital status as preference options to riders and ride seekers. Rider and ride seekers had the freedom of choosing the importance of each preference, i.e. the preference level. The system would match users according to their preferences, but this did not imply a correspondence between the preference and the actual profile of the partner. In addition, no experimental evaluation of this approach was provided. 
\\
Moreover, the ride-sharing problem was formulated in \cite{r11} as a multi source-destination path planning problem, where each driver could generate sub-optimal paths according to their own requirements by suitably adjusting the weights of some factors such as time, occupancy, social strength or closeness. In this approach, a social closeness model was proposed, which incorporated the personalized preferences and used current social interests to bring the most similar people closer, raising the chance of matching. Feedback and reliability scores in the form of reputation were also incorporated in the model. 
\\
In the same breath, the stable roommates problem was adapted for the matching problem of one-on-one passengers (non-vehicle owners) \cite{r12}. A matching model was performed based on ridesharing preferences, which included personal preferences and travel cost savings. The factors of user preferences were personality and steadiness of user personality. Added to that, a web-based software tool for the management of a carpooling service called PoliUniPool was developed in \cite{r13}. The system allowed some social network functionalities; e.g., drivers were able to create "pre-arranged crews", and users might specify individuals they preferrd or disliked.
\\
Other systems have been linked with social networks like the SRSS system described by \cite{r14}. The latter assumed the existence of a social network data source in which users were connected by means of groups and interests and used the "strength of social connection" to prioritize matches. Mobile phones with positioning technologies were utilized for tracking and communication.
\\
Besides, an auction model was developed in \cite{r15} to tradeoff the minimization of vehicle kilometers travelled, i.e. greenhouse emissions, with the overall probability of successful rideshares. This model permitted users to set some preferences such as user ratings and social network status other than the travel distance and time.
\subsubsection{Optimization objectives}
With regard to the optimization objectives targeted by ridesharing systems, most aim to reduce the total distance traveled or more precisely the total trip time. On the other hand, some researchers are proposing alternative methods. Kleiner et al. \cite{r15} proposed a system based on auctions. Depending on the detour to be made, users set prices and drivers choose knowingly their match based on the price offered and detour to achieve. In this way, ridesharing alternatives were naturally regulated according to supply and demand. Another category of objectives is to perform a multi-objective optimization. This is to simultaneously optimize a list of objectives. In this case, not all objectives are optimal at the same time. Nevertheless, the idea is to be in a position where all the values of the objectives are at the so-called Pareto optimality \cite{c2r11}; this means that at this point, any attempt to improve one objective will automatically lead to the deterioration of another objective. For example, Herbawi et al. \cite{c2r9} proposed a ridesharing system whose three objectives to be minimized are the cost and the duration of the trip as well as the number of vehicles taken by a passenger to complete the trip.
\\
On the other hand, systems that fulfill social preferences, had incorporated preferences into a single objective function, called the weighted sum function. The optimization problem has been reduced to the optimization of this function. However, the weakness of this model stands in the process of summarizing the different criteria. Through combining various dimensions, and consequently multiple units, the weighted sum model assumption has been violated and the result has been equivalent to "adding apples and oranges" \cite{r16}. Since the satisfaction of user preferences has to deal with multi-constraints with different units, the weighted sum model cannot be used without major changes in its strategy. In our approach, we apply a Multi-Criteria Decision Making (MCDM) method to tackle this issue. To introduce the concept, a thorough study of the decision support process and its products in the case of multi-criteria decision support will be presented in the next section.
\\
To better optimize our system, we aim afterwards to maximize the objective of each individual participant. Our objective is to create a stable matching of drivers and passengers such that there does not exist any pair of a driver and a passenger who prefers each other to their current partners. This notion of stability is analogous to that defined in the well-known stable marriage problem, which will be highlighted in the last section of this chapter.

\section{Multi-Criteria Decision Making}
\label{sec:1}
In this section, we will focus on the first methodological tool that will be used in our approach for the optimization  of a ridesharing system. In order to build a multi-criteria evaluation model, we rely on Multi-Criteria Decision Making (MCDM) method which is considered one of the most important branches of operational research and decision theory  \cite{hamrouni2008succinct}.

\subsection{Decision support}
\label{sec:11}
It is common in a decision support study to have to take into account several points of view to compare the relative attractiveness of the various actions likely to solve the problem of decision considered.
Decision support uses techniques and methodologies from the field of applied mathematics such as optimization, statistics, and decision theory as well as less formal domain theories such as organizational analysis and cognitive science \cite{gasmi2007extraction}.
\\
Roy et al. \cite{c3r1} define decision support as: "The activity of the person who, based on models clearly explained but not necessarily completely formalized, helps to obtain elements of answer to the questions posed by  a decision maker, elements that help to inform the decision and normally to prescribe a behavior likely to increase the coherence between the evolution of the process on the one hand, the objectives and the value system in the service of which this decision maker is placed on the other hand. "
\\
Thus defined, decision support is only part of the search for a truth. Theories or, more simply, the methodologies, the concepts, the models, the techniques on which it is based, have, most often, a different ambition: to reason the change that prepares a decision-making process in order to increase its coherence with the objectives and the value system of the actor for whom or on whose behalf the decision-making aid is exercised \cite{bouker2012ranking}.
\\
Indeed, Martel \cite{c3r1} supports the fact that a decision-making activity "implies a minimum of insertion in the decision-making process: it is done essentially with the decision makers in the establishment of a real relation of help". For Roy et al. \cite{c3r1}, a decision issue is not an object that preexists; The wording given to it cannot, in general, be totally objective and cannot be considered independently of the relationship between the individual and reality. 
\\
In this sense, Landry \cite{c3r3} notes that the success of a decision-making process in an organization requires an understanding of the entire decision-making process in which this support is embedded, which implies an ability to adequately grasp the problem that justifies the origin and supplies this process later.

\subsection{Multi-criteria decision analysis}	
\label{sec:12} 
In a decision-making process, when constructing the evaluation model, it is rare to lead to only one criterion corresponding to a single point of view on which the decision-maker expresses his preferences \cite{ferjani2012formal} \cite{hamdi2013trust} \cite{yahia2004revisiting}. It is therefore necessary to consider several points of view (costs, human resources, safety, environment, etc.) in the subsequent construction of the evaluation model \cite{bouzouita2006garc}. The decision in the presence of multiple criteria is difficult because the criteria are often conflicting. Multi-criteria decision support was then developed to offer both an approach and tools of solutions to complex decision-making problems \cite{hamrouni2010generalization}. Multi-criteria analysis is now considered one of the most important branches of operations research and decision theory \cite{brahmi2012omc} \cite{jelassi2014efficient}.
\\		
Technically, multi-criteria decision support is developed to deal with several classes of decision problems (choice, sorting, description, ranking ...) while considering several criteria, often conflicting and not commensurable, while seeking to best model the preferences and the values of the decision maker \cite{ayouni2011extracting} \cite{hamrouni2008succinct}. Also, Vincke \cite{c3r7} defines multi-criteria decision support as: "Multi-criteria decision support is intended, as the name implies, to provide a decision-maker with tools to progress in resolving the decision-making problem, where several, often contradictory, points of view must be taken into account."

\subsection{MCDM in ridesharing systems} 
\label{sec:14}
Although MCDM methodology has gained more and more appreciation and popularity among transportation researchers, few studies proposed a multi-criteria approach to investigate the optimization problems related to ridesharing system \cite{DBLP:conf/ictai/YahiaN04}. Worthy of mention, the work of Filcek et al. \cite{r20} where a model of the joint problem of matching carpoolers and routes planning as a multiple criteria optimization problem was proposed. The AHP method is used to collect preferences of the carpoolers and involve it to compute aggregated cost function for common routes of the driver and passengers assigned to him, to obtain finally the best routes presented to drivers. A Multi-criteria Decision Support System MDSS is proposed in \cite{r21} in order to create an intelligent tool for carpooling. In the context of strategic decision making, AHP and ELECTRE methods are integrated to solve the problem that optimizes the total revenue of drivers based on the car's capability and the time schedule. Li et al. \cite{r22} employ a method combining AHP and GIS with big data to determine the optimal locations of future stations. The AHP method is used to determine the weights of the decision criteria which are potential users, potential travel demand, potential travel purposes and distances from existing stations. Moreover Awasthi and Chauhan \cite{r23} present a hybrid approach for evaluating environment-friendly transport solutions. AHP method is used to structure and rate the measures for transport sustainability evaluation. All the above mentioned approaches have used the AHP method. The TOPSIS method is also kenned as one of the most popular methods among MCDM methods. In our approach, we propose to apply it to rank possible matches for each user according to his social preferences. These preference lists are then used to compute the stable matching.

\section{Stable marriage}
\label{sec:2}
The ridesharing problem is treated as a matching problem where drivers are assigned to passengers. To address this problem, we proposed a strategy based on the Stable Marriage Problem SMP. So it is important to describe this theorem before presenting our approach.
\subsection{History}
The analysis of correspondence mechanisms is based on an abstract idea: If rational individuals, who know their interests and behave accordingly, simply engage in unrestricted mutual exchanges, then the result should be effective. If this is not the case, some people are developing new exchanges that would be more favorable to them. A correspondence where the individuals concerned perceive no interest or gain in making further exchanges is called "stable" \cite{c4r1} \cite{c4r2}. The notion of stability is a central concept in cooperative game theory that is considered an abstract area of mathematical economics that aims to determine how a group of rational individuals can choose a correspondence while cooperating with each other. Lloyd Shapley is considered the leading architect of this branch of game theory by developing his main concepts in the 1950s and 1960s \cite{c4r3}.
\\
The foundations for the theoretical framework were established in 1962, when David Gale and Shapley Lloyd published a short article on a class of correspondence problems \cite{c4r2}. They considered a model of two sets of agents: workers and firms, which must be matched with each other. If a worker is hired by employer A, but this worker would have preferred employer B, who would also have liked to hire this worker (but did not do so), then there are untapped gains from this exchange. If employer B hired this worker, both would have had a better arrangement. Gale and Shapley proposed a delayed acceptance procedure that always leads to a stable match. The procedure shows how agents on one side of the market (e.g. employers) make offers to those on the other side, who accept or reject these offers according to certain rules.
\\
The empirical relevance of this theory was later recognized by Alvin Roth in 1984 \cite{c4r1}, Roth and Vate \cite{c4r5}, Roth and Peranson \cite{c4r6}. Roth found that the US market for new resident doctors has always suffered from a series of failures due to the poor matching of residents to hospitals, and that a centralized clearinghouse has improved the situation by a procedure essentially equivalent to the delayed acceptance procedure of Gale and Shapley.
Subsequently, Roth and his colleagues used this theory, in combination with empirical studies, controlled laboratory experiments, and computer simulations, to examine how other markets work \cite{c4r7}. Their research has not only informed the functioning of these markets, but has also proven useful in designing institutions that help markets work well, by implementing a version or extension of the Gale and Shapley procedure. This led to the emergence of a new and vigorous branch of the economy called "Market Design".
The stable marriage problem and its variants have been extensively studied in combinatorial optimization and game theory \cite{c4r1}. In addition, as real applications, matching programs have been established in several areas, we can mention: organ donors to patients \cite{c4r8}, CARMS (Canadian Resident Matching Service) in Canada \cite{c4r9} and JRMP (Japan Residency Matching Program) in Japan \cite{c4r10}.

\subsection{Principle}
\begin{figure}[h!]
\centering
\includegraphics[scale=0.5]{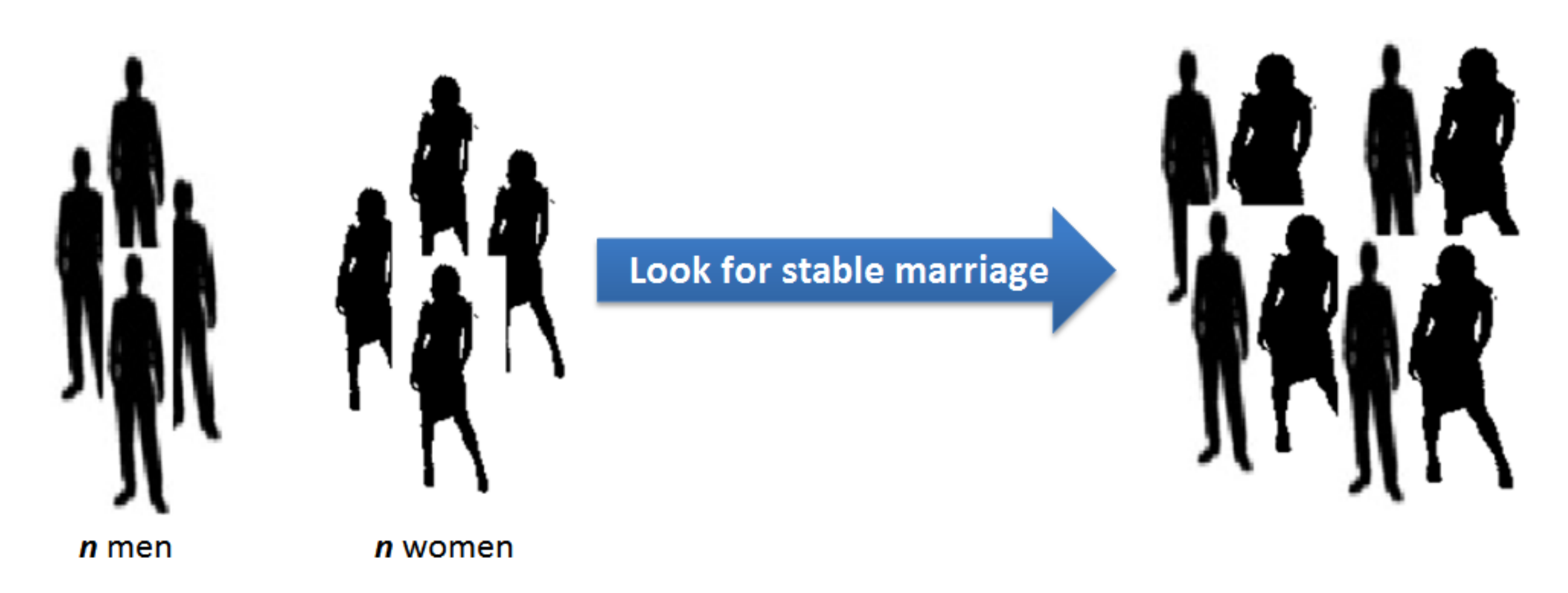}
\caption{Stable marriage problem} \label{f18}
\end{figure}

\begin{figure}[h!]
\centering
\includegraphics[scale=0.5]{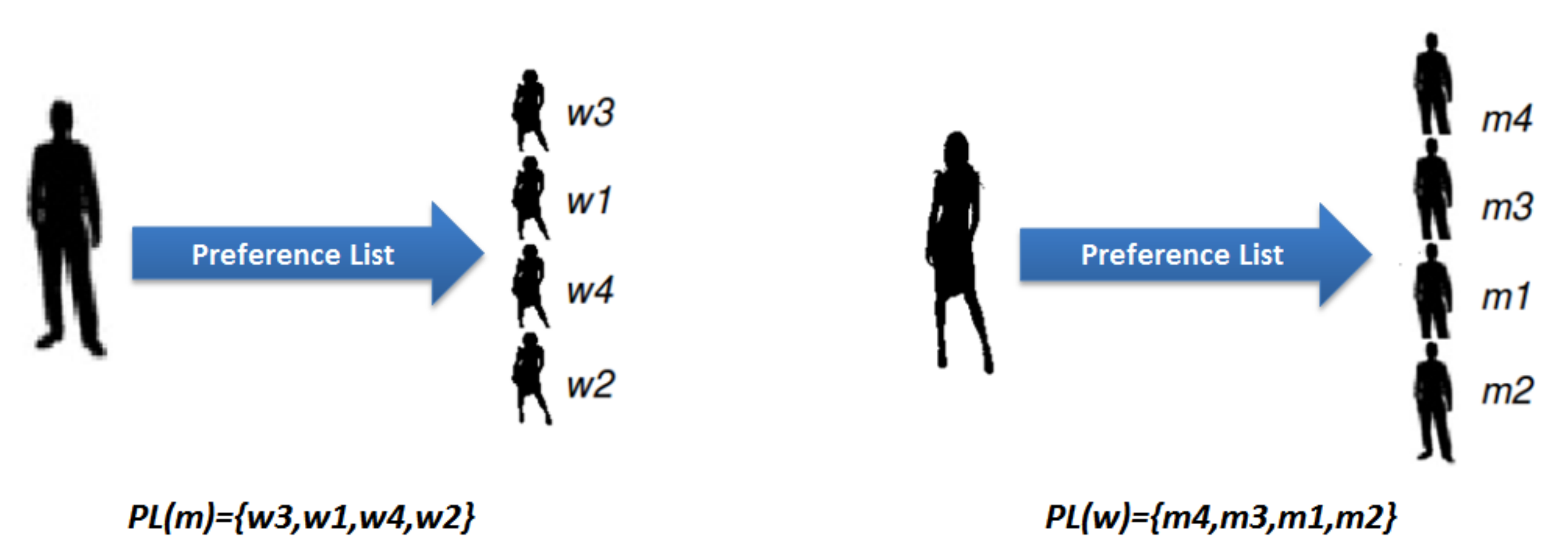}
\caption{Preference list} \label{f19}
\end{figure}

\begin{figure}[h!]
\centering
\includegraphics[scale=0.5]{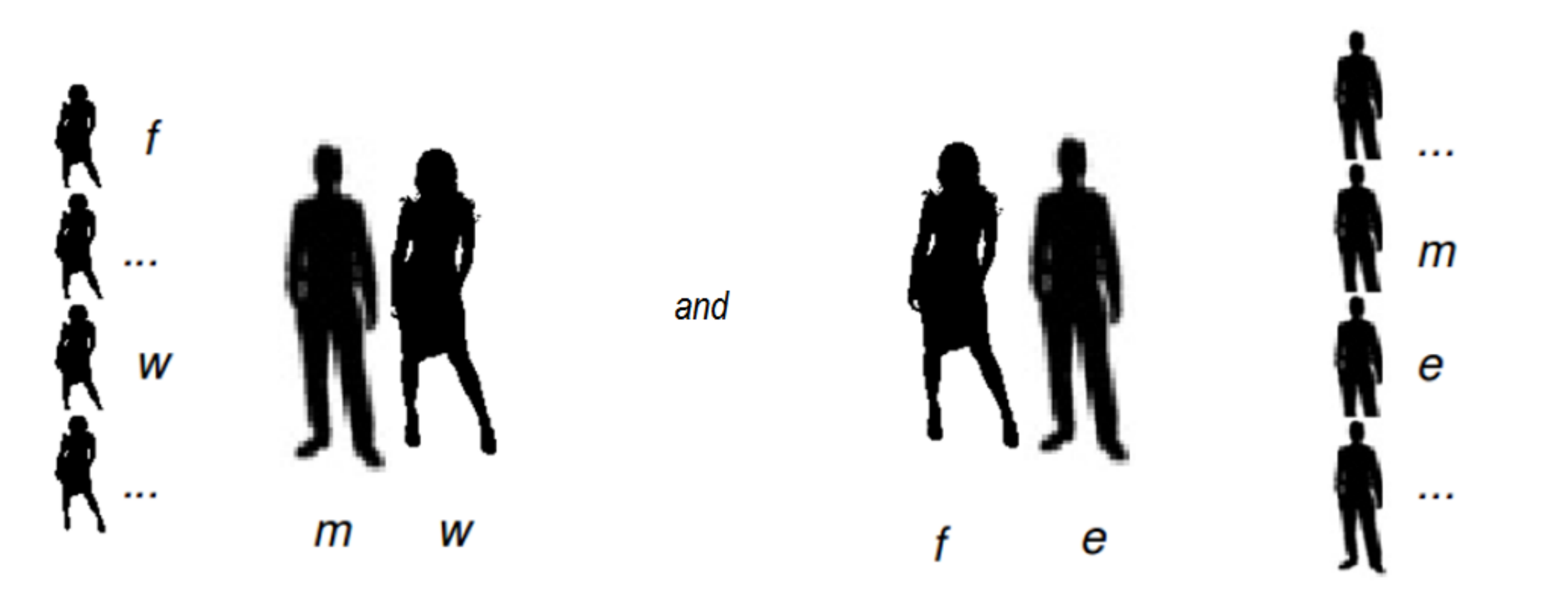}
\caption{Blocking pair (m,f)} \label{f20}
\end{figure}

An SMP is a combinatorial problem. It consists in looking for a set of stable marriages between $n$ men $M = \{m_1, m_2, ..., m_n\}$ and $n$ women $W = \{w_1, w_2, ..., w_n\}$ (Fig.\ref{f18}). Each individual has his or her Preference List $PL$ where all members of the opposite sex appear sorted according to their affinity with them. If man prefers woman $w$ to woman $w'$, we write $w \succ_{m} w'$ (Fig.\ref{f19}). 
\paragraph{Problem:}
Create the couples man-woman the best possible according to their preferences.
\\
A match $\mu$ is a subset of the $M*W$ product where each person appears once and only once. If $(m, w) \in \mu$, then $m$ and $w$ form a matched pair in $\mu$ and $\mu(m)=w$ and $\mu(w)=m$. If $m$ and $w$ are not matched in $\mu$, then $(m, w)$ is an unpaired couple. $m$ and $w$ form a blocking pair if $m$ prefers $w$ to $\mu(m)$ and $w$ prefers $m$ to $\mu(w)$, but $m$ and $w$ form an unpaired couple in $\mu$ (Fig.\ref{f20}).
\\
If the match contains no blocking pair, $\mu$ is a stable match. 
\paragraph{Example}
~~\\
\begin{em}
An example of SMP of size $n= 4$ appears in Table \ref{tab5} where men are labeled $m_1, m_2, m_3$ and $m_4$, women are labeled $w_1, w_2, w_3$ and $w_4$ and the following preference lists. Preference lists are in decreasing order, the most-preferred partner is on the left.
\begin{table}[h!]
    \begin{center}

    \begin{tabular}{| >{\centering\arraybackslash}m{2in} | >{\centering\arraybackslash}m{2in} |}

    \hline

    {\bf Men} & {\bf Women}   \\ \hline
    $PL(m_1): w_2, w_4, w_1, w_3$ & $PL(w_1): m_2, m_1, m_4, m_3$ \\ \hline
    $PL(m_2): w_3, w_1, w_4, w_2$ & $PL(w_2): m_4, m_3, m_1, m_2$ \\ \hline
    $PL(m_3): w_2, w_3, w_1, w_4$ & $PL(w_3): m_1, m_4, m_3, m_2$ \\ \hline
    $PL(m_4): w_4, w_1, w_3, w_2$ & $PL(w_4): m_2, m_1, m_4, m_3$ \\ \hline
  \end{tabular}
  \caption{A SMP instance of size $n=4$ }\label{tab5}
  \end{center}
\end{table}

The matching $(m_1, w_4), (m_2, w_3), (m_3, w_2), (m_4, w_1)$ is stable. This matching is defined as a set of ordered pairs (man, woman) and its stability can be verified by considering each man in turn as being a member of a blocking pair. The man $m_1$ can form a blocking pair with the woman $w_2$ he prefers to his partner $w_4$, but $w_2$ prefers his current partner $m_3$ to $m_1$. $m_2$ and $m_3$ are matched to their favorite women, so none of them can be in a blocking pair. Finally, $m_4$ can form a blocking pair with $w_4$, but she prefers to stay with her current partner $m_1$.
\\
Another possible stable match is $(m_1, w_4), (m_2, w_1), (m_3, w_2), (m_4, w_3)$. The stability of this matching can be verified in the same way. On the other hand, the correspondence $(m_1, w_1), (m_2, w_2), (m_3, w_3), (m_4, w_4)$, for example, is unstable because of the blocking pair $(m_1, w_4)$. $m_1$ prefers $w_4$ to his current partner and reciprocally, $w_4$ prefers $m_1$ to his current partner. Other unstable matches may have more than one blocking pair; for example, the matching $(m_1, w_1), (m_2, w_2), (m_3, w_4), (m_4, w_3)$ has six blocking pairs: $(m_1, w_2)$, $(m_1, w_4)$, $(m_2, w_1)$, $(m_2, w_4)$, $(m_3, w_2)$, $(m_4, w_4)$.
\end{em}

\paragraph{Definition:}
~~\\
Formally, we say that a marriage is stable iff:
\\
$\forall i$, $1 \leqslant i \leqslant n$\\
$M = \{m_i\}$ set of men\\
$W = \{w_i\}$ set of women\\
$\forall m_i,  \exists  PL (m_i) = W$\\
$\forall w_i,  \exists  PL (w_i) = M$\\
\paragraph{}
$\nexists (m_i,w_i) \in M \cup W $ with $\mu(m)\neq w, w \succ_{m} \mu(m)$ and $m \succ_{w} \mu(w)$.\\

David Gale and Lloyd Shapley proved that, for any equal number of men and women, it is always possible to solve the SMP and make all marriages stable. They proposed the Gale-Shapley algorithm \cite{c4r2} (see Algorithm \ref{alg:gs}) to do so.
 \begin{algorithm}
\caption{Gale-Shapley algorithm}
\label{alg:gs}
\begin{algorithmic}[1]
\State Initialize all $m \in M$ and $w \in W$ to free
   \While {$\exists$ free man $m$ who still has a woman $w$ to propose to}
    \State $w = $ first woman on $m$'s list to whom $m$ has not yet proposed
    \If{$w$  is free}
          \State $(m, w)$ become engaged
	\Else{ some pair $(m', w)$ already exists}
		\If{$w$ prefers $m$ to $m'$}
          \State $m'$  becomes free
          \State $(m, w)$ become engaged
	      \Else
		  \State $(m', w)$ remain engaged
	    \EndIf
	\EndIf
    \EndWhile
\end{algorithmic}
\end{algorithm}

\subsection{Stable marriage problem in ridesharing}
With regard to ridesharing systems, the problem of stable marriage is integrated in only one approach presented by Wang in \cite{c4r11}.  Wang assumed that the number of passengers equals that of drivers and relied on the simplest approach of stable marriage defined by Gale and Shapley in \cite{c4r2}. However, in real life, considering such assumption is the furthest from the reality. For that, in our approach we associate the problem of stable marriage with the notion of stability in the matching problem but we rely on stable marriage assignment for unequal sets presented in \cite{c4r12}. Furthermore, the preference in his assumption solely depends on the potential financial benefits, i.e., the cost savings as compared to driving alone, where a higher saving implied a higher preference. In our approach, we incorporate personal preference to ensure social comfort when sharing a private space with others.

\section{Conclusion}
An exploratory study was conducted on ridesharing to better understand this alternative transportation by making a state of the art of these services, and focusing on some innovative academic work for modeling and optimization of such systems. This study has led us to define the major axes leading to the implementation of the foundations of our approach. These main axes were, in fact, inspired by the problems and gaps that emerged from the study of existing approaches. A detailed description of our approach is then provided in the following chapter.

\chapter{ SMRM: A Stable Multi-Criteria Rideshare Matching}
\section{Introduction}
In the previous chapter, we presented the theoretical foundations on which our contribution is based. In this chapter, we introduce SMRM, a Stable Multi-Criteria Rideshare Matching system. Indeed, SMRM selects the matches that promise the satisfaction of drivers and passengers' social preferences, considering the notion of stability for rideshare matches.\\
In the second section of this chapter we illustrate the global idea of our approach. The entire process followed from the receipt of the first rideshare request to the generation of judgment matrices is detailed through the third section. The implementation details of the multi-criteria concept in our approach are then outlined in the level of the fourth section. Subsequently, the stable matching problem is described and formalized in section ~\ref{sec:c5}. Finally, section ~\ref{sec:c6} summarizes and concludes this chapter.
\section{System overview}
\label{sec:c2}
\subsection{Problematic}	
Although there are many attempts to provide a ridesharing system, they do not meet the expected success, given the lack of motivation of individuals to go to such systems. This lack of enthusiasm for this practice is explained in particular by the boredom of traveling with an unknown person. Indeed, for some, the car is a personal and intimate space, where he is free to do what he wants. In this sense, listening to the radio, singing or phoning via a headset is akin to private activities, which we cannot or hardly share with others. Therefore, sharing a private space as the vehicle is not simple, especially if tastes and habits are different. Passengers and drivers often aspire to have a pleasant ridesharing time and this is not always the case if they come across people with different social characters. Everyone has their own personality and their moods, and instead of ensuring participants' compatibility, classical ridesharing systems mainly focus on the improvement of their potential and performance in ridesharing to fulfill spatio-temporal constraints and to maximize certain objectives. In addition, a rideshare matching solution aiming to maximize the total system objective or the total number of matches may not necessarily maximize the objective of each individual participant.
\subsection{Proposed approach}
In this work, we introduce a new ridesharing system that we call SMRM, an acronym for Stable Multi-Criteria Rideshare Matching. SMRM avoids the drawbacks of previous approaches. Indeed, it selects the matches that promise the satisfaction of drivers and passengers' social preferences, considering the notion of stability for rideshare matches. A matching is stable if there is no pair of participants who both prefer each other to their partners who are matched to.
\subsubsection{Business case}
\begin{figure}[H]
\centering
\fbox{
\includegraphics[scale=0.8]{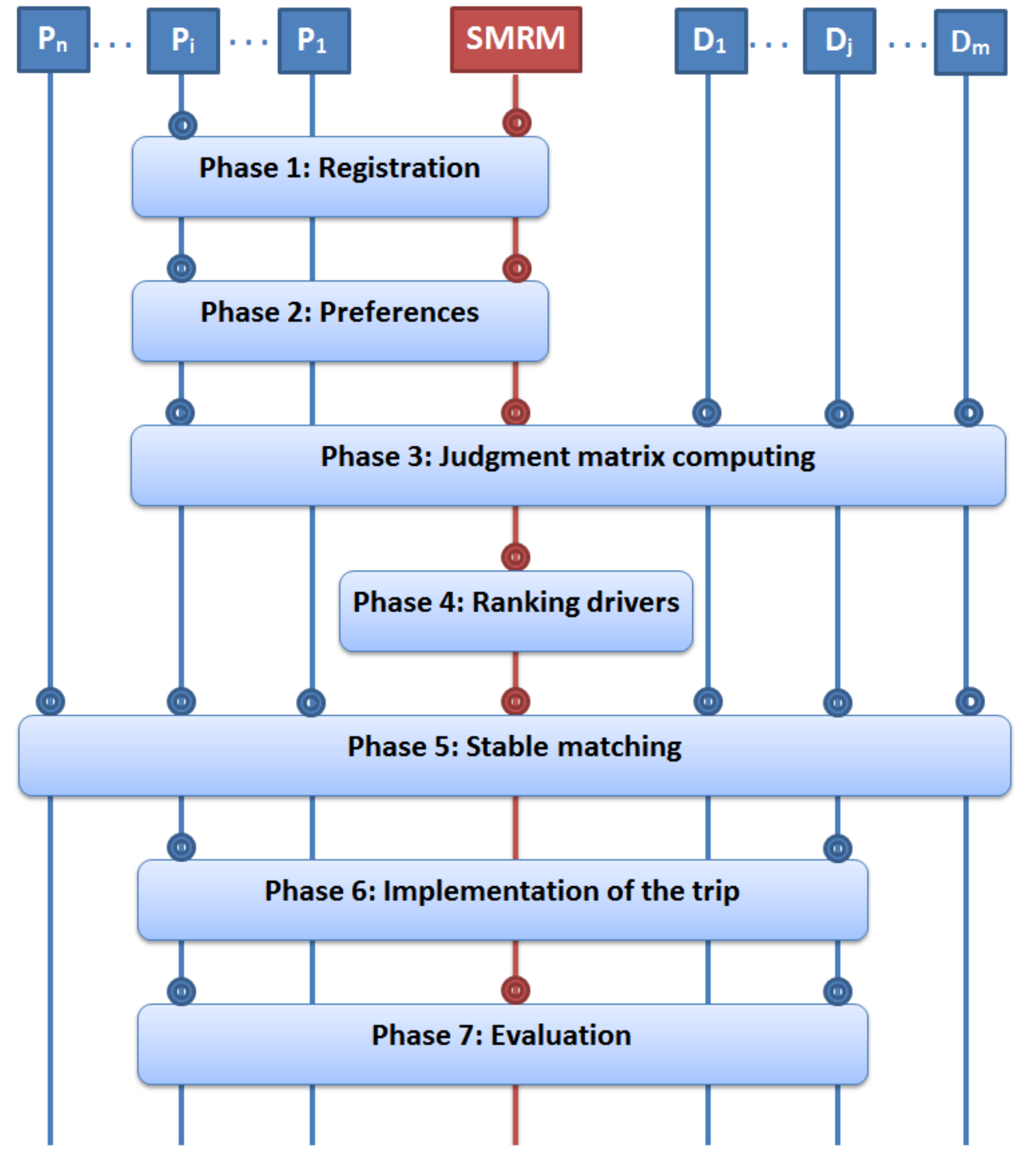}}
\caption{High-level SMRM scenario} \label{f24}
\end{figure}

A high-level scenario corresponding to the business case is shown in Fig.\ref{f24}.\\
The business case assumes that a person wishing to reach a destination calls SMRM service, seeking for a correspondent being directed towards the same place subject to given personal constraints. 
The scenario can be either passenger-, or driver-initiated and evolves in seven main phases. Examining the passenger $P_i$ initiated scenario (for brevity, the description of the scenario initiated by the driver is omitted), in the first phase, he is prompted to register himself/herself providing their personal details such as name, date of birth, gender, status, telephone number... as well as additional personal profile attributes such as smoking, music, pet friendly and vehicle range.  Afterward, $P_i$ is asked to complete a form indicating his personal preferences and the weight of each preference.
\paragraph{Remark:}
~~\\
\begin{em}
In the case the user is already registered, he can immediately submit a trip offer or a trip request. On the other hand, a change of user's profile is possible, the system will have to be able to update the appropriate parameters and adapt to the needs of the user. For example, in the case where he wants to change his address or telephone number, he can update this information by himself.
\end{em}
\paragraph{}
SMRM computes in the first step the judgment matrices for all drivers that satisfy the spatio-temporal constraints of $P_i$. To do that, it accesses to the latter's personal preferences, the weight of each preference, and the personal profile of each driver. \\
Subsequently, according to these judgment matrices, SMRM evaluates the multiple constraints simultaneously, ranks possible matches for this passenger and provides their preference list where all drivers appear sorted according to their affinity with them.\\
In the fifth phase, based on the preferences lists of all concerned users (passengers and drivers), the system returns the best possible pair according to their preferences. A pair of a stable matching in which there is no unmatched passenger and driver both prefer each other to their current correspondent. We suppose here that SMRM returns the driver $D_j$ as a correspondent of $P_i$. \\
The sixth phase refers to the implementation of the trip, while the final phase, which takes place after the completion of the trip, consists in the feedback provided by both parties for future reference.

\subsubsection{System Architecture}
\label{sec:c22}
\begin{figure}[H]
\centering
\fbox{
\includegraphics[scale=0.6]{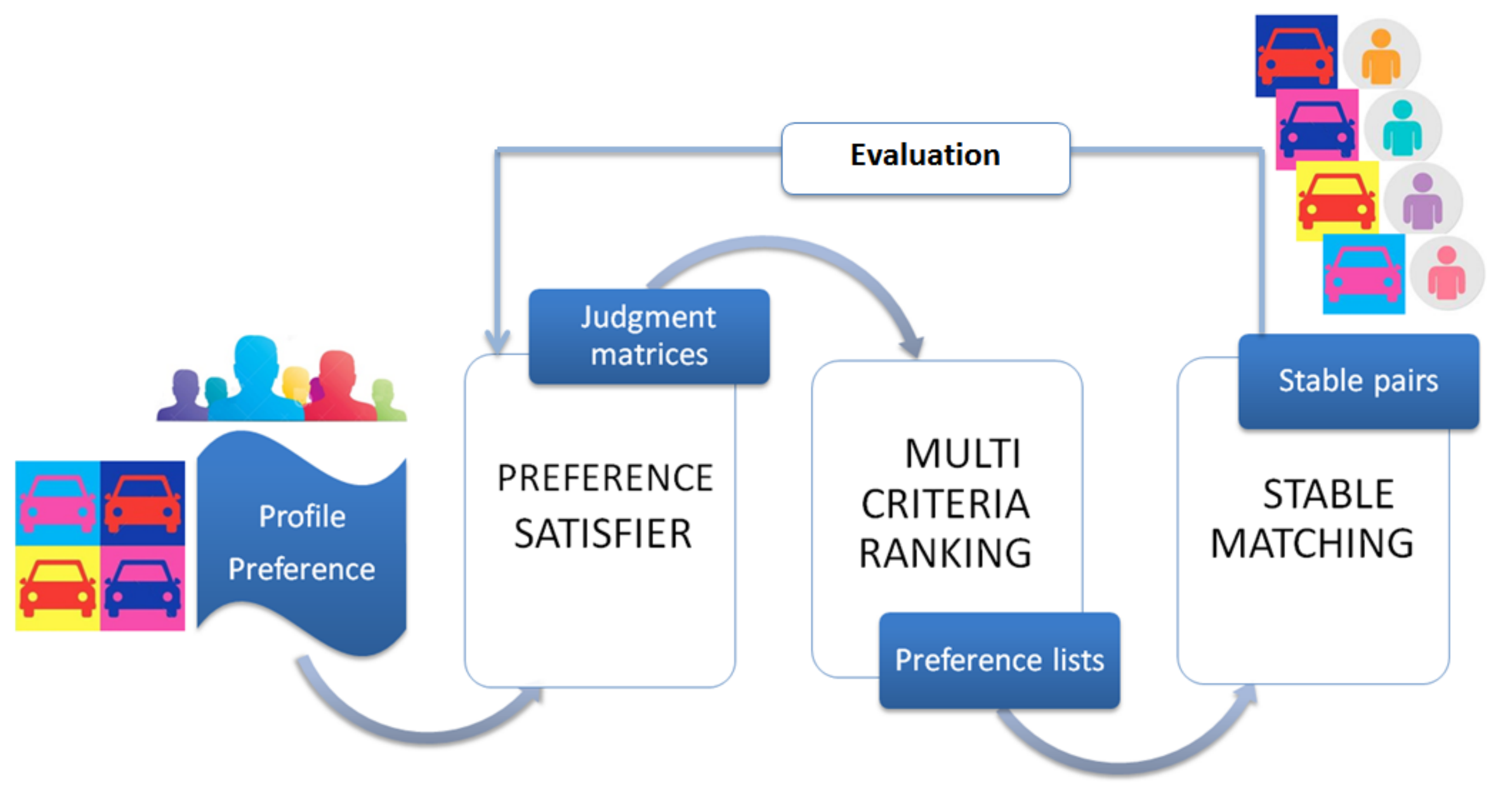}}
\caption{SMRM architecture at a glance} \label{f22}
\end{figure}

Fig.\ref{f22} shows the architecture of the SMRM system, which consists of three main components: Preference Satisfier; Multi-Criteria Ranking; and Stable Matching. The three latter components are thoroughly described in the following.

\paragraph{Preference Satisfier:}
This component receives drivers and passengers' preferences and then generates a judgment matrix for each user. This matrix represents the evaluation obtained from each correspondent with respect to the preferences of this utilizer. These matrices are computed based on preferences and weights of the user as well as on the profile and received evaluations of each correspondent.
\paragraph{Multi-Criteria Ranking:}
This component takes as an input the judgment matrices and matches after that each driver (resp. passenger) to a preferred list of passengers (resp. drivers). We formulate the problem as an MCDM problem and we adapt the Technique for Order Preference Similar to an Ideal Solution (TOPSIS) method as a correspondents ranking tool to evaluate the multiple constraints simultaneously.
\paragraph{Stable Matching:}
This component takes as an input a set of drivers $D$ and a set of passengers $P$ such that each driver $d \in D$ has a list of preferred passengers from $P$ and each passenger $p \in P$ has a list of preferred drivers from $D$. It returns a stable matching where there is no pair of participants who both prefer each other to the partners who they are matched to. After the completion of a ride, passengers and drivers could leave mutual feedback. The latter influences the computation of judgment matrices in the first component.
\section{Preference Satisfier}
\label{sec:c3}
This section develops the first component presented in Section ~\ref{sec:c22} and shown in Fig.\ref{f22}, as well as its operation and interactions. 
\subsection{User profile attributes}	
The attributes that represent the data associated with the user profile include identification data and matching judgment data. Identification data contain personal information about the user that makes it unique and identifiable by the system like CIN, surname, first name, phone, email, address, etc. Judgment data are necessary to judge the matching (provided in Table \ref{tab10}), they contain information concerning the gender, the age, the marital status, the vehicle range, whether the user is a smoker or not, whether he takes animals with him or not and whether he hears music or not. \\
Other judgmental data are the social behavior, driving skills and reliability. This information is inferred through the evaluation procedure in the system. This procedure is carried out by the drivers concerning passengers and vice versa at the end of each trip. It offers users (the driver and the passenger) the opportunity to complete a short questionnaire, evaluating the correspondent. The questions aim to extract the user's opinion of these parameters.\\

\begin{table}[h!]
    \begin{center}

    \begin{tabular}{| >{\centering\arraybackslash}m{1in} | >{\centering\arraybackslash}m{1in} | >{\centering\arraybackslash}m{1in} | >{\centering\arraybackslash}m{1in} | >{\centering\arraybackslash}m{2in} |} \hline
    {\bf attribute} & {\bf Notation} & {\bf Source} & {\bf Value} & {\bf User}   \\ \hline
    Gender	& Gender & Input & M / F &	Passenger \& Driver \\ \hline
    Age	&Age&	Input	&Integer $>18$&	Passenger \& Driver \\ \hline
Marital status&	Status	&Input&	Married / single	&Passenger \& Driver \\ \hline
Vehicle range&	VHrange&	Input&	Basic / Comfort / Luxury	&Driver \\ \hline
Pet friendliness	&Pets	&Input&	Yes / No&	Passenger \& Driver \\ \hline
Listening to music	&Music	&Input&	Yes / No	&Passenger \& Driver \\ \hline
Smoking habit&	Smoking	&Input	&Yes / No&	Passenger \& Driver \\ \hline
Social behavior	&SocialBehavior	&Feedback	&$[0..10]$	&Passenger \& Driver \\ \hline
Driving skills&	DrivingSkills	&Feedback&	$[0..10]$	&Driver \\ \hline
Reliability	&Reliability&	Feedback	&$[0..10]$&	Passenger \& Driver \\ \hline
  \end{tabular}
  \caption{Judgment data}\label{tab10}
  \end{center}
\end{table}
It is important to make evaluation easily understandable and factual for users, and it is the system that converts each evaluation category into a score. In this way, after each trip, the system updates this information by assigning the score of the evaluation and calculating the new obtained average. This information makes it possible to create more successful matches in the future.
The evaluation attributes, their potential values and their corresponding scores are summarized in Table \ref{tab11}.

\begin{table}[h!]
    \begin{center}
    \begin{tabular}{| >{\centering\arraybackslash}m{1in} | >{\centering\arraybackslash}m{2in} | >{\centering\arraybackslash}m{2in} |} \hline
    {\bf attribute} & {\bf Value} & {\bf Score}    \\ \hline
\multirow{3}{*}{SocialBehavior}&	Friendly &	10 \\ \cline{2-3}
	&Polite&	5\\ \cline{2-3}
	&Rude&	0\\ \hline
\multirow{3}{*}{DrivingSkills}&	Efficient	&10\\ \cline{2-3}
	&Acceptable&	5\\ \cline{2-3}
	&Dangerous	&0\\ \hline
\multirow{3}{*}{Reliability}&	Extremely reliable &	10\\ \cline{2-3}
	&moderately Reliable&	5\\ \cline{2-3}
	&Not reliable&	0\\ \hline
  \end{tabular}
  \caption{Evaluation attributes}\label{tab11}
  \end{center}
\end{table}

\paragraph{Remark:}
~~\\
Vehicle range and driving skills data are optional, since a system user may not own a car, but may only use the SMRM service as a passenger.
\subsection{User preference attributes}		
Social constraints are important to define and agree between users. In this respect, passengers and drivers are asked to indicate their personal preferences. The attributes that represent the preferences associated with the correspondent are the age, the gender, the marital status, the vehicle range, smoking habit, pet friendliness, listening to music.\\
More specifically, each user indicates the preferred age, as well as the age tolerance range in which he wishes his correspondent to be. Usually, the gender of the user makes no difference, but sometimes the gender can help people to have common interests so the chance of having something to discuss increases and makes the trip more enjoyable. Being married or single affects at a large extent whether the user will accept a match or not. For example, mothers may have many common interests in their children, which they could not have done by traveling with a single person. Smoking habit can be a very important parameter for a user and being in agreement on this point creates higher possibility for successful journey.\\
Shared music preferences create and intensify social bonds and finally leads to the social attraction. Indeed, listening to music and talking about one's favorite music is a great conversation starter when meeting new people. Pet friendliness is also a factor that influences the users' correspondence and their ability to make a pleasant trip. Participant's concern about vehicle range is because people tend to be in a comfortable and convenient atmosphere when traveling which affects the acceptance of correspondence. \\
With regard to social behavior, driving and reliability attributes, the user also specifies his weight but he does not attribute a value of preference. This means that the user is able to indicate the level of interest on this attribute, or it does not matter to him.
\\
A summary of the user preference attributes is provided in \ref{tab12}.

\begin{table}[h!]
    \begin{center}

    \begin{tabular}{| >{\centering\arraybackslash}m{1in} | >{\centering\arraybackslash}m{2in} | >{\centering\arraybackslash}m{1in} | >{\centering\arraybackslash}m{2in} |} \hline
    {\bf Attribute} & {\bf Notation} & {\bf Value} & {\bf User}   \\ \hline
    Gender	& GenderPref &  M / F &	Passenger \& Driver \\ \hline
    Age	& AgePref &	Integer $>18$ &	Passenger \& Driver \\ \hline
 Age tolerance	& AgeTolerance &	$[1..20]$ &	Passenger \& Driver \\ \hline
Marital status&	StatusPref	&	Married / single	& Passenger \& Driver \\ \hline
Vehicle range &	VHrangePref &	Basic / Comfort / Luxury	& Passenger \\ \hline
Pet friendliness	& PetsPref	&	Yes / No&	Passenger \& Driver \\ \hline
Listening to music	& MusicPref	&	Yes / No	&Passenger \& Driver \\ \hline
Smoking habit&	SmokingPref	& Yes / No &	Passenger \& Driver \\ \hline
Social behavior&	SocialBehaviorPref	& Not indicated &	Passenger \& Driver \\ \hline
Driving skills&	DrivingSkillsPref	& Not indicated &	Passenger \\ \hline
Reliability &	ReliabilityPref	&	Not indicated  & Passenger \& Driver \\ \hline
  \end{tabular}
  \caption{User preference attributes}\label{tab12}
  \end{center}
\end{table}

\subsection{User weight attributes}								
In addition to the above attributes, a SMRM user must specify the importance that he assigns to each of his preferences. This is achieved by assigning each attribute with a certain weight. Of course, it is possible that the attributes have the same weight for the user. For example, a user may consider just as important that his passenger listens to music, as well as whether he is married or not. Practically, the user assigns each attribute a value between $0$ and $10$, $0$ meaning that the user ignores this preference and $10$ pointing for great importance. In the case where a attribute has the same weight with another (or more than one) attribute, it is deduced that the user has an equal interest in these attributes.\\
The attributes that represent the weight vector are the age, the gender, the marital status, the vehicle range, smoking habit, pet friendliness, listening to music,social behavior,driving Skills and reliability.\\
The following table presents these attributes.
\begin{table}[h!]
    \begin{center}

    \begin{tabular}{| >{\centering\arraybackslash}m{1.5in} | >{\centering\arraybackslash}m{1.5in} | >{\centering\arraybackslash}m{2in} |} \hline
    {\bf Attribute} & {\bf Notation} & {\bf User}   \\ \hline
    Gender	& GenderW &	Passenger \& Driver \\ \hline
    Age	& AgeW  &	Passenger \& Driver \\ \hline
Marital status &	StatusW	& Passenger \& Driver \\ \hline
Vehicle range &	VHrangeW &	 Passenger \\ \hline
Pet friendliness	& PetsW	&	Passenger \& Driver \\ \hline
Listening to music	& MusicW	&Passenger \& Driver \\ \hline
Smoking habit&	SmokingW	&	Passenger \& Driver \\ \hline
Social behavior&	SocialBehaviorW	&	Passenger \& Driver \\ \hline
Driving skills&	DrivingSkillsW	&	Passenger \\ \hline
Reliability &	ReliabilityW	&	Passenger \& Driver \\ \hline
  \end{tabular}
  \caption{User weight attributes}\label{tab13}
  \end{center}
\end{table}

\subsection{Judgment matrix}	
The preference satisfier component computes driver/passenger profiles and preferences then represents them into a judgment matrix for each user. It takes as input  a set $D$ of $n$ drivers  and a set $P$ of $m$ passengers. Each driver $d \in D$ (resp. passenger $p \in P$) has a profile vector $prof(d)$ (resp. $prof(p)$) and a preference vector $pref(d)$ (resp. $pref (p)$), in addition to the corresponding weight of each  preference $w(d)$ (resp.  $w(p)$). The attributes of $prof$, $pref$ and $w$ vectors are illustrated in Table \ref{tab14}.\\
\begin{table}
\begin{center}
\caption{Attributes of $prof$, $pref$ and $w$ vectors of drivers and passengers}\label{tab14}
\begin{tabular}{|l|m{12cm}|}
\hline
Vector &  Attributes \\
\hline
$prof(d)$ & Gender, Age, Status, VHrange, Pets, Smoking, Music, SocialBehavior, DrivingSkills, Reliability\\
\hline
$prof(p)$ & Gender, Age, Status, Pets, Smoking, Music, SocialBehavior, Reliability\\
\hline
$pref(d)$ & GenderPref, AgePref, StatusPref, PetsPref, SmokingPref, MusicPref,  SocialBehaviorPref, ReliabilityPref\\
\hline
$pref(p)$ & GenderPref, AgePref, StatusPref, VHrangePref, PetsPref, SmokingPref, MusicPref, SocialBehaviorPref, DrivingSkillsPref, ReliabilityPref\\
\hline
$w(d)$ & GenderW, AgeW, StatusW, PetsW, SmokingW, MusicW, SocialBehaviorW, ReliabilityW\\
\hline
$w(p)$ & GenderW, AgeW, StatusW, VHrangeW, PetsW, SmokingW, MusicW, SocialBehaviorW, DrivingSkillsW, ReliabilityW\\
\hline
\end{tabular}
\end{center}
\end{table}
The judgment matrix of a driver $d$ is noted $X(d) = (x_{ij})$ $ m*k $ (in our case $k = 8$) where  rows represent passengers, columns represent driver $d$ preferences and $x_{ij}$ is the score of passenger $p_i$ with respect to the driver preference $j$. Respectively, the judgment matrix of a passenger $p$ is noted $X(p) = (x_{ij})$ $n*l $ (in our case $l = 10$) where rows represent drivers, columns represent passenger $p$ preferences and $x_{ij}$ is the score of driver $d_i$ with respect to the passenger preference $j$.

The values $x_{ij}$ of $X(d)$ and $X(d)$ are computed using the following functions:

\begin{itemize}
    \item \textbf{Binary\_Score(i,j)}: if the preference $j$ is a GenderPref, StatusPref, PetsPref, SmokingPref, MusicPref or VHrangePref.
    \item \textbf{Age\_Score(i,j)}: if the preference $j$ is an AgePref.
    \item \textbf{Feedback\_Score(i,j)}: if the preference $j$ is a SocialBehaviorPref, DrivingSkillsPref or ReliabilityPref.
\end{itemize}

\begin{definition}
\textsc{\textbf{(Binary\_Score)}}: Let $Pref_j$ be a preference of a driver $d$  (resp. passenger $p$)  and $Prof_j$ be the correspondent profile input of a passenger $p_i$ (resp. driver $d_i$), we define \textsc{Binary\_Score} as follows:

\newcommand{\reels}{\mathbb{R}}
\begin{alignat*}{2}
Binary\_Score(i,j)=&
\begin{cases}
    \textsc{1}       & \quad \text{if } \text{$Pref_j=Prof_j$}\\
    \textsc{0}       & \text{otherwise}\\
\end{cases}
\end{alignat*} 
\end{definition}

\begin{definition}
\textsc{\textbf{(Age\_Score)}}: 
 Let $AgePref$ be  the age preference of a driver $d$  (resp. passenger $p$) and $AgeTolerance$ its age tolerance.   Let  $Age$ be the age of a passenger $p_i$ (resp. driver $d_i$). We define \textsc{Age\_Score} as follows:

\begin{equation*}
   Age\_Score(i,j)= \frac{AgeTolerance}{|Age-AgePref| + T}
\end{equation*}

\end{definition}

\begin{definition} 
\textsc{\textbf{(Feedback\_Score)}}:  is  the average  evaluations $e_i$ $(i=1,...,n)$ received from drivers or passengers according to their previous experiences. 
\begin{equation*}
  Feedback\_Score(i,j)= \frac{\sum\limits_{i=1}^n e_{i}}{n} 
\end{equation*}
\end{definition}

\paragraph{Illustrative example}
~~\\
\begin{em}
We consider a passenger $P_1$ and six drivers $D_1,...,D_6$ making a ridesharing  requests that fulfill the spatio-temporal constraints of $P_1$ request. $P_1$ is supposed to have already filled his personal profile and his preferences about preferred drivers and also stating the weights of each preference. Table \ref{tabf26} draws $P_1$ preferences and their respective weights. Similarly, we suppose that all drivers have already provided their personal details or profile as shown in Table \ref{tabf27}.
\end{em}
\begin{table}[h]
\begin{minipage}{3in}
\begin{center}

\begin{tabular}{| >{\centering\arraybackslash}m{1.5in} | >{\centering\arraybackslash}m{0.8in} |} \hline
    {\bf Preferences} & {\bf Pref(P1)}  \\ \hline
   GenderPref	& M \\ \hline
AgePref	& 30 \\ \hline
StatusPref & Single \\ \hline
VHrangePref & Comfort \\ \hline
PetsPref & Yes \\ \hline
MusicPref &	No \\ \hline
SmokingPref & No \\ \hline
SocialBehaviorPref & -  \\ \hline
DrivingSkillsPref &	- \\ \hline
ReliabilityPref &	- \\ \hline
 \end{tabular}
  
    \end{center}
\end{minipage}
\begin{minipage}{3in}
\begin{center}
    \begin{tabular}{| >{\centering\arraybackslash}m{1.5in} | >{\centering\arraybackslash}m{0.8in} |} \hline
    {\bf Weights} & {\bf w(P1)}  \\ \hline
   GenderW	& 4 \\ \hline
AgeW	& 9 \\ \hline
StatusW	& 6 \\ \hline
VHrangeW &	5 \\ \hline
PetsW & 0 \\ \hline
MusicW	& 8 \\ \hline
SmokingW & 8 \\ \hline
SocialBehaviorW & 6 \\ \hline
DrivingSkillsW &	7 \\ \hline
ReliabilityW & 0 \\ \hline
  \end{tabular}
  \end{center}    
  \end{minipage}
 
 \paragraph{}

  \begin{center}
   \begin{tabular}{| >{\centering\arraybackslash}m{1.5in} | >{\centering\arraybackslash}m{0.8in} |} \hline
  AgeTolerance &	5 \\ \hline
  \end{tabular}
   \end{center} 
   \caption{Preferences and weights vector of passenger $P_1$}\label{tabf26}
\end{table}
 \paragraph{}
 ~~\\
\begin{table}[h]
    \begin{center}
\begin{small}
    \begin{tabular}{| >{\centering\arraybackslash}m{0.5in} | >{\centering\arraybackslash}m{0.3in} 
	| >{\centering\arraybackslash}m{0.3in} | >{\centering\arraybackslash}m{0.5in} |
	>{\centering\arraybackslash}m{0.5in} |>{\centering\arraybackslash}m{0.3in} |
	>{\centering\arraybackslash}m{0.3in} |>{\centering\arraybackslash}m{0.5in} |
	>{\centering\arraybackslash}m{0.5in} |>{\centering\arraybackslash}m{0.5in} |
	>{\centering\arraybackslash}m{0.5in} |} \hline

 {\bf Profiles} & {\bf Gen-der} & {\bf Age} & {\bf Status} & {\bf VH range} & {\bf Pets} & {\bf Mu-sic} & 
{\bf Smo-king} & {\bf Social Behavior} & {\bf Driving Skills} & {\bf Relia-bility} \\ \hline 
Prof(D1) & M & 26 & Married & Luxury & No & Yes & Yes & 4,77 & 4,12 & 7,78 \\ \hline
Prof(D2) & F & 44 & Married & Basic & Yes & No & Yes & 1,06 & 5,94 & 6,39 \\ \hline
Prof(D3) & F & 34 & Single & Basic & No & No & Yes & 5,58 & 9,3 & 6,46 \\ \hline
Prof(D4) & M & 65 & Single & Luxury & No & No & No & 0,34 & 4,34 & 0,23 \\ \hline
Prof(D5) & M & 38 & Single & Comfort & Yes & Yes & No & 4,37 & 1,63 & 8,65 \\ \hline
Prof(D6) & F & 49 & Married & Comfort & Yes & Yes & Yes & 3,08 & 8,91 & 1,88 \\ \hline
  \end{tabular}
  \end{small}
  \caption{Profile vectors of drivers}\label{tabf27}
  \end{center}
\end{table}
\begin{em}
In the sequel, the PreferenceSatisfier component is responsible for computing the judgment matrix of passenger $P_1$, $X(P_1)= (x_{ij})$ $6*10$, based on  his preferences presented in Table \ref{tabf26} and the profile of each driver provided in Table \ref{tabf27}.\\

Table \ref{tabf28} depicts the generated judgment matrix of passenger $P_1$. In this matrix, $x_{11} = Binary\_Score(1,1)= 1$ since the gender preference of passenger $P_1$ (Male) is equal to the gender profile of the driver $D_1$ (Male). Also, $x_{12}= Age\_Score(1,2)=\frac{AgeTolerance}{|Age-AgePref| + AgeTolerance}= \frac{5}{| 26-30| + 5} = 0.55$. It is important to note that SocialBehaviorPref, DrivingSkillsPrefand and ReliabilityPref derive from the evaluations the driver $D_1$ has received in the previous matches and their values do not depend on passenger preferences.
\end{em}

\begin{table}[h]
    \begin{center}
\begin{small}
    \begin{tabular}{| >{\centering\arraybackslash}m{0.5in} | >{\centering\arraybackslash}m{0.3in} 
	| >{\centering\arraybackslash}m{0.3in} | >{\centering\arraybackslash}m{0.5in} |
	>{\centering\arraybackslash}m{0.5in} |>{\centering\arraybackslash}m{0.3in} |
	>{\centering\arraybackslash}m{0.3in} |>{\centering\arraybackslash}m{0.5in} |
	>{\centering\arraybackslash}m{0.5in} |>{\centering\arraybackslash}m{0.5in} |
	>{\centering\arraybackslash}m{0.5in} |} \hline
{\bf Driver-ID} & {\bf Gen-der} & {\bf Age} & {\bf Status} & {\bf VH range} & {\bf Pets} & {\bf Mu-sic} & 
{\bf Smo-king} & {\bf Social Behavior} & {\bf Driving Skills} & {\bf Relia-bility} \\ \hline 
D1 & 1 & 0,55 & 0 & 0 & 0 & 0 & 0 & 4,77 & 7,78 & 4,12 \\ \hline
D2 & 0 & 0,26 & 0 & 0 & 1 & 1 & 0 & 1,06 & 6,39 & 5,94 \\ \hline
D3 & 0 & 0,55 & 1 & 0 & 0 & 1 & 0 & 5,58 & 6,46 & 9,3 \\ \hline
D4 & 1 & 0,12 & 1 & 0 & 0 & 1 & 1 & 0,34 & 0,23 & 4,34 \\ \hline
D5 & 1 & 0,38 & 1 & 1 & 1 & 0 & 1 & 4,37 & 8,65 & 1,63 \\ \hline
D6 & 0 & 0,2 & 0 & 1 & 1 & 0 & 0 & 3,08 & 1,88 & 8,91 \\ \hline
  \end{tabular}
  \end{small}
  \caption{The generated judgment matrix of passenger $P_1$}\label{tabf28}
  \end{center}
\end{table}

\section{Multi-Criteria Ranking}
\label{sec:c4}
MCDM is a branch of operation research models, which is suitable for solving an issue featuring a high number of decision criteria, different forms of information, multi-interests and perspectives, and conflicting objectives \cite{r16}. In the dedicated literature, there are dozens of methods used for solving MCDM problems such as the analytical hierarchy process , TOPSIS, the elimination and choice translating reality, the preference ranking organization method for enrichment evaluation, the compromise programming, and the multi-attribute utility theory, to cite but a few.
\\
The TOPSIS method can be considered as one of the most widely accepted variants. The basic concept of TOPSIS is to find the best compromise solution according to the designer's objective weights. This method attempts to choose the alternatives that simultaneously have the shortest Euclidean distance from the positive ideal solution and the furthest Euclidean distance from the negative ideal solution. The ideal solution is composed of all attainable best attribute values; the negative ideal solution is composed of all attainable worst attribute values. TOPSIS, therefore, provides a cardinal ranking for all the alternatives by taking the relative closeness to the ideal solution.
\\
In our case we adapt the TOPSIS method to rank possible matches for each user (driver and passenger) according to their preferences.
\\
Formally, for each driver $d$ (passenger $p$), we have a ranking problem with $m$ alternatives $Pi (i = 1, ..., m)$ evaluated on $k$ criteria $Cj (j = 1, ..., k)$ ($n$ alternatives $Di (i = 1, ..., n)$ evaluated on $l$ criteria $Cj (j = 1, ..., l)$). The proposed procedure can be expressed for each driver $d$ in the following steps \cite{r32}.\\
The input is the judgment matrix $X(d) = (x_{ij})$ $m*k$ generated by the first component and defined as follows:
\[
    \mathbf{X(d)} = 
\bordermatrix{ & \textrm{Pref}_\textrm{1} & \textrm{Pref}_\textrm{2} & \dots & \textrm{Pref}_\textrm{j} & \dots & \textrm{Pref}_\textrm{k} \cr
      \textrm{p}_\textrm{1} & \textrm{x}_\textrm{11} & \textrm{x}_\textrm{12} & \dots & \textrm{x}_\textrm{1j} & \dots & \textrm{x}_\textrm{1k}  \cr
      \textrm{p}_\textrm{2} & \textrm{x}_\textrm{21} & \textrm{x}_\textrm{22} & \dots & \textrm{x}_\textrm{2j} & \dots & \textrm{x}_\textrm{2k} \cr
      \vdots & \vdots & \vdots &  & \vdots &  & \vdots \cr
      \textrm{p}_\textrm{i} & \textrm{x}_\textrm{i1} & \textrm{x}_\textrm{i2} & \dots & \textrm{x}_\textrm{ij} & \dots & \textrm{x}_\textrm{ik} \cr
      \vdots & \vdots & \vdots &  & \vdots &  & \vdots \cr
      \textrm{p}_\textrm{m} & \textrm{x}_\textrm{m1} & \textrm{x}_\textrm{m2} & \dots & \textrm{x}_\textrm{mj} & \dots & \textrm{x}_\textrm{mk} } \qquad
\]
\\
The proposed procedure can be expressed for each driver $d$ in the steps presented in the following figure \cite{r32}.
 \begin{figure}[h!]
\centering
\fbox{
\includegraphics[scale=0.7]{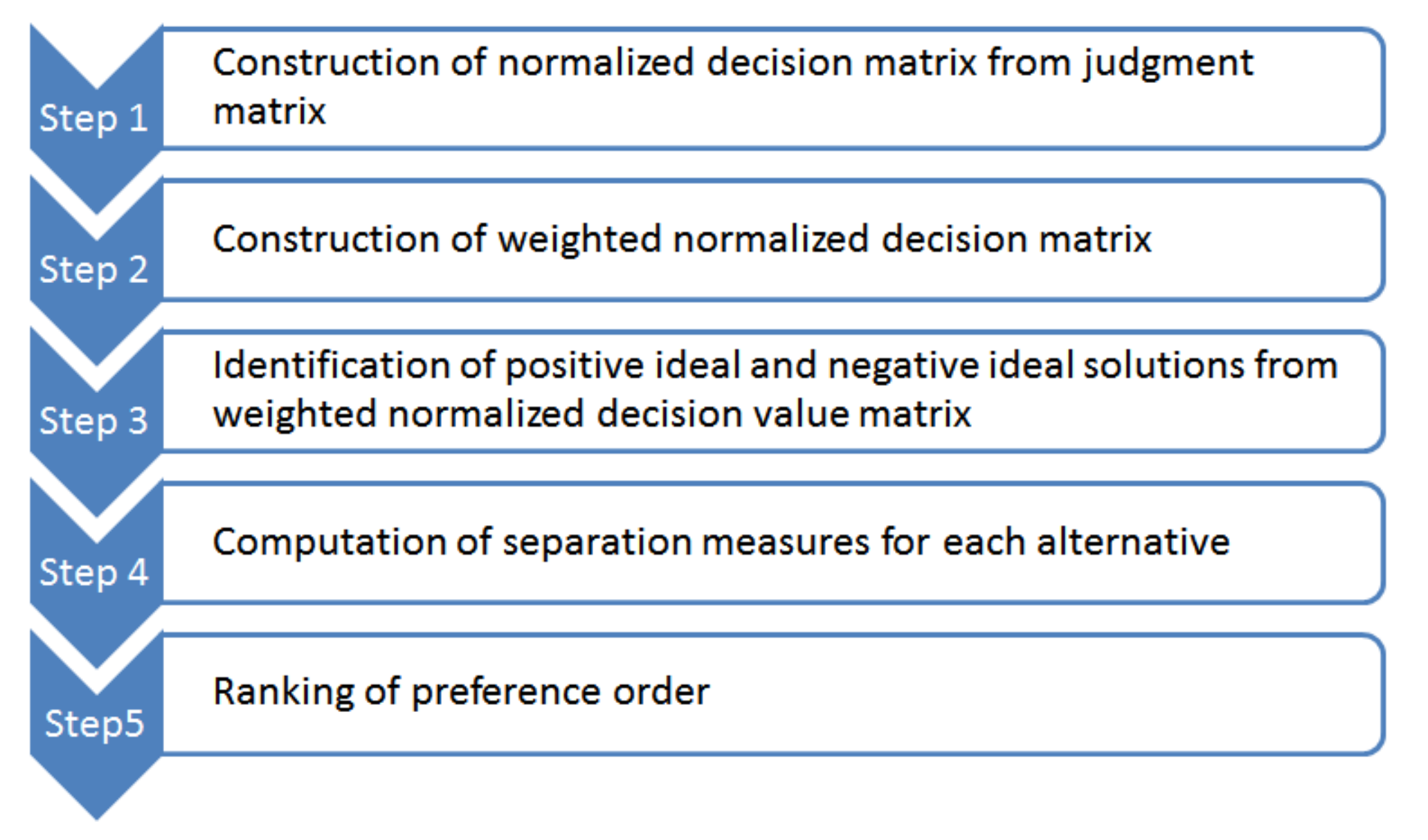}}
\caption{Stepwise procedure for performing TOPSIS methodology} \label{f29}
\end{figure}

\subsection{Step 1: Normalized decision matrix}
\begin{figure}[h!]
\centering
\fbox{
\includegraphics[scale=0.6]{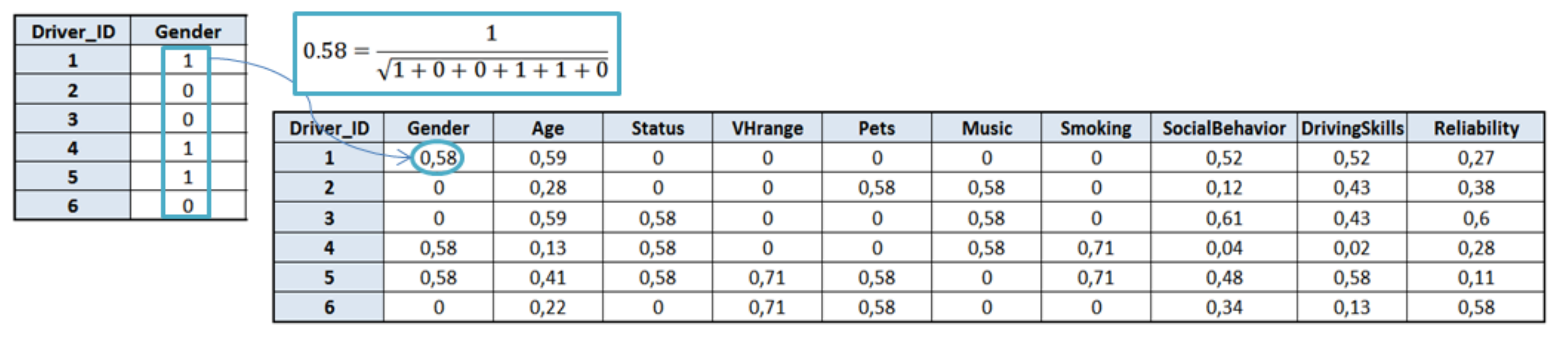}}
\caption{Construction of normalized decision matrix from judgment matrix} \label{f30}
\end{figure}

This process transforms the attribute dimensions into non dimensional attributes, which allows comparing across attributes.
The normalized decision matrix $R(d)$ can be computed with the help of Eq.\ref{eq1}.
\begin{equation}
\label{eq1}
   r_{ij} = \frac{x_{ij}}{\sqrt{\sum\limits_{i=1}^m (x_{ij}^2)} } \ \ \ \text{ for } i = 1, ..., m; j = 1, ..., k
\end{equation}
\\
The pseudo code of this step is shown in Algorithm \ref{alg:s1}.\\
\begin{algorithm}
\caption{TOPSIS}
\label{alg:s1}
\begin{algorithmic}[1]
\Function{STEP\_1}{$X, m, k$} \label{alg:s1}
\State Initialize T = \text{Totals vector of size k}
\State Initialize R = \text{Normalized decision matrix of size $m*k$}
\For {$j = 1 \to k$}
         \For{ $i = 1 \to m$ }
            \State $T[j] + = X[i][j]^{2}$
         \EndFor
\EndFor  

\For {$i = 1 \to m$}
         \For{ $j = 1 \to k$ }
            \State $R[i][j] = X[i][j]/\sqrt{T[j]}$
         \EndFor
\EndFor 
 \State \Return $R$
\EndFunction
\end{algorithmic}
\end{algorithm}
Fig.\ref{f30} shows the result of this step on the matrix example expressed in Table \ref{tabf28}.

\subsection{Step 2: Weighted normalized decision matrix}
The set of weights $w(d) = (w_{1}, w_{2},...,_{l})$ from the driver, as presented in Table \ref{tab1}, is accommodated to the decision matrix in this step. The weighted normalized decision matrix is computed by multiplying each column of the matrix R with its associated weight $wj$ as given in Eq.\ref{eq2}.
\begin{equation}
\label{eq2}
   v_{ij} = w_{j} * r_{ij}  \ \ \ \text{ for } i = 1, 2, ..., m; j = 1,..., k
\end{equation}

\begin{figure}[h!]
\centering
\fbox{
\includegraphics[scale=0.65]{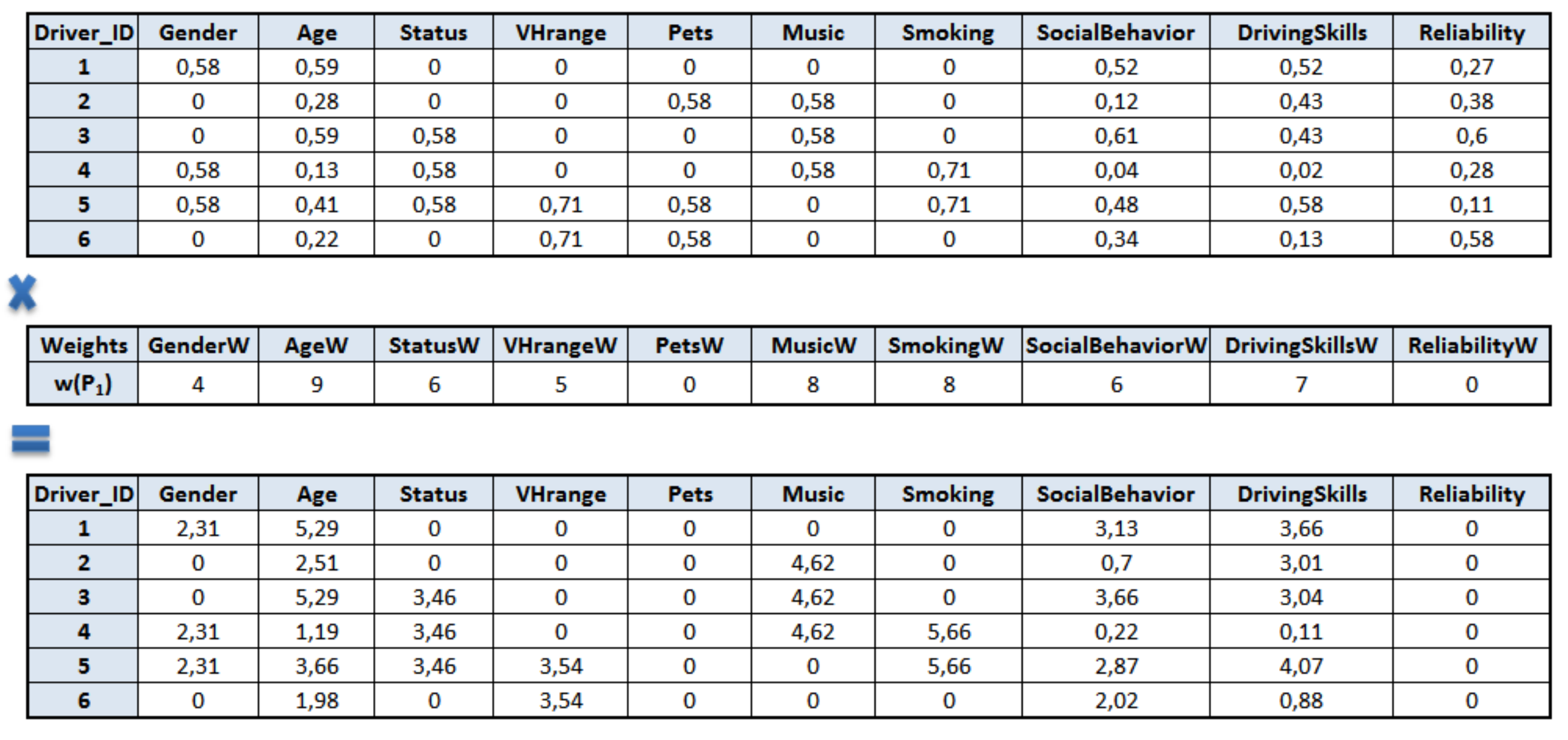}}
\caption{Construction of weighted normalized decision matrix} \label{f31}
\end{figure}

The pseudo code of this step is shown in Algorithm \ref{alg:s2}.\\
\begin{algorithm}
\caption{TOPSIS}
\label{alg:s2}
\begin{algorithmic}[1]
\Function{STEP\_2}{$R, W, m, k$} \label{alg:s1}
\State Initialize V = \text{Weighted normalized decision matrix of size $m*k$}
\For {$i = 1 \to m$}
         \For{ $j = 1 \to k$ }
            \State $V[i][j] = R[i][j]*W[j]$
         \EndFor
\EndFor 
 \State \Return $V$
\EndFunction
\end{algorithmic}
\end{algorithm}
Fig.\ref{f31} shows the result of this step on the matrix example expressed in Fig.\ref{f30}.
\subsection{Step 3: Positive ideal and negative ideal solutions}
Two artificial alternatives A* and A- indicating the most preferable alternative (positive ideal solution) and the least preferable one (negative-ideal solution) are respectively defined as:
\begin{equation}
\label{eq3}
   A^{*} = \{v_{1}^{*},v_{2}^{*},...,v_{j}^{*},...,v_{k}^{*}\} 
   = \{max_{\forall i} (v_{ij}) | \ \ \ i = 1, ..., m; \ j = 1,..., k\}
\end{equation}
\begin{equation}
\label{eq4}
   A^{-} = \{v_{1}^{-},v_{2}^{-},...,v_{j}^{-},...,v_{k}^{-}\} 
   = \{min_{\forall i} (v_{ij}) | \ \ \ i = 1, ..., m; \ j = 1,..., k\}
\end{equation}

The pseudo code of this step is shown in Algorithm \ref{alg:s3}.\\

\begin{figure}[h!]
\centering
\fbox{
\includegraphics[scale=0.5]{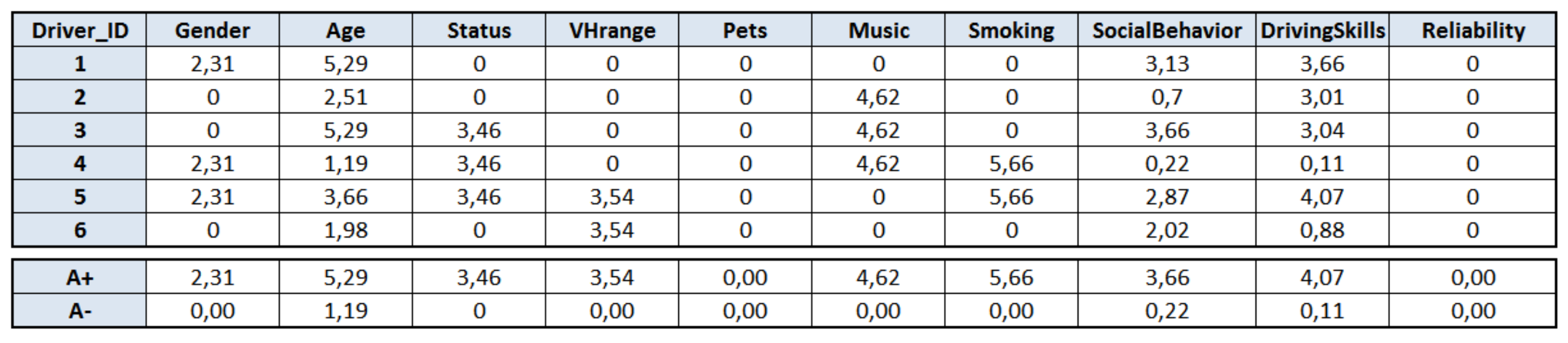}}
\caption{Identification of positive ideal and negative ideal solutions} \label{f32}
\end{figure}

\begin{algorithm}
\caption{TOPSIS}
\label{alg:s3}
\begin{algorithmic}[1]
\Function{STEP\_3}{$S, V, m, k$} \label{alg:s3}
\If{$S==0$} \Comment{Positive Solution}
\State Initialize P = \text{Positive Ideal vector of size $k$}
\For {$j = 1 \to k$}
\State Initialize $P[j] = 0$
 \For{ $i = 1 \to m$ }
\If{$V[i][j] > P[j] $}
 \State $P[j] = V[i][j]$;
\EndIf   
\EndFor
\EndFor 
\State \Return $P$
\Else \Comment{Negative Solution}
\State Initialize N = \text{Negative Ideal vector of size $k$}
\For {$j = 1 \to k$}
\State Initialize $N[j] = 10$
 \For{ $i = 1 \to m$ }
\If{$V[i][j] < N[j] $}
 \State $N[j] = V[i][j]$;
\EndIf   
\EndFor
\EndFor 
\State \Return $N$
\EndIf
\EndFunction
\end{algorithmic}
\end{algorithm}
Fig.\ref{f32} shows the result of this step on the matrix example expressed in Fig.\ref{f32}.

\subsection{Step 4: Separation measures}
The separation between each alternative can be measured by the Euclidean distance. The separation of each alternative from the ideal one is then given by:
\begin{equation}
\label{eq5}
   S_{i}^{*} = {\sqrt{\sum\limits_{j=1}^k (v_{ij}-v_{j}^{*})^2} } \ \ \ \text{ for } i = 1,..., m
\end{equation}
Similarly, the separation from the negative-ideal one can be derived using the following equation:
\begin{equation}
\label{eq6}
   S_{i}^{-} = {\sqrt{\sum\limits_{j=1}^k (v_{ij}-v_{j}^{-})^2} } \ \ \ \text{ for } i = 1,..., m
\end{equation}

\begin{figure}[h!]
\centering
\fbox{
\includegraphics[scale=0.5]{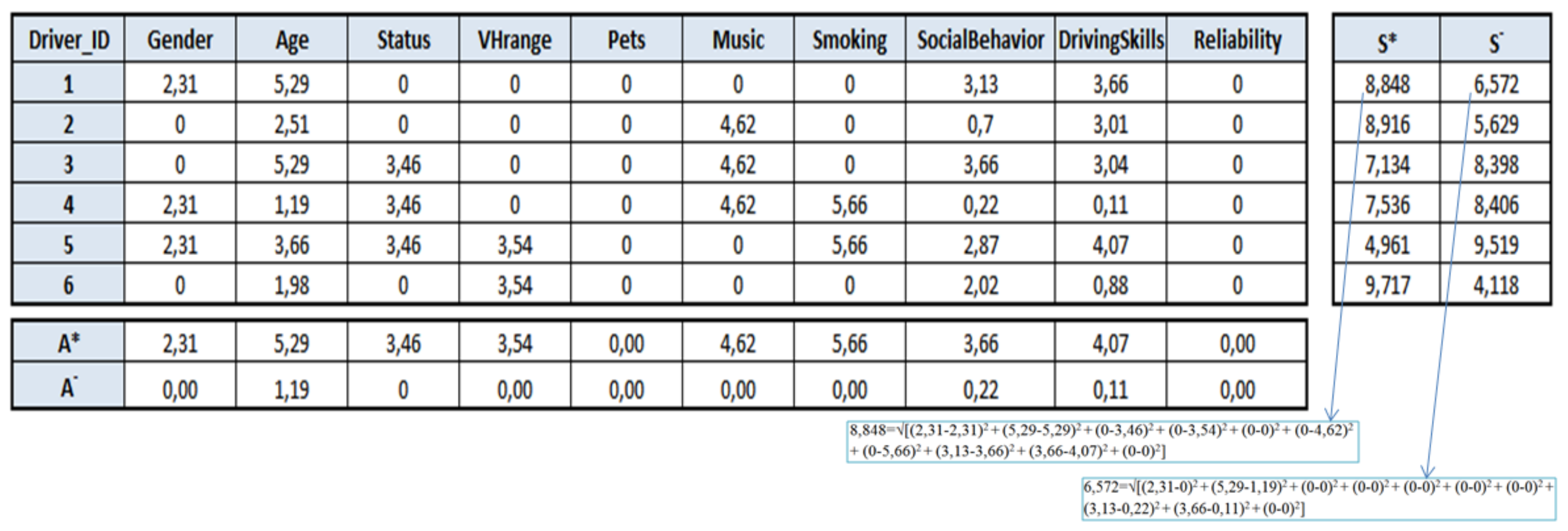}}
\caption{ Computation of separation measures for each alternative} \label{f33}
\end{figure}

The pseudo code of this step is shown in Algorithm \ref{alg:s4}.\\
\begin{algorithm}
\caption{TOPSIS}
\label{alg:s4}
\begin{algorithmic}[1]
\Function{STEP\_4}{$V,P, m, k$} \label{alg:s4}
\State Initialize PS = \text{Positive Separation vector of size $m$}
\For {$i = 1 \to m$}
\For{ $j = 1 \to k$ }
 \State $PS[i] + = (V[i][j]-P[j])^{2}$;   
\EndFor
\State $PS[i]  = \sqrt{S[i]}$;
\EndFor 
\State \Return $PS$
\EndFunction
\end{algorithmic}
\end{algorithm}
We call the same function with the parameter N instead of parameter P to compute the separation from the negative-ideal alternative.
Fig.\ref{f33} shows the result of this step on the solution example expressed in Fig.\ref{f32}.

\subsection{Step 5: Preference order}
\label{sec:po} 
A set of alternatives can now be ranked by preference through the computation of the relative closeness to the ideal value solution $C_{i}^{*}$ utilizing the following equation: 
\begin{equation}
\label{eq7}
   C_{i}^{*} = \frac{S_{i}^{-}}{(S_{i}^{*}+S_{i}^{-})}  \ \ \ \text{ for } i = 1,..., m
\end{equation}
\\
These preference lists are afterwards used to compute the stable matching.\\
The pseudo code of this step is shown in Algorithm \ref{alg:s5}.\\

\begin{figure}[h!]
\centering
\fbox{
\includegraphics[scale=0.7]{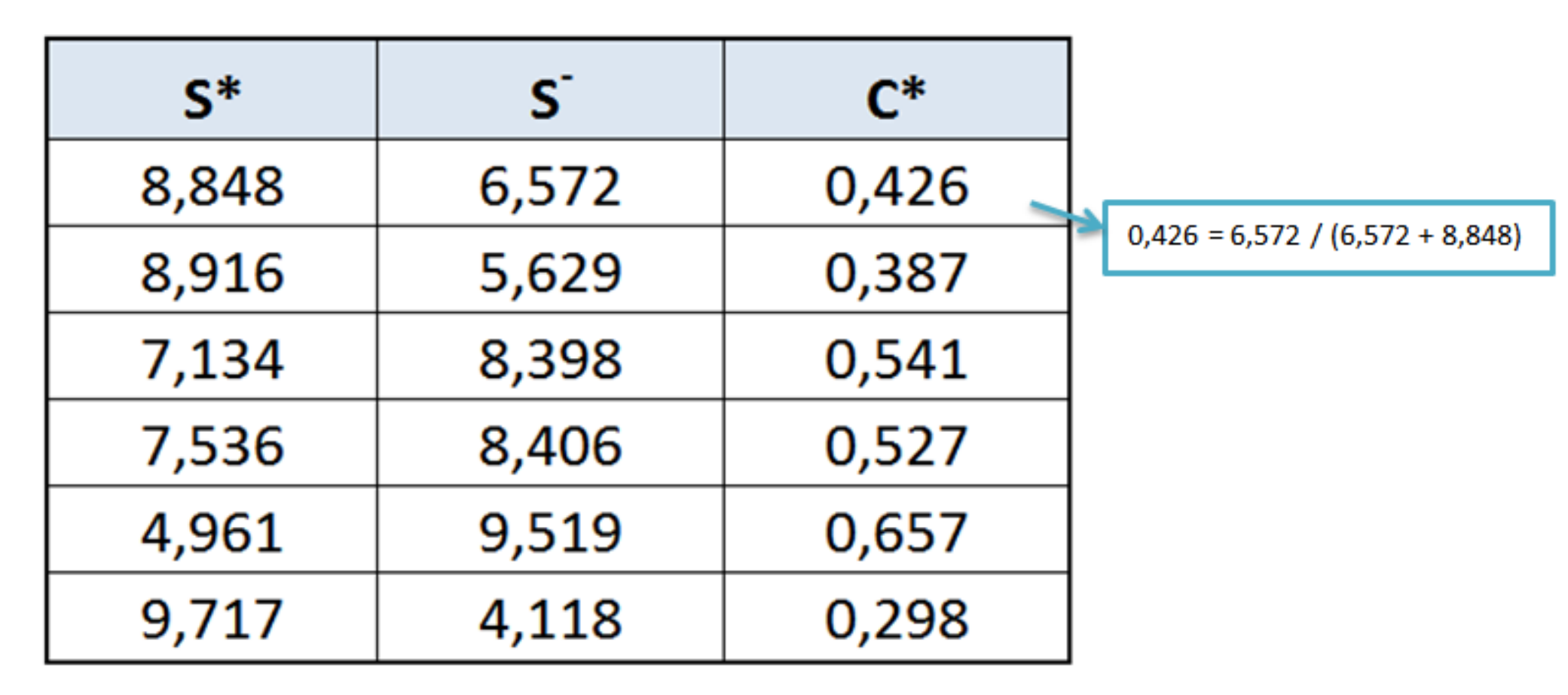}}
\caption{ Computation of the relative closeness to the ideal value solution} \label{f34}
\end{figure}

\begin{algorithm}
\caption{TOPSIS}
\label{alg:s5}
\begin{algorithmic}[1]
\Function{STEP\_5}{$PS,NS, m$} \label{alg:s5}
\State Initialize C = \text{Closeness to the ideal solution vector of size $m$}
\For {$i = 1 \to m$}
 \State $C[i]  = NS[i]/(NS[i]+PS[i])$;   
\EndFor 
\State \Return $C$
\EndFunction
\end{algorithmic}
\end{algorithm}
Fig.\ref{f34} shows the result of this step on the solution example expressed in Fig.\ref{f33}.
Finally, through ranking these values, we obtain the preference list of passenger P1. $PL(P_1)= D_5, D_3, D_4, D_1, D_2, D_6$.

\section{Stable Matching}
\label{sec:c5}
The ridesharing problem is treated as a matching problem where drivers are assigned to passengers. Our objective, in this phase, is to create a perfect matching of drivers and passengers such that there does not exists any pair of a driver and a passenger who prefers each other to their current partners. If a passenger and a driver form such a pair, they are called a blocking pair. If there are no blocking pairs in the matching solution, we call it a stable rideshare matching.
\subsection{Satisfaction of users' preferences } 
Classical ridesharing systems mainly focus on the improvement of their potential and performance in ridesharing to maximize certain objectives. Indeed, they propose a matching solution that maximizes the total system objective or the total number of matches. Nevertheless, such a solution may not necessarily maximize the objective of each individual participant, and may subsequently be rejected by the participants. Even if a proposed match satisfies a participant's constraints, a passenger and/or driver may not accept a match if they believe they can establish a better one. 
\paragraph{Example}
~~\\
\begin{em}
Table \ref{tabf25stable} and Table \ref{tab15stable} show an illustrative example of 3 passengers and 6 drivers. We assume that after running the multi-criteria ranking component, we obtain the result shown in Table \ref{tabf25stable} where $C_i^*$ (the relative closeness to the ideal value solution) is computed for each pair driver / passenger. Through ranking these values, we obtain also the preference list of each user as shown in Table \ref{tab15stable}.\\
\begin{table}[h]
\begin{minipage}{0.6\linewidth}
\begin{center}

\begin{tabular}{| >{\centering\arraybackslash}m{0.23in} |>{\centering\arraybackslash}m{0.23in} |>{\centering\arraybackslash}m{0.23in} |>{\centering\arraybackslash}m{0.23in} |>{\centering\arraybackslash}m{0.23in} |>{\centering\arraybackslash}m{0.23in} | >{\centering\arraybackslash}m{0.32in} |} \hline
{\bf $C^*$} & {\bf D1} & {\bf D2} & {\bf D3} & {\bf D4} & {\bf D5} & {\bf D6} \\ \hline
{\bf P1} & 0,5 & 0,7 & 0,31 & 0,12 & 0,48 & 0,25 \\ \hline
{\bf P2} & 0,15 & 0,9 & 0,72 & 0,38 & 0,14 & 0,19 \\ \hline
{\bf P3} & 0,22 & 0,17 & 0,77 & 0,57 & 0,24 & 0,81 \\ \hline
 \end{tabular}
 \paragraph{} 
~~\\
  $C^*$ of the passengers
    \end{center}
\end{minipage}
\begin{minipage}{0.3\linewidth}
\begin{center}
    \begin{tabular}{|>{\centering\arraybackslash}m{0.25in} |>{\centering\arraybackslash}m{0.25in} |>{\centering\arraybackslash}m{0.25in} | >{\centering\arraybackslash}m{0.25in} |} \hline
{\bf $C^*$} & {\bf P1} & {\bf P2} & {\bf P3} \\ \hline
{\bf D1} & 0,89 & 0,55 & 0,68 \\ \hline
{\bf D2} & 0,9 & 0,7 & 0,57 \\ \hline
{\bf D3} & 0,38 & 0,42 & 0,54 \\ \hline
{\bf D4} & 0,29 & 0,19 & 0,17 \\ \hline
{\bf D5} & 0,25 & 0,15 & 0,28 \\ \hline
{\bf D6} & 0,16 & 0,18 & 0,23 \\ \hline
  \end{tabular}
\paragraph{} 
~~\\ 
  $C^*$ of the drivers
  \end{center}    
  \end{minipage}
 
   \caption{$C^*$ of the users}\label{tabf25stable}
\end{table}


\begin{table}[h]
    \begin{center}

    \begin{tabular}{| >{\centering\arraybackslash}m{2.5in} | >{\centering\arraybackslash}m{2in} |}

    \hline

    {\bf Passengers} & {\bf Drivers}   \\ \hline
    $PL(P_1): D_2, D_1,D_5,D_3,D_6,D_4$ & $PL(D_1): P_1,P_3,P_2$ \\ \hline
    $PL(P_2): D_2,D_3,D_4,D_6,D_1,D_5$ & $PL(D_2): P_1,P_2,P_3$ \\ \hline
    $PL(P_3): D_6,D_3,D_4,D_5,D_1,D_2$ & $PL(D_3): P_3,P_2,P_1$ \\ \hline
     & $PL(D_4): P_1,P_2,P_3$ \\ \hline
     & $PL(D_5): P_2,P_1,P_3$ \\ \hline
     & $PL(D_6): P_3,P_2,P_1$ \\ \hline
  \end{tabular}
  \caption{User's preference list}\label{tab15stable}
  \end{center}
\end{table}

The system-optimal solution is to assign passenger $P_1$ to driver $D_1$, passenger $P_2$ to driver $D_2$ and passenger $P_3$ to driver $D_3$. This would result in a system-wide objective of $4.3$, an individual objective of $0.5$ for $P_1$ and an individual objective of $0.7$ for $D_2$.\\

However, $P_1$ prefers $D_2$ to his current partner ($D_1$) and $D_2$ prefers $P_1$ to his current partner ($P_2$).  So, $(P_1, D_2)$ form a blocking pair as they would both prefer to be matched together instead of to their current partners. If they were matched, it would increase their individual goals by $0.2$.
\end{em}
\subsection{Problem Formulation}
Formally \cite{r33}, let $D = \{d_{1}, d_{2},...,d_{n}\}$ be a set of drivers. Let $P = \{p_{1}, p_{2},...,p_{m}\}$ be a set of passengers. 
A matching is a one-to-one mapping $\mu$ from $D \cup P$ to itself, such that:
\begin{enumerate}
\item $\mu(d)=p$ if and only if $\mu(p)=d$, in where case d is matched to p;
\item If $\mu(d)$ is not in P, then $\mu(d)=d$, in which case d is unmatched;
\item If $\mu(p)$ is not in D, then $\mu(p)=p$, where case d is unmatched.
\end{enumerate}
We also define the notation for preference as the form $p1 \succ_{d} p2$ denotes that driver $d$ prefers passenger $p1$ to $p2$. 

A matching $\mu$ is a stable matching if it contains no blocking pairs. A blocking pair is defined as a pair $(d,p) \in D \cup P$ with $\mu(d)\neq p, p \succ_{d} \mu(d)$ and $d \succ_{p} \mu(p)$. Let $A$ denote the set of acceptable pairs. 
The incidence vector of a matching $\mu$ is a vector $x \in \{0,1\}^{|D|\times |P|}$ such that $x_{d,p} = 1$ if $\mu(d)=p$. Otherwise, $x_{d,p} = 0$. We identify each matching with its incidence vector. A vector  $x \in \mathbb{N}^{|D|\times |P|}$ is a stable matching if and only if it is an integer solution of the following system of linear equations:
    \begin{alignat}{2}
        \sum\limits_{j \in P} x_{d,j} & \leq 1 &\quad& \text{for each $d \in D$}\\
        \sum\limits_{i \in D} x_{i,p} & \leq 1 &\quad& \text{for each $p \in P$}\\
        x_{d,p} & \geq 0 &\quad& \text{for each $(d,p) \in D\times P$}\\
        x_{d,p} & = 0 &\quad& \text{for each $(d,p) \in (D\times P) \setminus A$}\\
        \sum\limits_{j \succ_{d} p} x_{d,j} + \sum\limits_{i \succ_{p} d} x_{i,p} + x_{d,p}  & \geq 0 &\quad& \text{for each $(d,p) \in  A$}        
    \end{alignat}
Constraints (8), (9), and (10) represent matching constraints. Constraint (11) is called \textit{individual rationality constraint}. Constraint (12) defines the stability constraints. It ensures that for each acceptable pair $(d,p)$, either driver $d$ is corresponding to someone they prefer to passenger $p$, or $p$ is corresponding to someone they prefer to $d$, or $d$ and $p$ are corresponding to each other.
\subsection{Stable marriage solutions}
\begin{figure}[h!]
\centering
\includegraphics[scale=0.9]{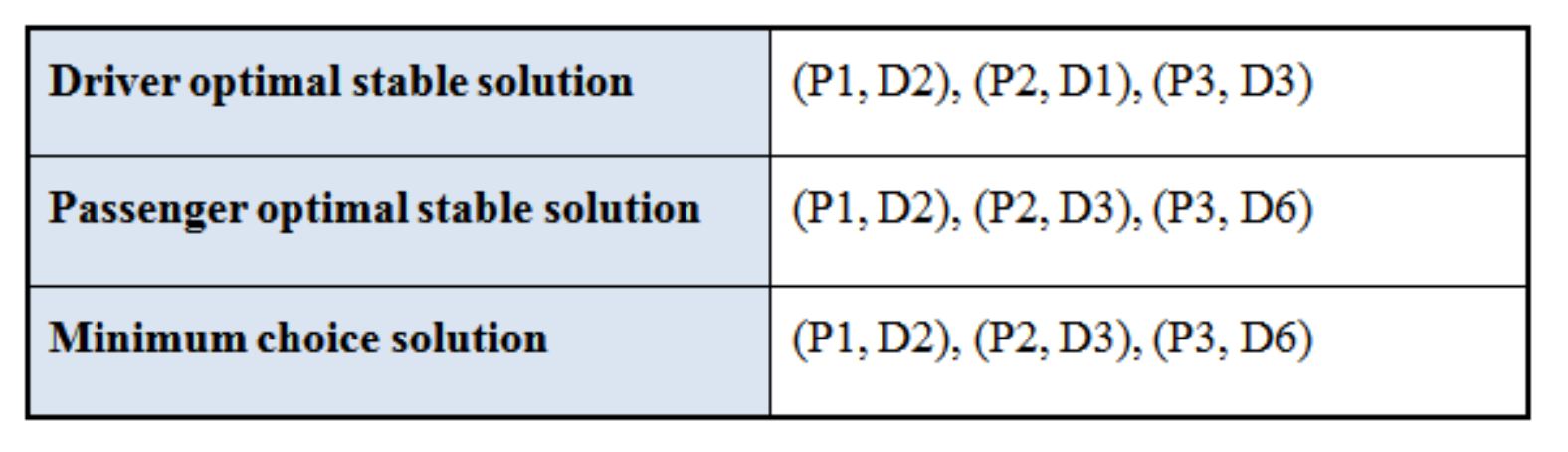}
\caption{ Stable marriage solutions} \label{f35}
\end{figure}

For any given sets $P$ and $D$ there are in general several stable marriage solutions. According to the work of \cite{c4r12}, the stability of the matching is defined on one of the following three solutions:
\begin{enumerate}
\item Driver optimal stable solution, which is optimal from the drivers' point of view. This is the stable matching when there are no other stable solutions in which each driver is matched with the same passenger or with a passenger they prefer to their partner.
\item Passenger optimal solution, which is optimal from the passengers' point of view. This is the stable matching when there are no other stable solutions in which each passenger is matched with the same driver or with a driver they prefer to their partner.
\item The minimum choice solution, which is optimal from the less numerous set's point of view. This is the stable matching when the less numerous set get their best possible choices. Thus, if there are fewer drivers than passengers, the driver optimal solution is obtained. However, if there are more drivers than passengers, the passenger optimal solution is obtained.
\end{enumerate}
\paragraph{Example}
~~\\
\begin{em}
	
Fig.\ref{f35} presents stable marriage solutions for the lists of preferences given in Table \ref{tab15stable}. The matching $(P_1, D_2), (P_2, D_1), (P_3, D_3)$ give a stable marriage since only one driver $D_1$ would consider another matching an improvement $(P_1$ or $P_3)$, however passenger $P_1$ prefers driver $D_2$ to driver $D_1$ and passenger $P_3$ prefers driver $D_3$ to driver $D_1$. This is in fact the driver optimal stable solution. The other stable marriage is given by the matching $(P_1, D_2)$, $(P_2, D_3)$, $(P_3, D_6)$ and this is the passenger optimal stable solution. Only one passenger $P_2$ would consider the matching with driver $D_2$ an improvement, but driver $D_2$ prefers passenger $P_1$ to passenger $P_2$.
The minimum choice solution is the passenger optimal stable solution since there are fewer passengers than drivers.
\end{em}
\subsection{Proposed algorithm } 
The driver optimal solution works well when there are fewer drivers than passengers. However, when the drivers are more numerous, several drivers will have to exhaust their preference lists since they will not be chosen by any passenger and wilt be the drivers who end up unmatched. With the same reasoning, the passenger optimal solution works well when there are fewer passengers than drivers. \\
We opt then for the minimum choice solution which is clearly a more efficient solution for finding a stable matching. Thus, if there are fewer drivers than passengers, the drivers' optimal solution is reached. Nevertheless, if there are more drivers than passengers, the passengers' optimal solution is obtained. To find this solution and make all matches stable, we rely on the SM algorithm given in Algorithm \ref{alg:sm}.\\
\\

\begin{algorithm}
\caption{SM algorithm}
\label{alg:sm}
\begin{algorithmic}[1]

\Procedure{SM}{$driverchoice, passengerchoice, matching, n, k$} 
\State Initialize $menlarger = n>k$
\If{$menlarger$}
          \State $max=k$ 
          \State $min=n$ 
	\Else
	\State $max=n$ 
	\State $max=n$ 
 \EndIf
\State Initialize $chno$ matrix of integer of size $min*max$
\State Initialize $counter$ vector of $0$ of size $min$
\For {$1 = 1 \to min$}
\For{ $j = 1 \to max$ }
\If{$menlarger$}
\State $chno[i][passengerchoice[i][j]] = j$ 
\Else 
\State $chno[i][driverchoice[i][j]]= j$
   \EndIf
\EndFor
\EndFor

\For{ $i = 1 \to min$ }
\If{$menlarger$}
\State $PROPOSAL( i, passengerchoice, matching, counter, chno)$ 
\Else 
\State $ PROPOSAL(i, driverchoice,matching, counter,chno)$
   \EndIf
\EndFor
\EndProcedure
\end{algorithmic}
\end{algorithm}
Procedure $SM (driverchoice, passengerchoice, matching, n, k)$ finds a single stable matching (SM). There are $n$ drivers and $k$ passengers, and the smaller set proposes. The optimal stable solution for the smaller set is obtained. The result is left in the integer array matching. Thus $matching[i]$ is the driver whom the $i-th$ passenger is matched to if $n < k$, but if there are less passengers then $marriage[i]$ is the passenger whom the $i-th$ drivers is matched to. If $matching[i] = 0$ at the end then that person is unmatched. There will be $|n- k|$ elements of matching zero. $driverchoice$ and $passengerchoice$ are the choice matrices for the drivers and the passengers respectively, i.e. $driverchoice[i,j]$ is the $j-th$ choice of the $i-th$ driver. The formal integer arrays should have the following sizes, $driverchoice [1 : n, 1 : k]$,  $passengerchoice [1 : k, 1 : n ]$, $matching [1 :max(n, k)]$ ;
\\
Procedure $PROPOSAL(i, choice)$ (Algorithm \ref{alg:sm2}) makes the next proposal for driver/ passenger $i$, and calls the procedure $REFUSAL$ to see what effect this proposal will have. The procedure does nothing if driver/ passenger is the dummy $0$.
\\
\begin{algorithm}
\caption{SM algorithm}
\label{alg:sm2}
\begin{algorithmic}[1]
\Procedure{PROPOSAL}{$i, choice, matching, counter,chno$} 
\If{$i<> 0$} 
\State $counter [i] ++ $
\State $REFUSAL(i, choice[i][j], choice, matching, counter, chno)$
 \EndIf
 \EndProcedure
\end{algorithmic}
\end{algorithm}
Procedure $REFUSAL(i, j, choice)$ (Algorithm \ref{alg:sm3}) decides which of the two proposals, the one being kept in suspense or the one just received, should be retained. Whichever is rejected goes back to the procedure $PROPOSAL$ to make the next proposal.\\
\begin{algorithm}
\caption{SM algorithm}
\label{alg:sm3}
\begin{algorithmic}[1]
\Procedure{REFUSAL}{$i,j, choice, matching, counter, chno$} 
\State Initialize $l$ integer
\If{$matching[j]== 0$}
          \State $matching[j]= i$  
\Else
	\If{$chno[j][matching[j]] > chno[j][i]$}
      \State $l = matching[j]$  
      \State $matching[j]= i$
      \State $PROPOSAL(l, choice, matching, counter, chno)$
	\Else
	\State $PROPOSAL(i, choice, matching, counter,chno)$ 
 \EndIf
 \EndIf
\EndProcedure
\end{algorithmic}
\end{algorithm}

\begin{figure}[h!]
\centering
\fbox{
\includegraphics[scale=0.7]{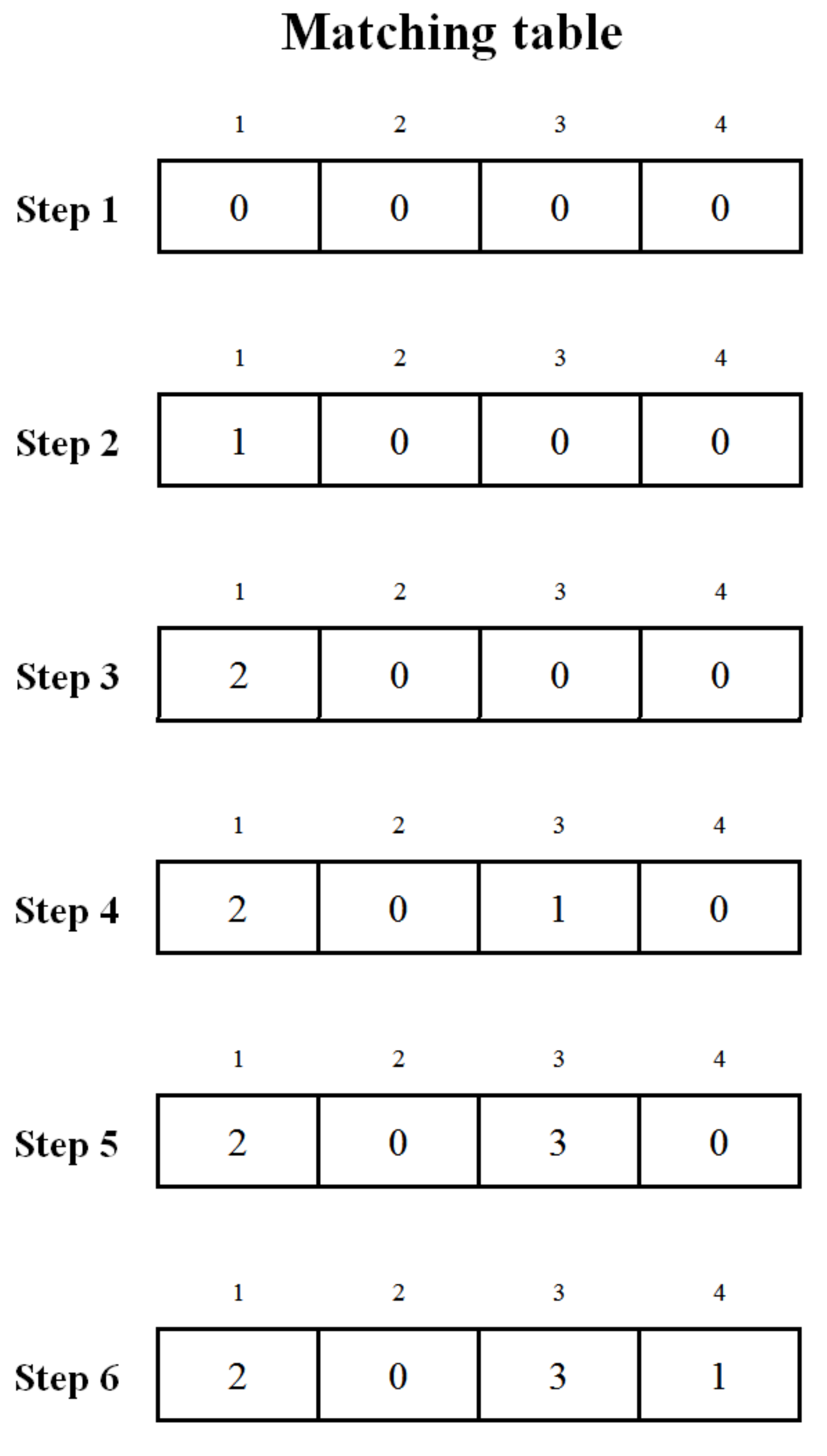}}
\caption{ Execution of the SM algorithm} \label{f39}
\end{figure}

Fig.\ref{f39} shows an illustrative example of the execution of this algorithm on the stable marriage instance of $4$ passengers and $3$ drivers presented in Table \ref{tab16} .

\begin{table}[h!]
    \begin{center}

    \begin{tabular}{| >{\centering\arraybackslash}m{2.5in} | >{\centering\arraybackslash}m{2in} |}

    \hline

    {\bf Passengers} & {\bf Drivers}   \\ \hline
    $PL(P_1): D_2, D_1,D_3$ & $PL(D_1): P_1,P_3,P_4,P_2$ \\ \hline
    $PL(P_2): D_2,D_3,D_1$ & $PL(D_2): P_1,P_4,P_2,P_3$ \\ \hline
    $PL(P_3): D_3,D_2,D_1$ & $PL(D_3): P_3,P_4,P_2,P_1$ \\ \hline
    $PL(P_4): D_1,D_2,D_3$ & \\ \hline
  \end{tabular}
  \caption{Stable marriage instance of 4 passengers and 3 drivers}\label{tab16}
  \end{center}
\end{table}

\section{Conclusion}
\label{sec:c6}
In this chapter, we have suggested SMRM, a system that promises the satisfaction of drivers and passengers' social preferences, considering the notion of stability for rideshare matches. The tasks associated with the SMRM system have been divided into three components each of which has a specific role for the optimization of the service as a whole.\\
The following chapter details the different tools used in the development of our system. The performed experimentations to assess our system are then discussed.

\chapter{Experimental Results}
\section{Introduction}
The main concepts related to our proposal have been defined, this chapter concretizes all these concepts thus providing the pragmatic level required for the validation of our approach.
\\
All basic concepts and fundamentals related to the development phase, including implementation and testing, are described. We will first present a brief summary of the aspects related to our method of resolution. A description of the experimental environment will be the subject of the third section. The last section of this chapter will discuss the different execution results.

\section{Areas of scientific interests}
In order to restore the balance between the improvement aspect and the satisfaction requirements from a practical point of view, we were able to consider several theoretical concepts as well as the necessary tools to benefit from them. A methodological and strategic choice was thus the result of efforts expended in this direction by a research team for the elaboration of this master in the LIPAH laboratory. Our work was then oriented in this direction and resulted in a combination of scientific fields.\\
We thus classify our work as a meeting point of varied and highly evolved domains. Indeed, the approach we have reached uses different concepts. These latter are implemented by the different steps developed for the achievement of our objectives, each step involves one or more areas of interest. \\

\begin{figure}[h!]
\centering
\includegraphics[scale=0.7]{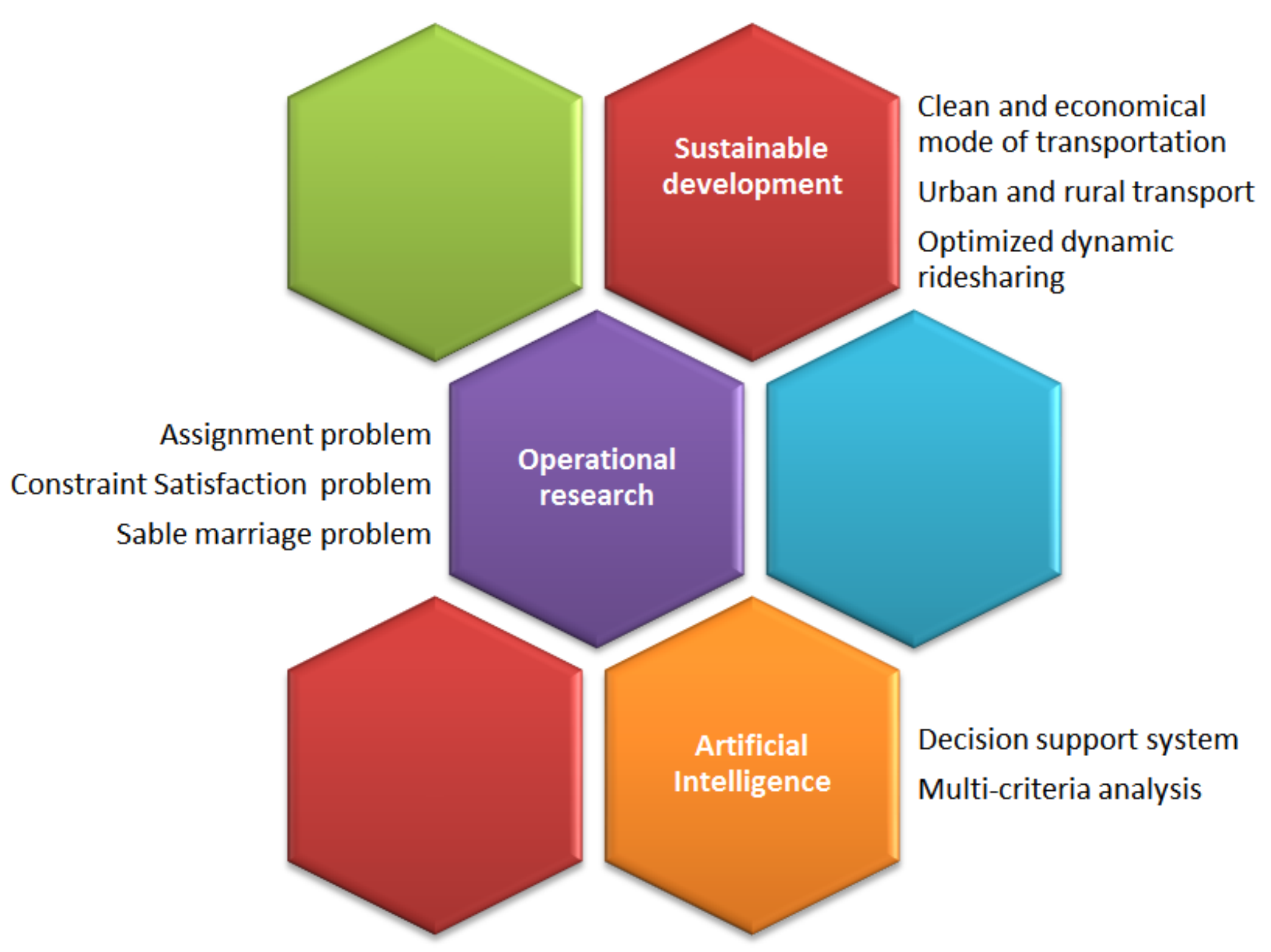}
\caption{SMRM: A wide range of scientific fields} \label{f36}
\end{figure}

These lead on the whole to a decision support system making use of artificial intelligence to search for stable matching in a dynamic ridesharing context in favor of the realization of the sustainable development project (Fig.\ref{f36}).
\section{Experimental environment}
In this section we detail the different tools and data used as well as the evaluation metrics for the testing of our application.
\subsection{Implementation}
To devaluate the impact of stability on system performance, we test two different implementations on an Intel Core i5 Linux machine with 8GB RAM: one with the SM algorithm implemented in Java, and the other, with CPLEX interface with Java as a linear and binary integer programming solver to find the optimal solution for unconstrained matching (A), i.e. relaxing constraint (3.12).
\paragraph{CPLEX}
~~\\
In order to implement an optimal linear assignment solution, we used CPLEX interface with Java.  The latter favors the resolution of real-world assignment problems. It contains a robust optimizer that handles the side constraints that are invariably found in all types of problems. For pure academic problems, it finds solutions that are comparable to solutions found by specialized algorithms.
Certain combinatorial optimization problems cannot be easily linearized and solved with traditional mathematical programming methods. To handle these problems, it provides a large set of arithmetic and logical constraints, as well as a robust optimizer that brings all the benefits of a model-and-run development process to combinatorial optimization.
\subsection{Evaluation metrics}
\label{eval}
To evaluate the choice of our solution for the multi-criteria evaluation, we compare the head of the preference list of our approach to that of the approach of our competitors. For this, we use a metric that compares the total weight of preferences where the head of the list exceeds its competitor of the other approach. We define the Weight Superiority of the user $U_n$ compared to the user $U_m$ as:

\begin{equation}
\label{eqws}
   W^{Superiority}_{U_n/U_m} = \sum\limits_{i=1}^{k} w_i \mid  x_{ni}>x_{mi}
\end{equation}

To evaluate the quality of our stable matching solution, we compute the following quality criteria as follows:
\\
Let $pr_{d}(p)$  (resp. $pr_{p}(d)$) denote the position of passenger $p$ (resp. driver $d$) in the preference list of driver $d$ (resp. of passenger $p$). The regret cost $r(A)$ of a stable matching $A$ is defined as:
\begin{equation}
\label{eq13}
   r(M)= \max\limits_{(d,p) \in A}\max\{pr_{d}(p),pr_{p}(d)  \}
\end{equation}
The egalitarian cost $c(M)$ is:
\begin{equation}
\label{eq14}
   c(M)= \sum\limits_{(d,p) \in A} pr_{d}(p) + \sum\limits_{(d,p) \in A} pr_{p}(d)
\end{equation}
The sex equality cost $d(M)$ is:
\begin{equation}
\label{eq15}
   d(M)= \Bigg\vert \sum\limits_{(d,p) \in A} pr_{d}(p) - \sum\limits_{(d,p) \in A} pr_{p}(d) \Bigg\vert
\end{equation}
Finally, to evaluate the impact of stability on system performance, we compare the value of the optimal objective function for unconstrained matching ($A$) ,i.e. relaxing constraint (12), and with that for stable matching ($A^{s}$). We define the price of stability $\delta$ as follows:
\begin{equation}
\label{eq16}
   \delta = \frac{A-A^{s}}{A}
\end{equation}

\subsection{Data sets}	
Since real world transport and social data sets are not available, in an attempt to evaluate the performance of the proposed approach, simulated data sets are derived and used. The source data utilized in testing are commercial products of Geomatic, a Danish company specializing in geo-demographic data and analysis for market segmentation, business intelligence, and direct marketing \cite{r34}. A few constraints are added so as to transfer them into multi preference instances.\\
Our data is stored in six tables DriverProfile Table, DriverPreferences Table, DriverWeight Table, PassengerProfile Table, PassengerPreferences Table and PassengerWeight Table.
\paragraph{DriverProfile Table}
~~\\
The DriverProfile table is described in Table \ref{tabt1}.
\begin{table}[h!]
\centering
\includegraphics[scale=0.7]{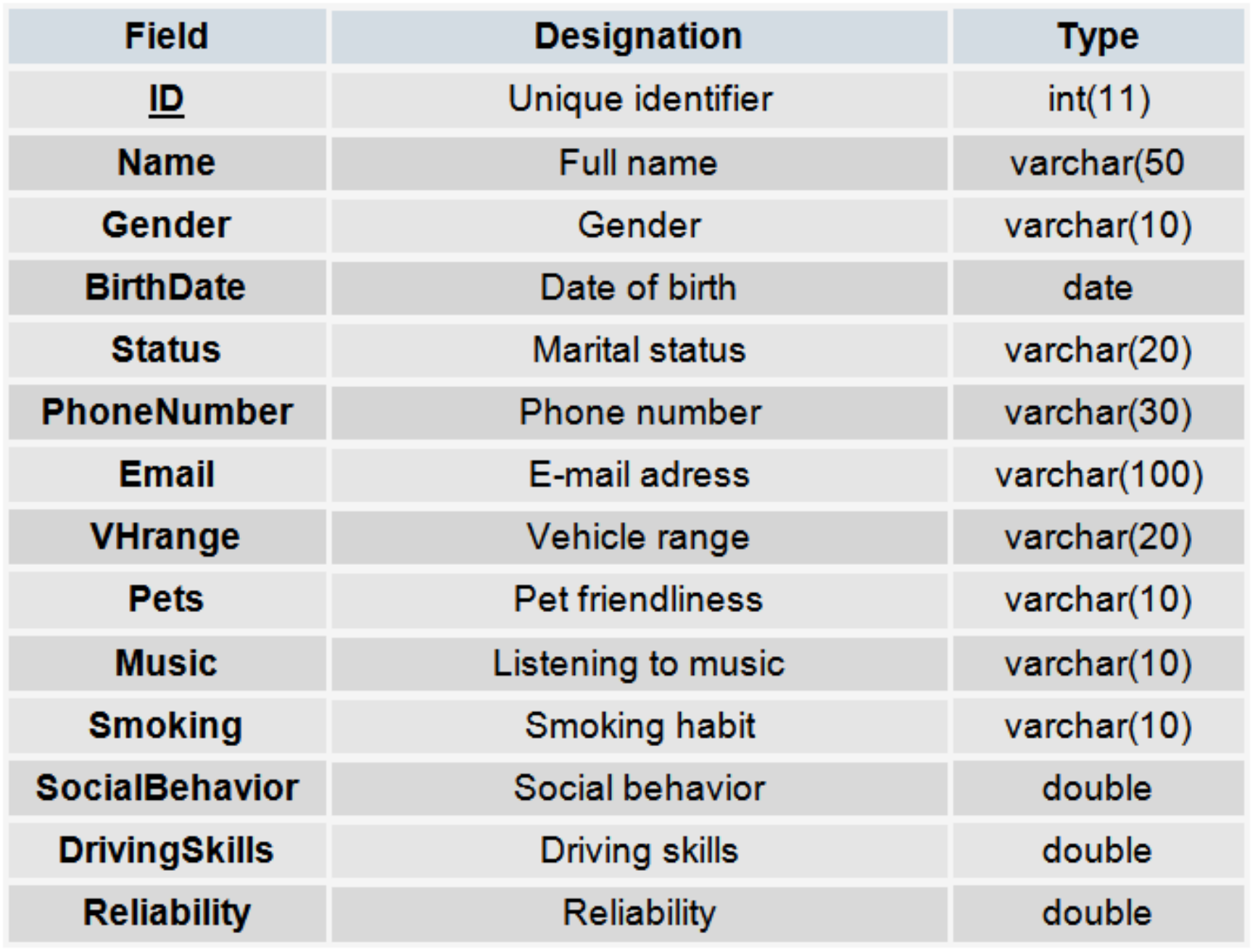}
\caption{The DriverProfile table}\label{tabt1}
\end{table}
\paragraph{DriverPreferences Table}
~~\\ 
The DriverPreferences table is described in Table \ref{tabt2}.
\begin{table}[h!]
\centering
\includegraphics[scale=0.7]{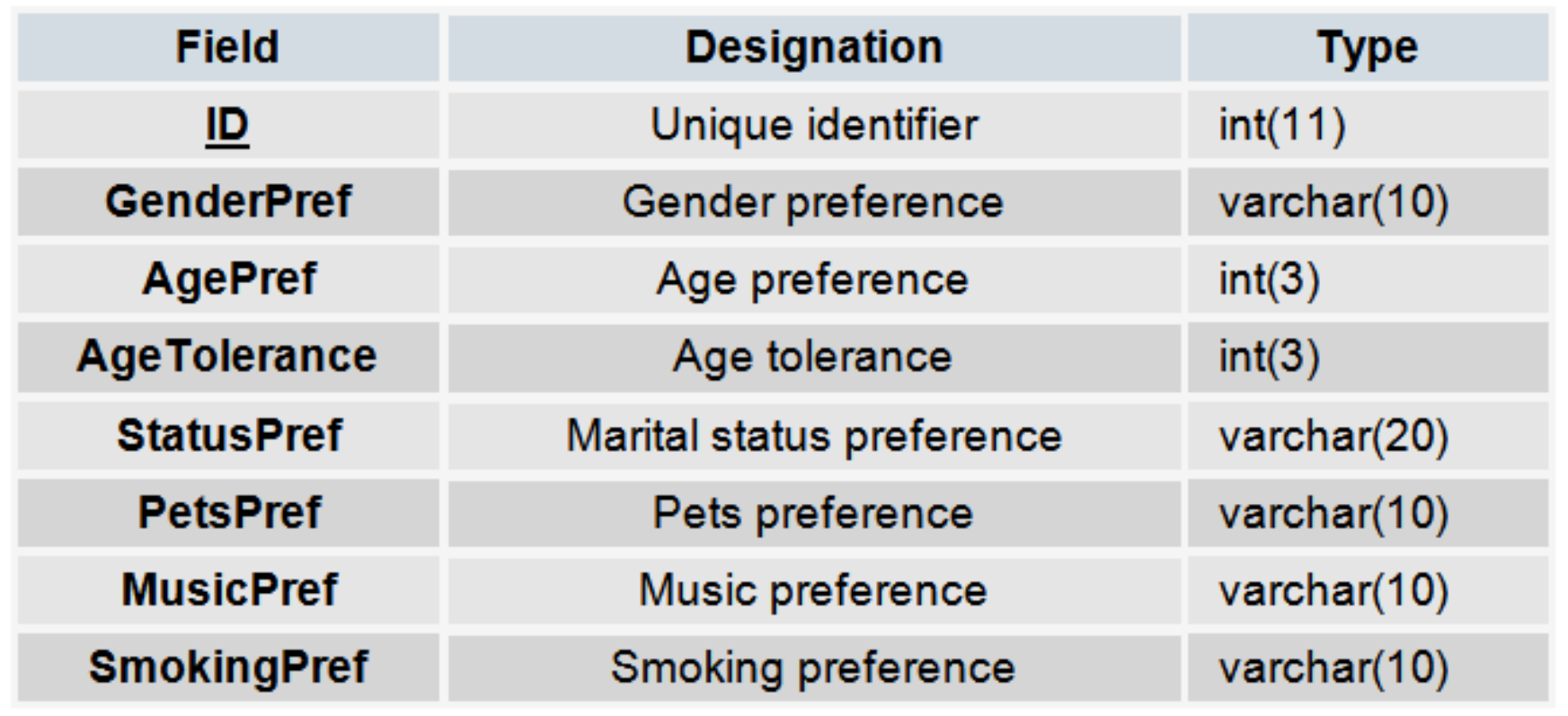}
\caption{The DriverPreferences table}\label{tabt2}
\end{table}
\paragraph{DriverWeight Table}
~~\\ 
The DriverWeight table is described in Table \ref{tabt3}.
\begin{table}[h!]
\centering
\includegraphics[scale=0.7]{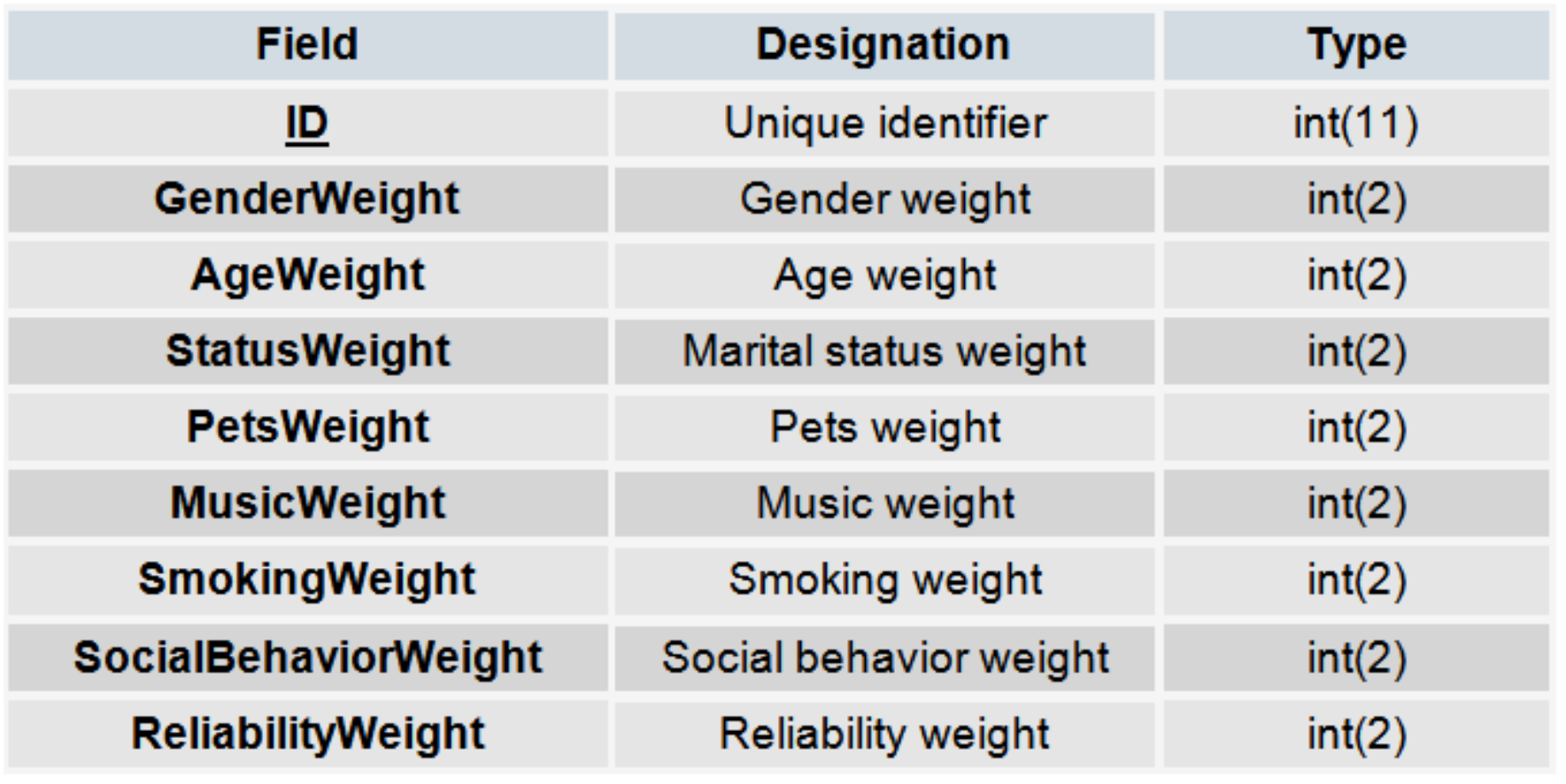}
\caption{The DriverWeight table}\label{tabt3}
\end{table}
\paragraph{PassengerProfile Table}
~~\\
The PassengerProfile table is described in Table \ref{tabt4}.
\begin{table}[h!]
\centering
\includegraphics[scale=0.7]{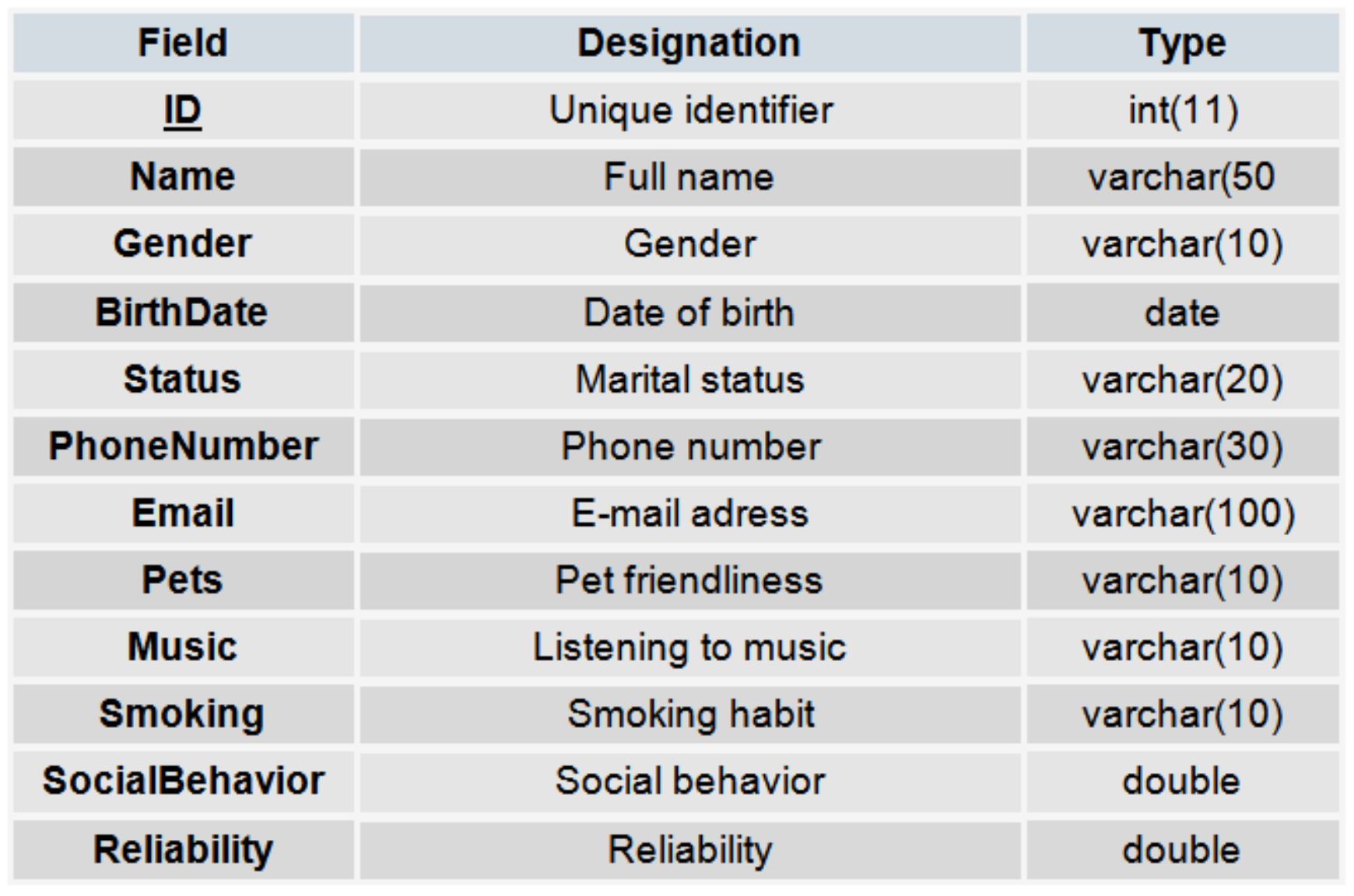}
\caption{The PassengerProfile table}\label{tabt4}
\end{table}
\paragraph{PassengerPreferences Table}
~~\\
The PassengerPreferences table is described in Table \ref{tabt5}.
\begin{table}[h!]
\centering
\includegraphics[scale=0.7]{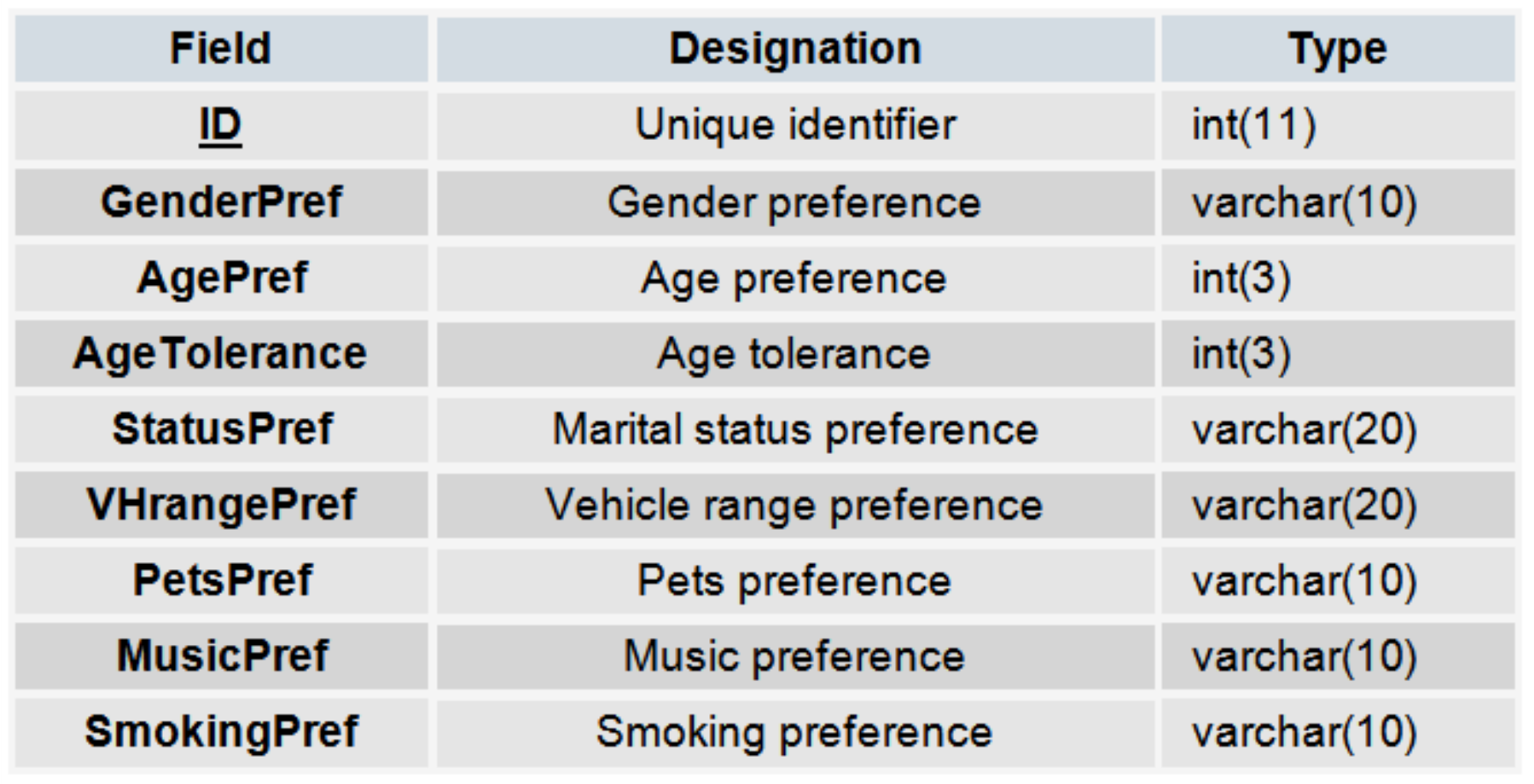}
\caption{The PassengerPreferences table}\label{tabt5}
\end{table}
\paragraph{PassengerWeight Table}
~~\\ 
The PassengerWeight table is described in Table \ref{tabt6}.
\begin{table}[h!]
\centering
\includegraphics[scale=0.7]{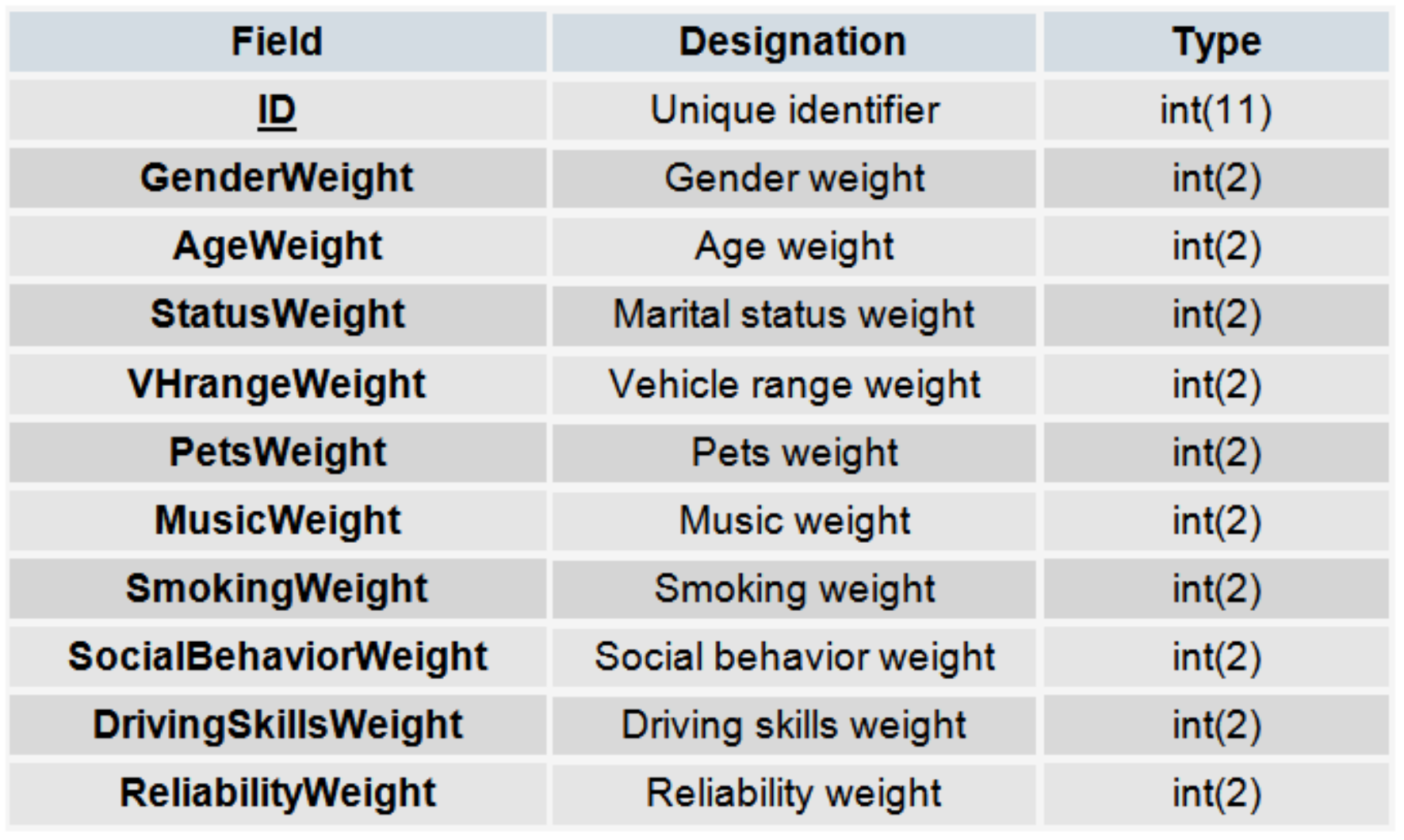}
\caption{The PassengerWeight table}\label{tabt6}
\end{table}
\section{Experiments} 
\subsection{Multi-criteria ranking}
This section contains indicative results that derive from the utilization of the second component of SMRM to a simulated ridesharing infrastructure environment. Some indicative, realistic everyday scenarios are presented. Our Multi-criteria ranking component is compared against the Weighted Sum Model WSM used in \cite{r11} which can lead to comprehensive results and thus showcases the efficiency of our proposed solution. Three scenarios will be presented. The scenarios are passenger-driven, in that they are differentiated based on passenger preferences and drivers profiles. In this respect, the first one presents a typical passenger service request. The second scenario presents a case where only one driver satisfies a high priority passenger preference. Finally, the third one serves as an example for the case when a passenger values extremely high a certain parameter (the vehicle range, in our case). 
We apply separately the two methods TOPSIS and WSM to rank drivers and we compare their respective results. To do so, we compute the respective $C^*$ and $A^{WSM-score}$ (as presented in formula \ref{forml}) values for every candidate driver. Through ranking these values, we obtain two different preference lists of the passenger.
\begin{equation}
\label{forml}
   A_i^{WSM-score} = \sum\limits_{j=1}^n w_j a_{ij},  \ \ \ \text{ for } i = 1, ..., m
\end{equation}
\subsubsection{Scenario 1: Regular ridesharing service request } 
\begin{figure}[h!]
\centering
\fbox{
\includegraphics[scale=0.7]{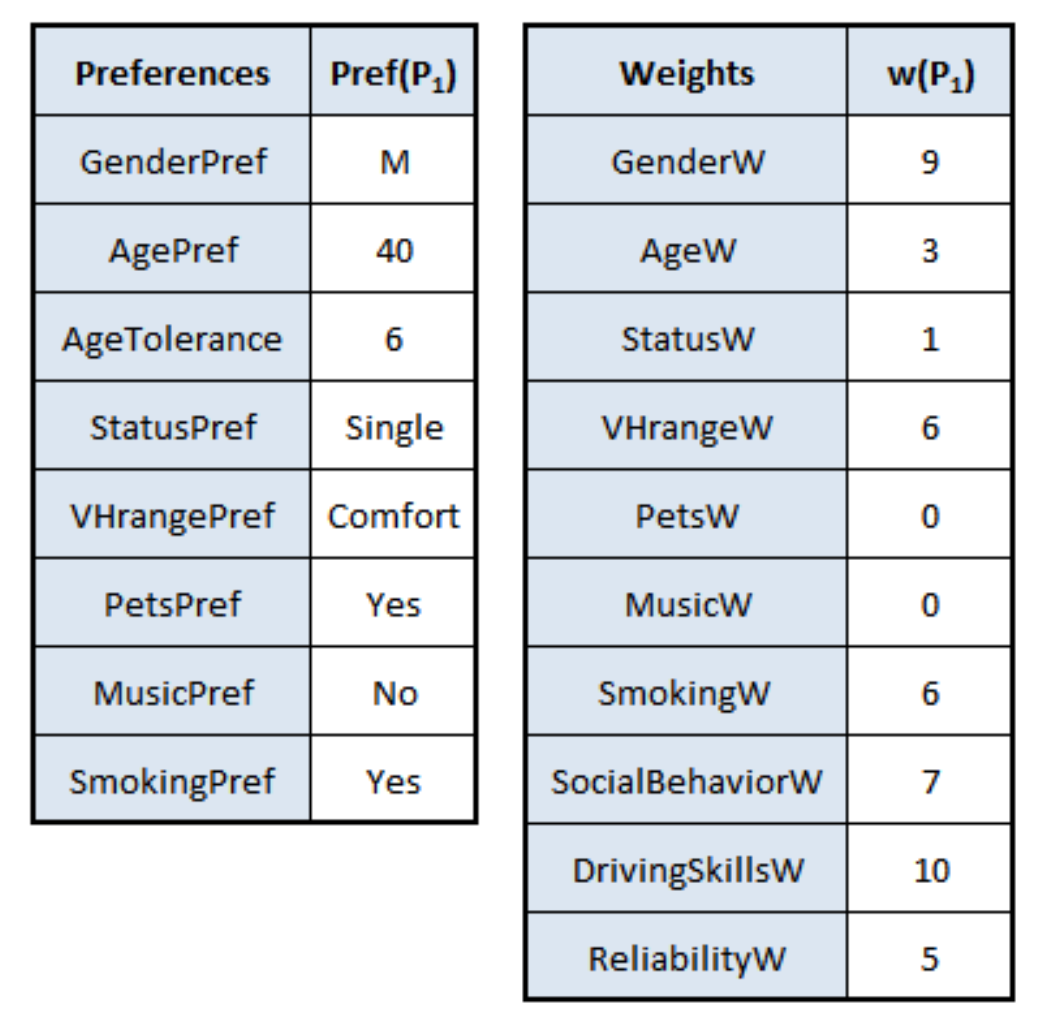}}
\caption{Scenario 1- The preferences and weights vectors of passenger P1} \label{f40}
\end{figure}

We consider a passenger $P1$ who has already registered on the SMRM and therefore disposes a unique identity in the system. He is supposed to have already filled his personal profile, stating his personal preferences on the driver and also depicting the weights of parameters. Fig.\ref{f40} presents the preferences and their respective weights. 

\begin{figure}[h!]
\centering
\includegraphics[scale=0.5]{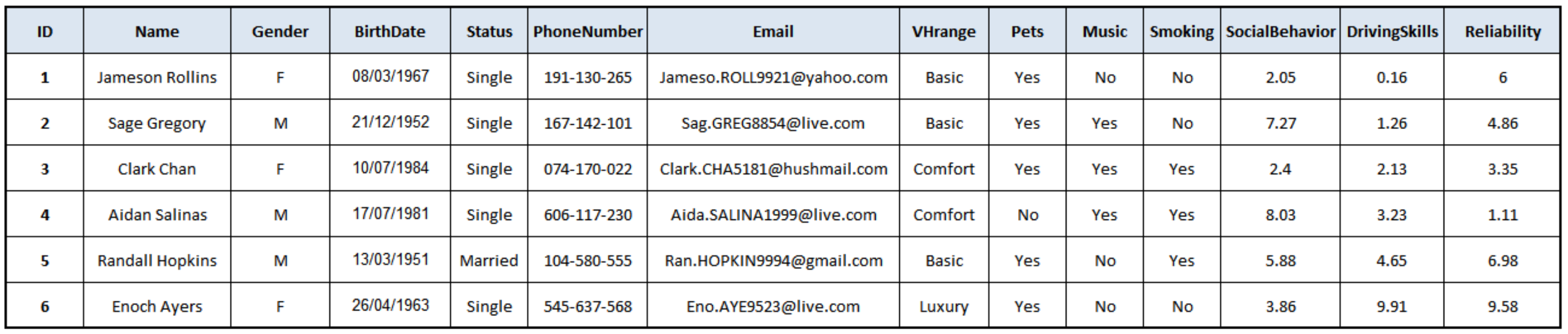}
\caption{Scenario 1- The drivers profiles} \label{f41}
\end{figure}

At the same time, six (6) drivers make a ridesharing system request, making the system aware that they fulfill the spatio-temporal constraints of the passenger $P1$. Fig.\ref{f41} presents the drivers profiles. 

\begin{figure}[h!]
\centering
\includegraphics[scale=0.9]{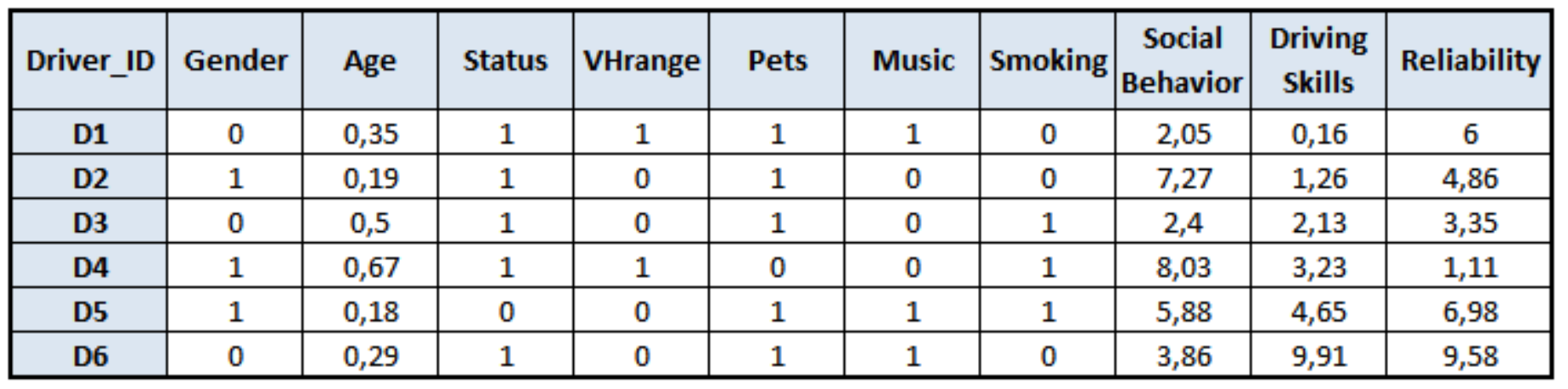}
\caption{Scenario 1- P1 judgment matrix} \label{f42}
\end{figure}

In the sequel, the two methods TOPSIS and WSM are separately responsible for evaluating simultaneously the multiple constraints and ranking possible matches for the passenger $P1$ according to his preferences. Specifically, taking into consideration the data provided in fig Fig.\ref{f40} and Fig.\ref{f41}, the judgment matrix of $P1$ presented in fig Fig.\ref{f42} is formed. 

\begin{figure}[h!]
\centering
\includegraphics[scale=0.5]{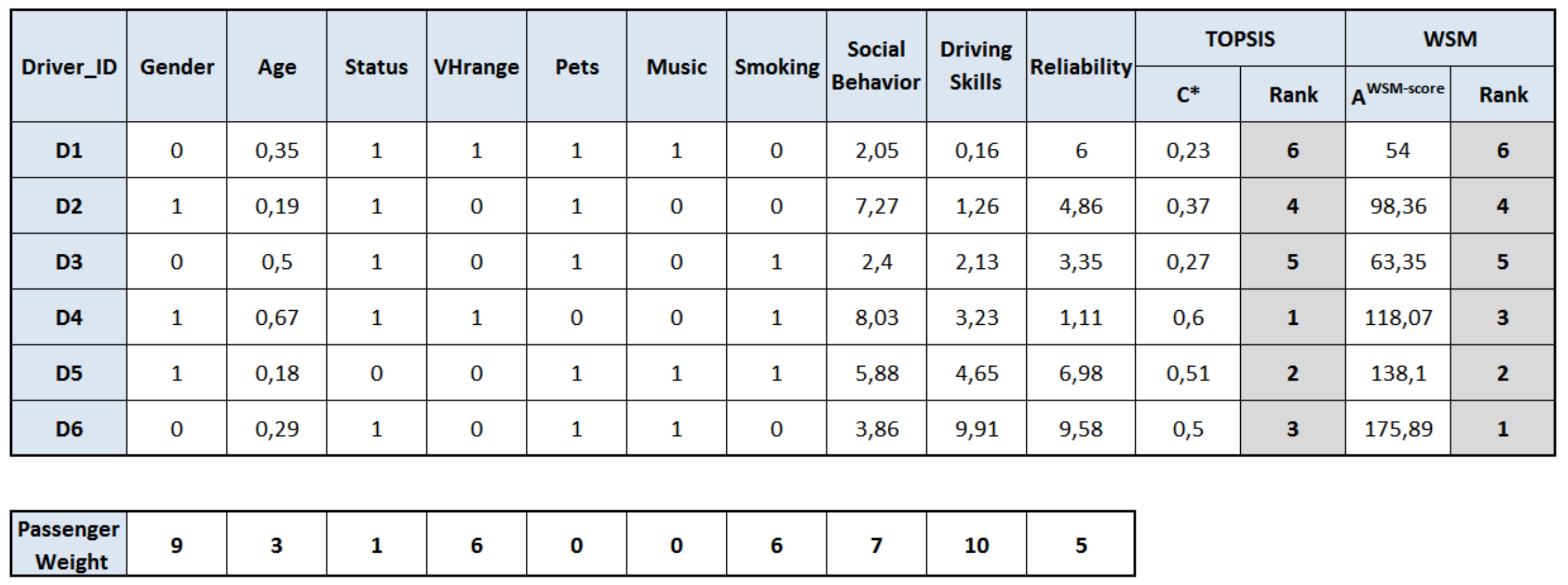}
\caption{Scenario 1- The result} \label{f43}
\end{figure}

Thereafter, for every candidate driver, we calculate his respective $C^*$ and SM values, based on the Judgment matrix and the passenger weights, as presented in fig Fig.\ref{f43}. Through ranking these values, we obtain two different preference lists of the passenger $P1$.
 
Based on the results, the preference list obtained by TOPSIS method is $PL_{TOPSIS} (P1) =D4 - D5 - D6 - D2 - D3 - D1$ and that obtained by the WS model is $PL_{WSM} (P1) = D6 - D5 - D4 - D2 - D3 - D1$. By the TOPSIS method, the first place is occupied by $D4$ as $D4$'s $C^*$ value is the highest, whereas it is occupied by D6 by the WS model as $D6$'s $A^{WSM-score}$ value is the highest. Observing these two drivers, we find that $D4$ complies with $P1$'s preferences regarding gendre, vhrange and smoking. In contrast, $D6$ does not meet any of these preferences. In addition, $D4$ exceeds $D6$ on the age and the social behavior preferences. All of these preferences have a total weight of $31$. On the other hand, $D6$ exceeds $D4$ only on the driving skills and the reliability preferences that have a total weight of $15$.
\subsubsection{Scenario 2: Single driver for a high priority preference }  
\begin{figure}[h!]
\centering
\includegraphics[scale=0.5]{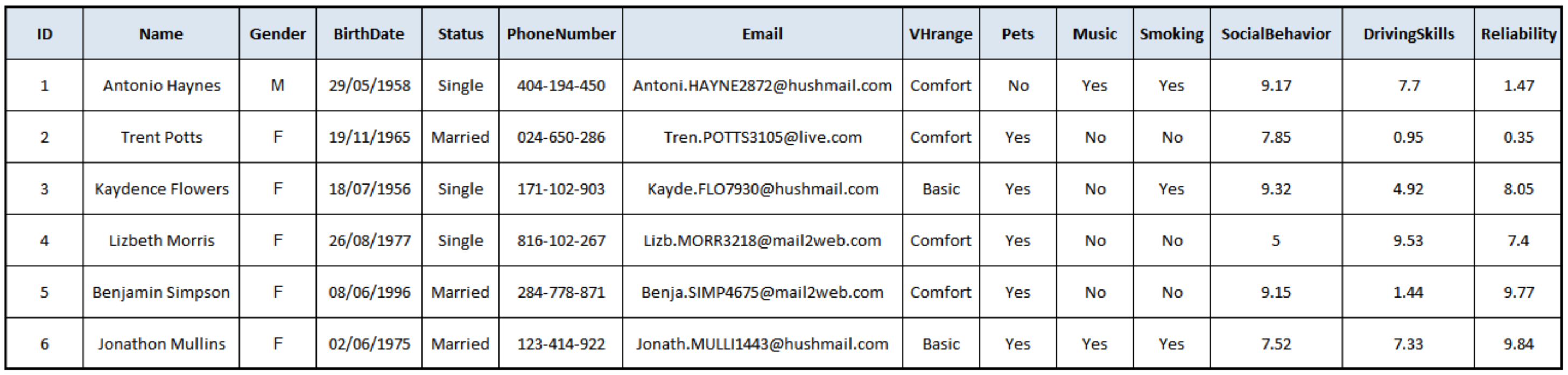}
\caption{Scenario 2- The drivers profiles} \label{f44}
\end{figure}

In this case, we consider the same passenger $P1$ with the same preferences and the same weights as shown in Fig.\ref{f40}. On the other hand, six (6) other drivers wish to make the same itinerary as $P1$ and seem to be close in time. Out of the 6 drivers, only one meets the preference of the gender that $P1$ considers with driving skills as more important factors to make the match. Fig.\ref{f44} presents the drivers profile.
 
\begin{figure}[h!]
\centering
\includegraphics[scale=0.8]{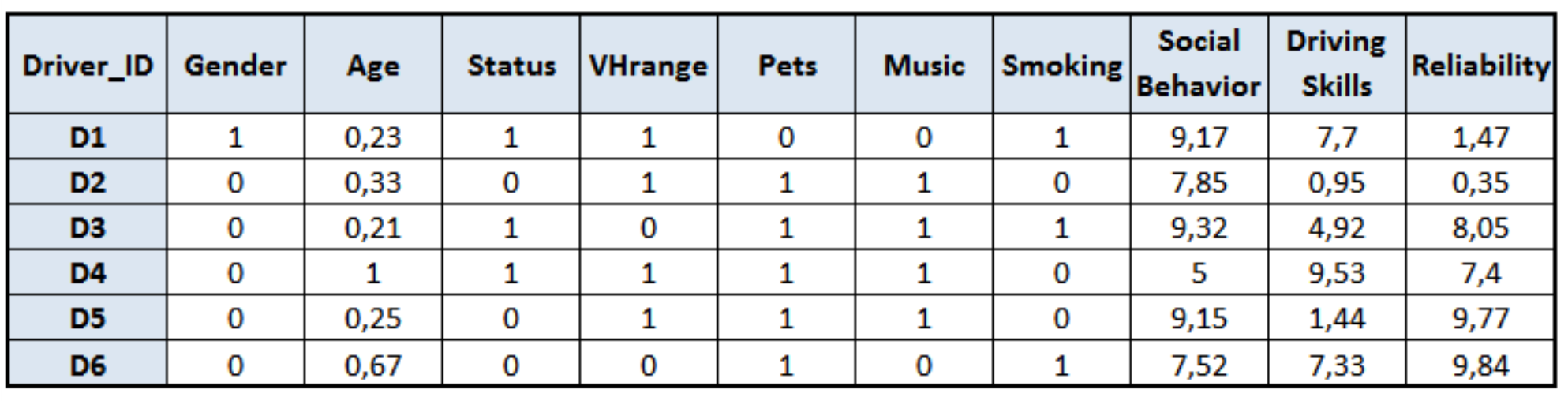}
\caption{ Scenario 2- P1 judgment matrix} \label{f45}
\end{figure}

\begin{figure}[h!]
\centering
\includegraphics[scale=0.5]{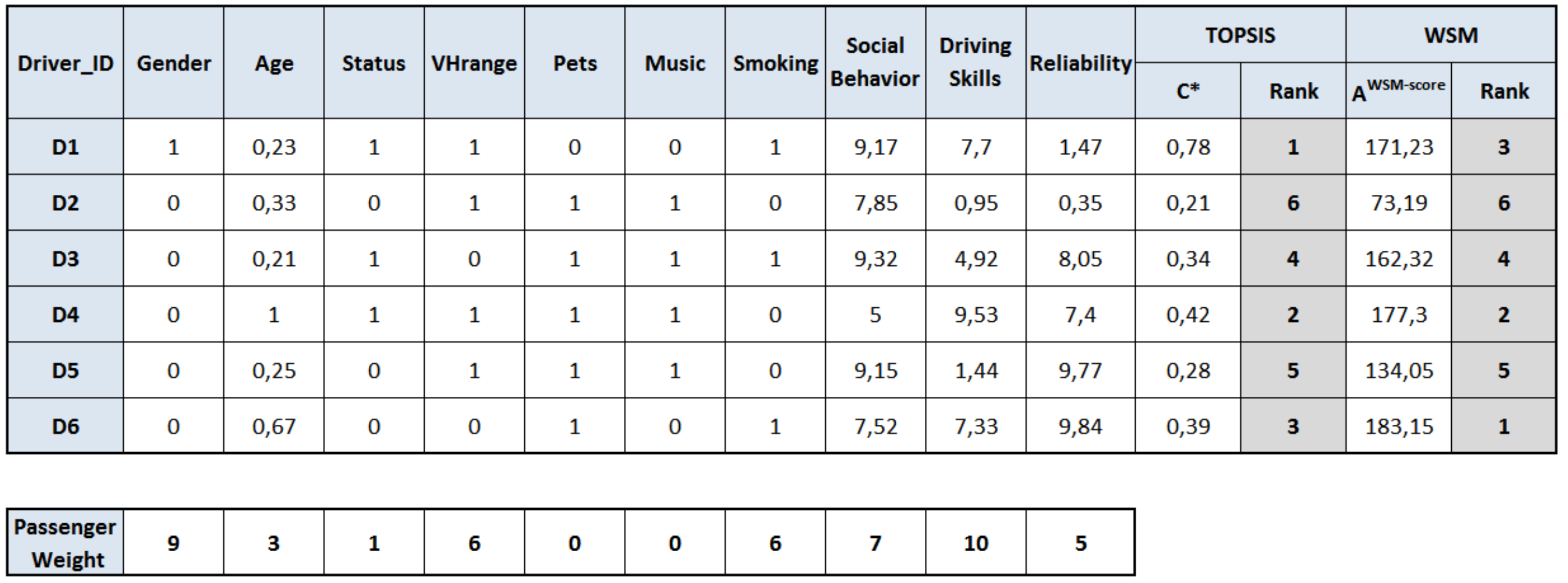}
\caption{ Scenario 2- The result} \label{f46}
\end{figure}

As explained in scenario 1, taking into consideration the data provided in Fig.\ref{f40} and Fig.\ref{f44}, the judgment matrix of $P1$ presented in fig Fig.\ref{f45} is formed. 
Then, based on the judgment matrix and the weights, we obtain the result shown in Fig.\ref{f46}.

As expressed in the figure above, the preference list obtained by TOPSIS method is $PL_{TOPSIS} (P1) = D1 - D4 - D6 - D3 - D5 - D2$ and that obtained by the WS model is $PL_{WSM} (P1) = D6 - D4 - D1 - D3 - D5 - D2$. TOPSIS method decides in favor of $P1$ and $D1$, as $D1$'s $C^*$ value is the highest, among the six candidate drivers. However, $D1$ occupies the third place and $D6$ the first place with the WS model. It is clear that $D1$ is the most appropriate match for $P1$, meeting him only the gender preference. It may also be observed that beside the satisfaction of the gender preference, $D$1 has better values than $D6$ in terms of social behavior and driving skills preferences. Furthermore, unlike $D6$, $D1$ fulfills the Status and the vhrange constraints. Therefore, $D1$ surpasses $D6$ with a total weight equal to $33$ $(9+1+6+7+10)$.
\subsubsection{Scenario 3: An extremely high preference } 
\begin{figure}[h!]
\centering
\fbox{
\includegraphics[scale=0.7]{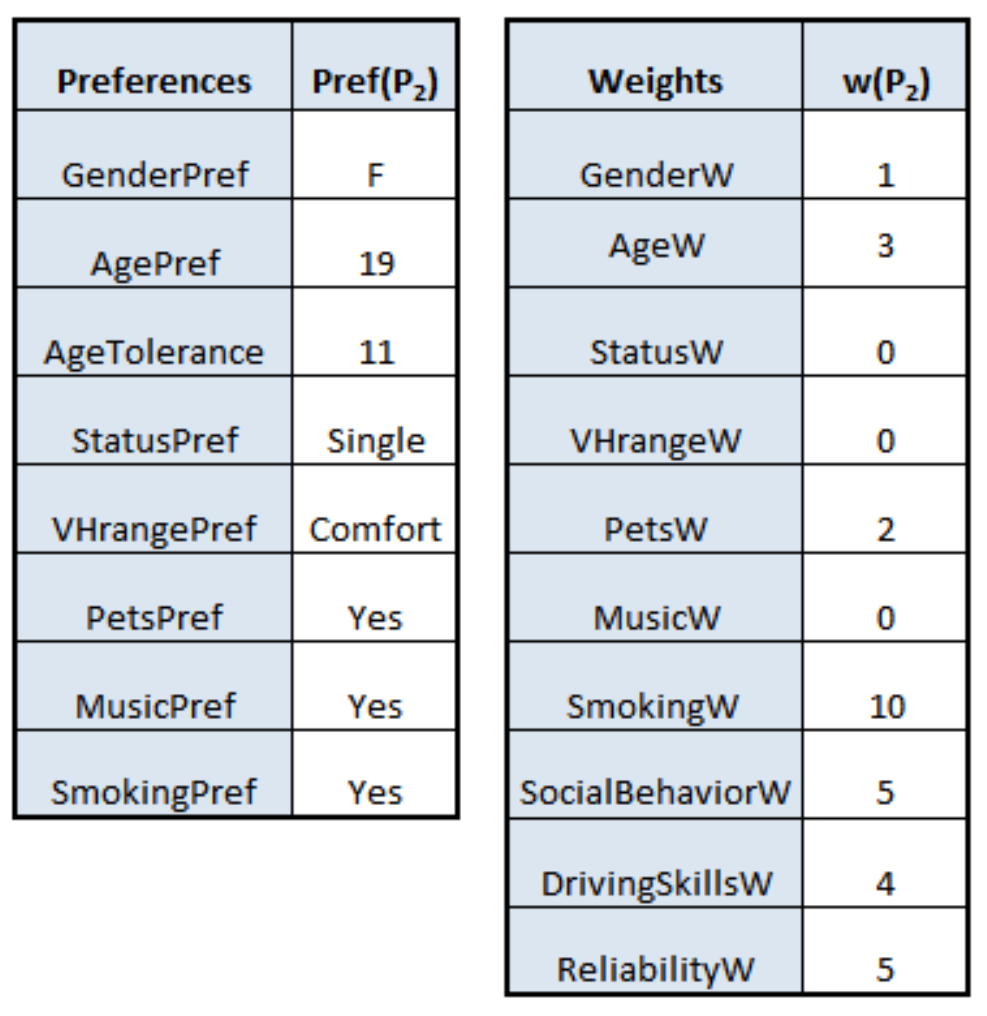}}
\caption{ Scenario 3- The preferences and weights vectors of passenger P2} \label{f49}
\end{figure}

In this case, stating the weight of Vhrange to $10$, passenger $P2$ wishes to be matched to a driver with a comfortable car.  Figure Fig.\ref{f49} describes $P2$'s preferences and their respective weights. We consider that the same six drivers of senario 1 (with profiles presented in Fig.\ref{f41}) fulfill also the spatio-temporal constraints of $P2$.

\begin{figure}[h!]
\centering
\includegraphics[scale=0.7]{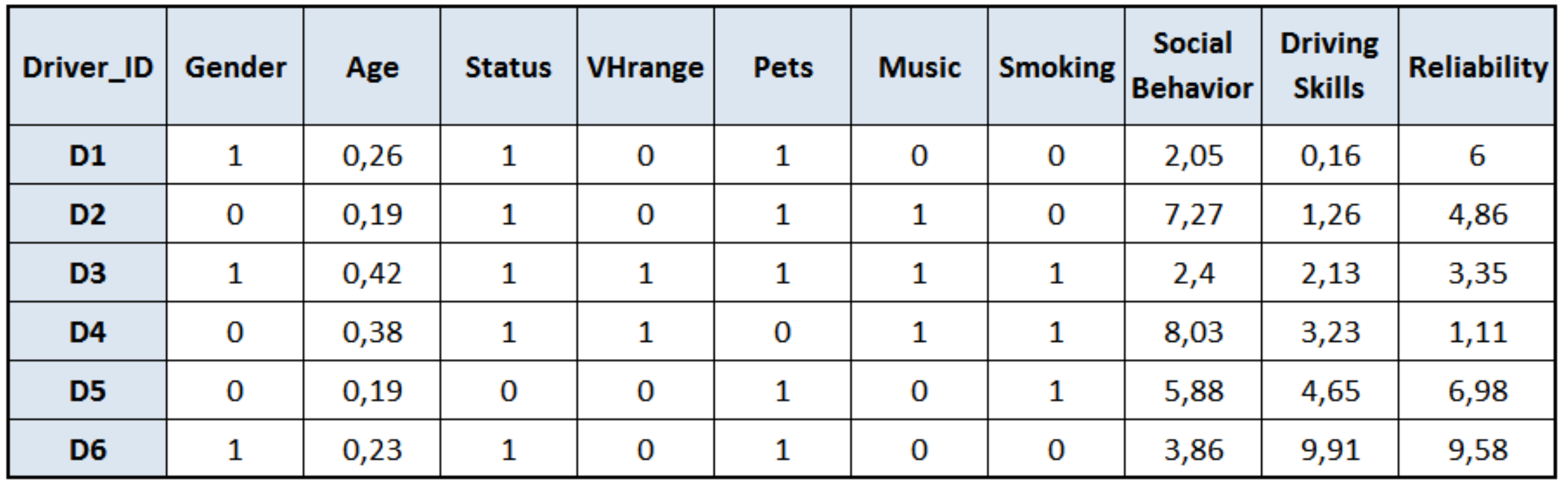}
\caption{ Scenario 3- P2 judgment matrix} \label{f47}
\end{figure}

\begin{figure}[h!]
\centering
\includegraphics[scale=0.5]{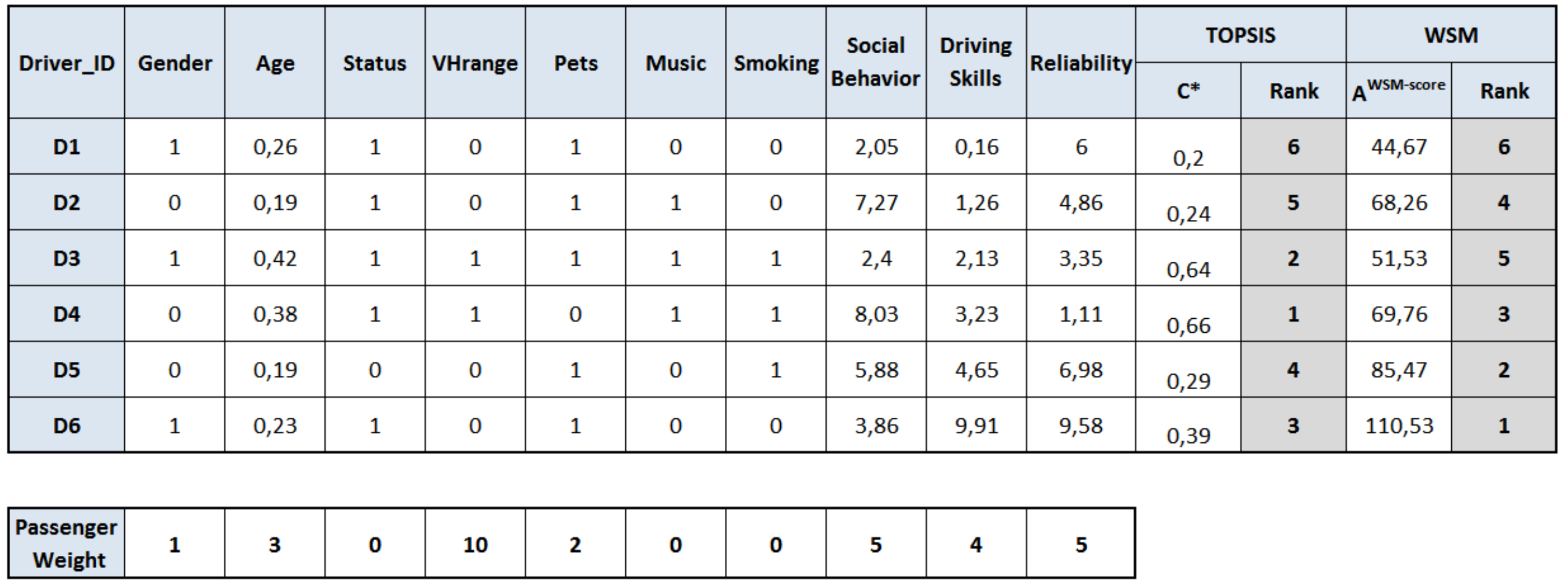}
\caption{ Scenario 3- The result} \label{f48}
\end{figure}

\begin{figure}[h!]
\centering
\fbox{
\includegraphics[scale=0.7]{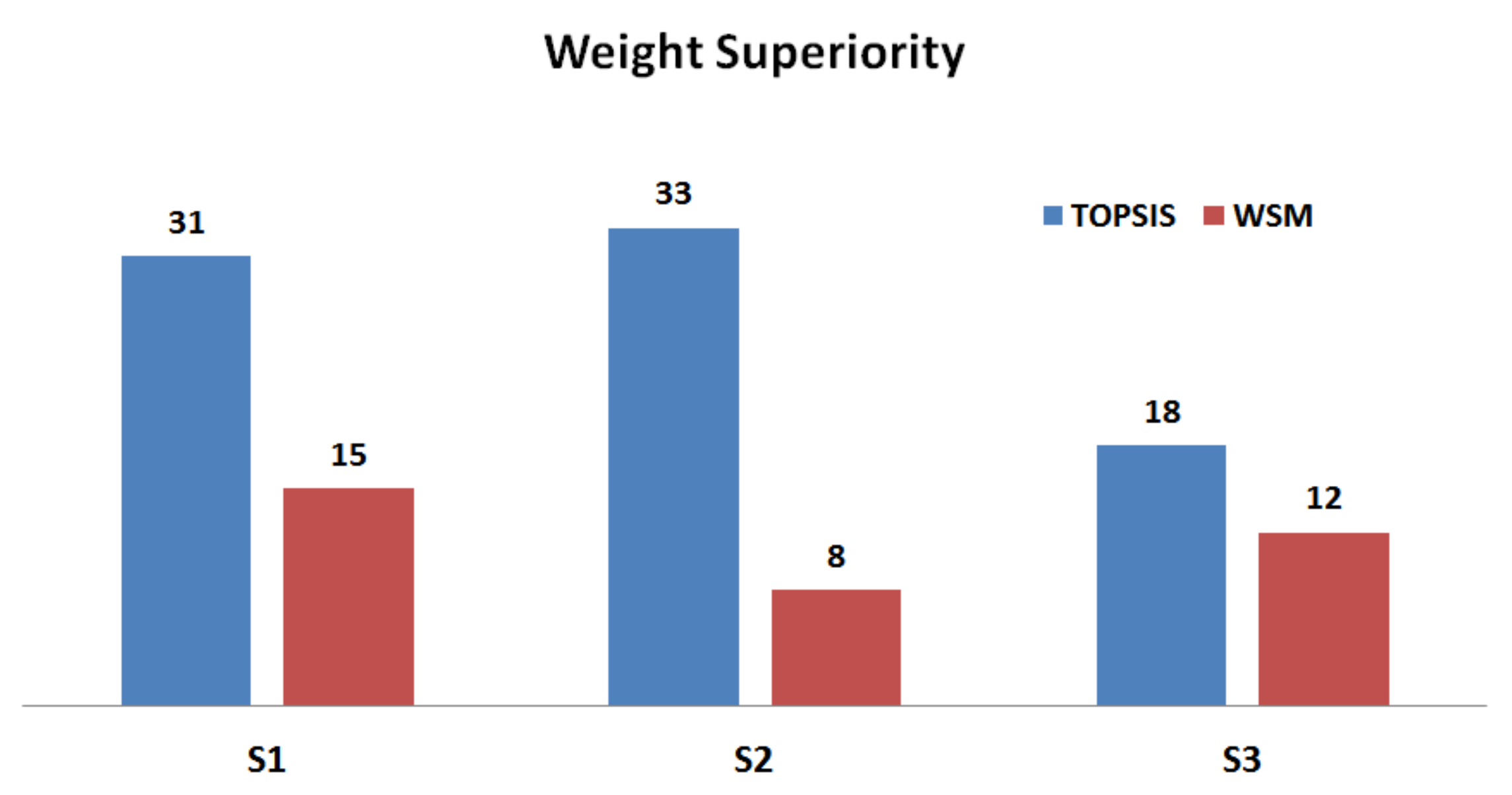}}
\caption{ Weight Superiority of the top lists} \label{f50}
\end{figure}

Following the examples of the previous scenarios, we compute the judgment matrix of $P2$ (presented in Fig.\ref{f47}). 
Then, based on the judgment matrix and the weights of P2, we obtain the result shown in Fig.\ref{f48}.
According to the results obtained in the figure above, TOPSIS method produces $PL_{TOPSIS}(P2) = D4 - D3 - D6 - D5 - D2 - D1$ whereas WS model produces $PL_{WSM}(P2) = D6 - D5 - D4 - D2 - D3 - D1$ as a preference list of $P2$.
As it may be observed, the passenger thinks of the vehicle range as a very important factor of the trip. Another point to underline is that only Drivers $D3$ and $D4$ meet this constraint.
These last two drivers occupy the first places of the $P2$'s preference list according to the TOPSIS method. However they occupy the third and the fifth place according to WS model. We also note that, beside selecting the most appropriate drivers regarding the highest preference, the head of the TOPSIS list ($D4$) transcends the head of the WSM list ($D6$) with a total weight equal to $18$ $(3+10+5)$.
\paragraph{}
In concluding, we manifest that our approach selects the most appropriate candidates with respect to the passenger preferences.
In the first scenario, the head of the TOPSIS list exceeds that of the WS model on preferences with a total Weight Superiority ($W^{Superiority}_{D4/D6}$) eqaul to 31. However, the head of the WS model list exceeds that of the TOPSIS on preferences with a Weight Superiority ($W^{Superiority}_{D6/D4}$) of 15.
In the second scenario, the first with TOPSIS surpasses that with WS with a Weight Superiority ($W^{Superiority}_{D1/D6}$) equal to 33. The latter only overtakes the top of the TOPSIS list on preferences with a Weight Superiority ($W^{Superiority}_{D6/D1}$) of 8. Finally, the same result is obtained in the third scenario. Indeed, the top of the TOPSIS list exceeds the top of the WS list with a Weight Superiority ($W^{Superiority}_{D4/D6}$) equal to 18. On the other hand, the top of the WSM list exceeds the top of the TOPSIS list with a Weight Superiority ($W^{Superiority}_{D6/D4}$) equal to 12.
Fig.\ref{f50} graphically presents the Weight Superiority obtained by the heads of the lists of the two methods in the three scenarios.

\subsection{Stable matching}
This section contains indicative results that derive from the utilization of the third component of SMRM to a simulated ridesharing infrastructure environment. 
\subsubsection{Test scenarios}
We choose to compare our Stable Matching $SM$ component in terms of the previously presented metrics against  the classical Gale-Shapley $GS$ algorithm since only the approach presented in \cite{c4r11} integrated the problem of stable marriage. The latter relied on the $GS$ algorithm and assumed that the number of passengers was equal to that of drivers. For this reason, we used a data set with equal sets of passengers and drivers. On the other hand, we used unequal sets data source to show the performance of our system without comparing its performance to that of $GS$ algorithm sine the latter does not support unequal sets. We use data sizes of up to 1,000, while we generate synthetic data of diverse skewness and types. In our experiments, the egalitarian cost and sex equality cost metrics are normalized, i.e. divided by $n$.
\\
\begin{figure}
    \centering
    \includegraphics[scale=0.5]{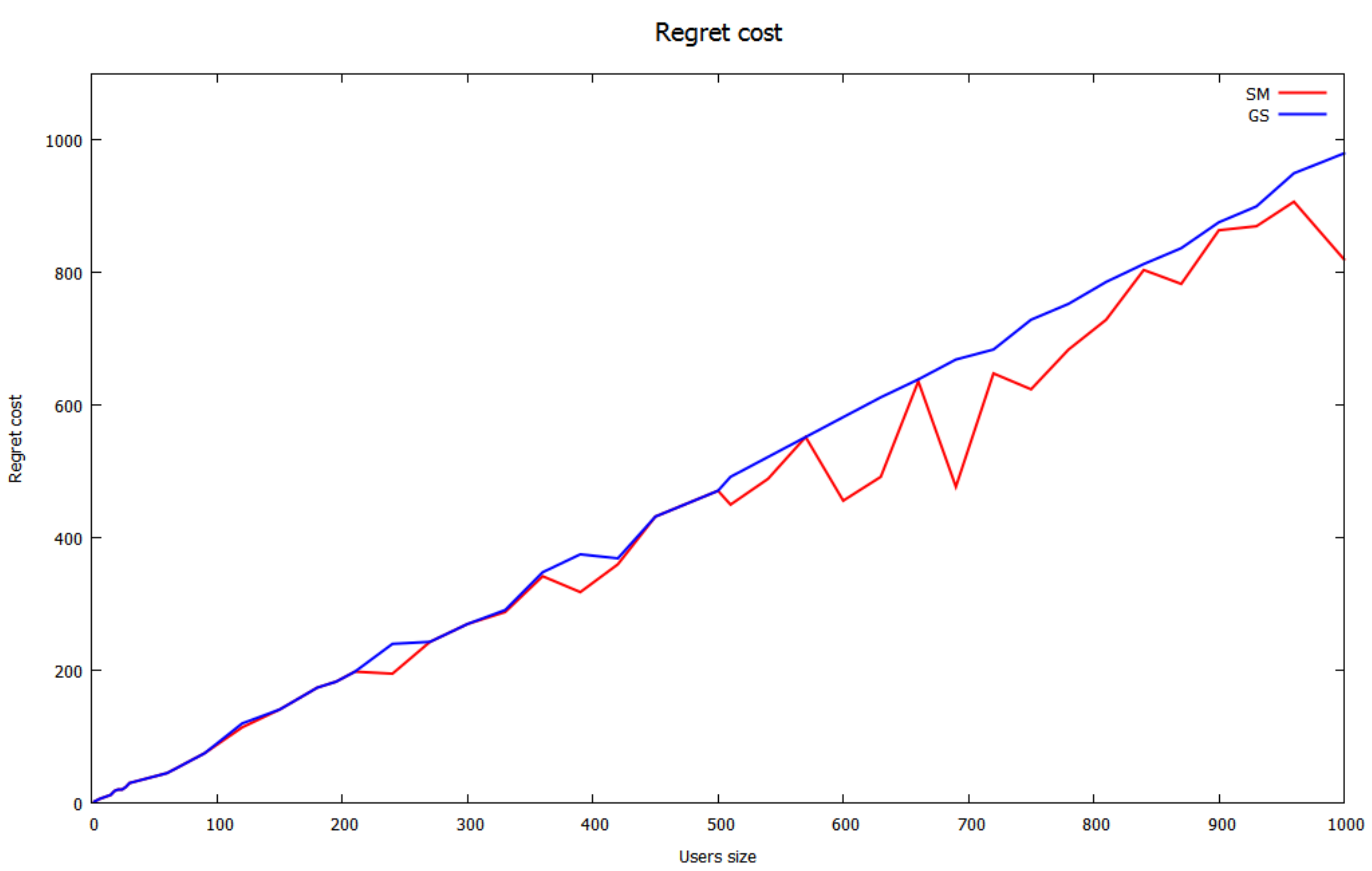}
    \caption{Regret cost of stable matching within equal sets}
    \label{sub1}
\end{figure}
\begin{figure}
    \centering
    \includegraphics[scale=0.5]{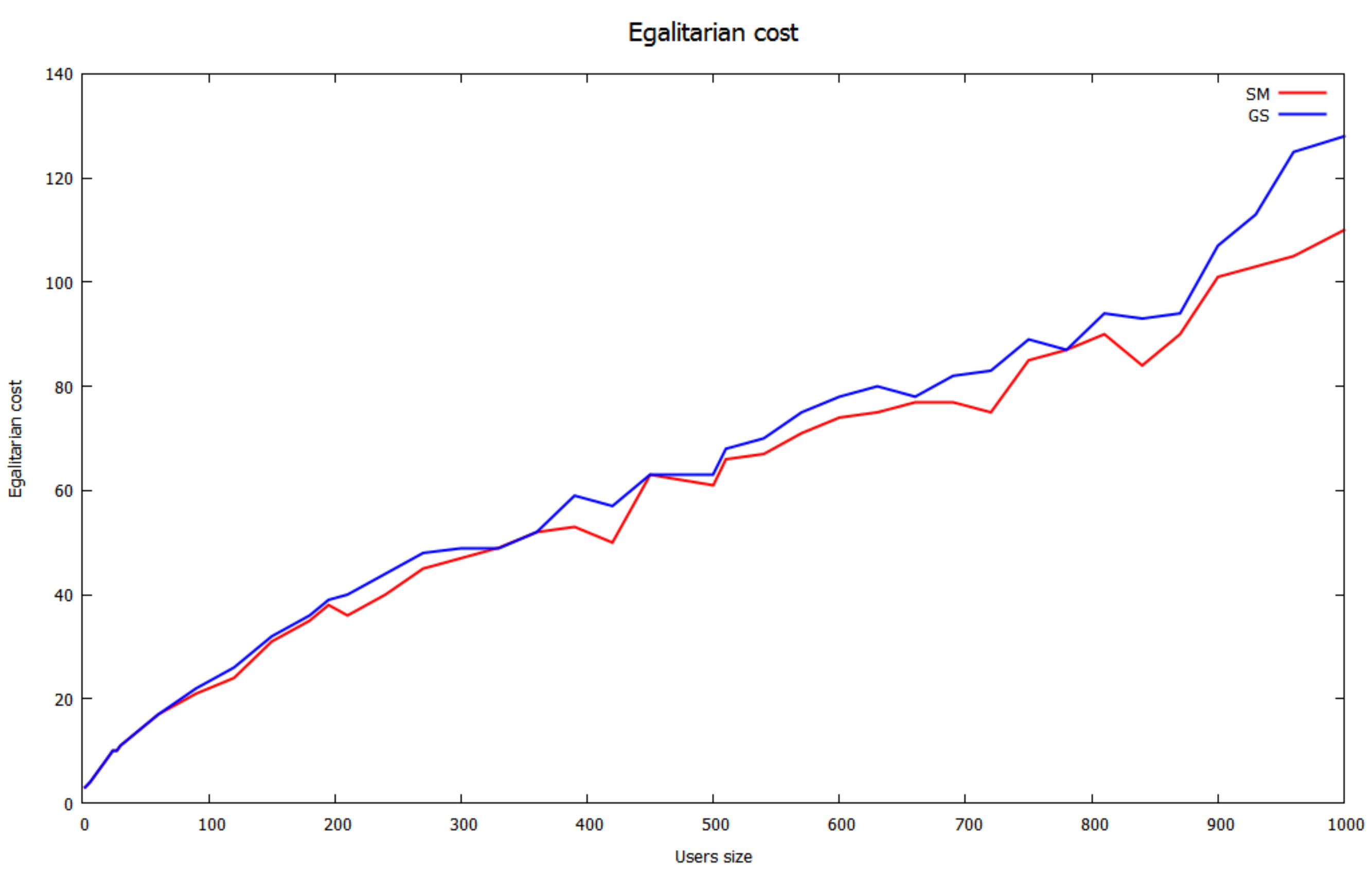}
    \caption{Egalitarian cost of stable matching within equal sets}
    \label{sub2}
\end{figure}
\begin{figure}
    \centering
    \includegraphics[scale=0.5]{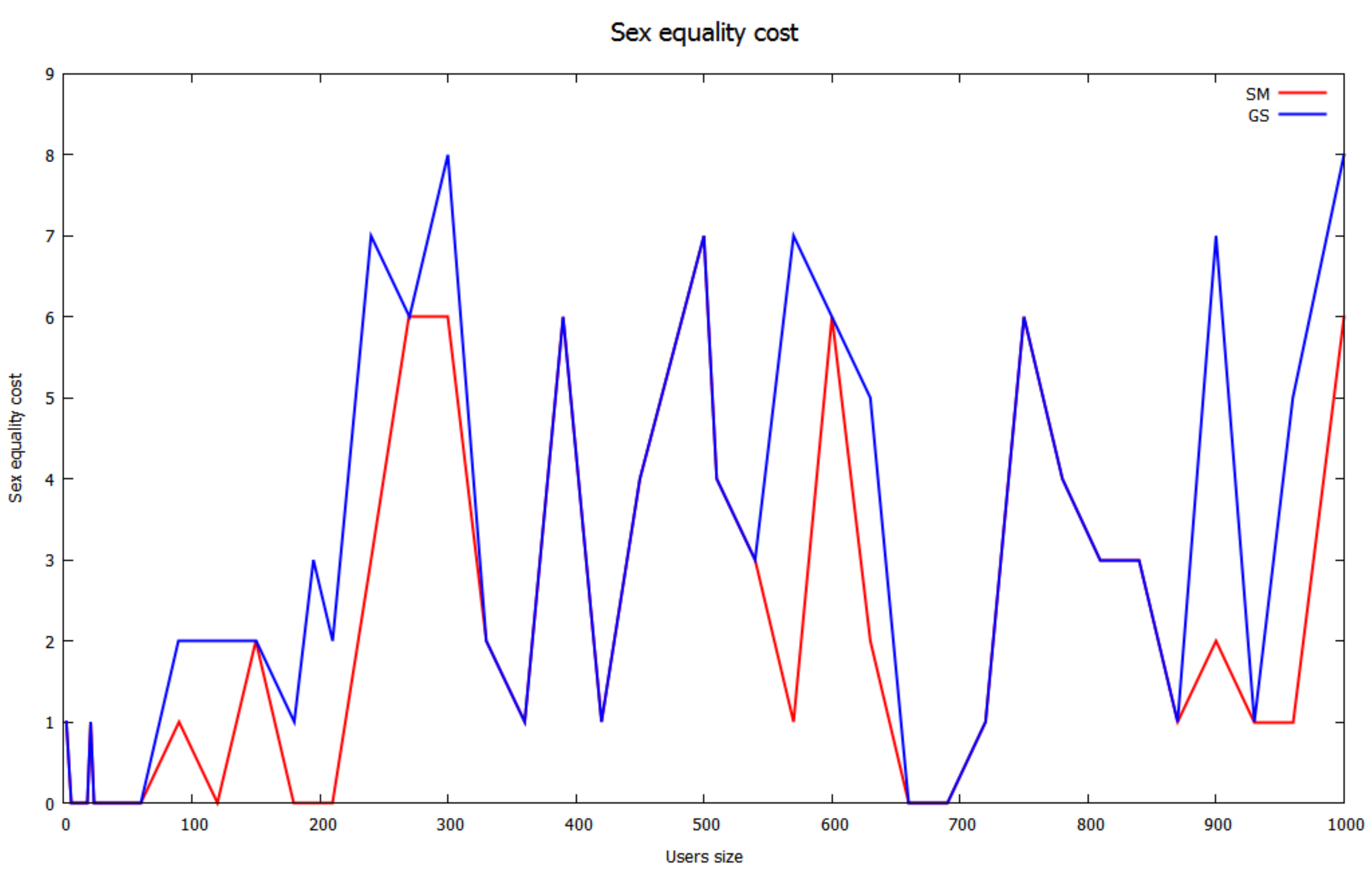}
    \caption{Sex equality cost of stable matching within equal sets}
    \label{sub3}
\end{figure}

\begin{figure}
    \centering
 \includegraphics[scale=0.5]{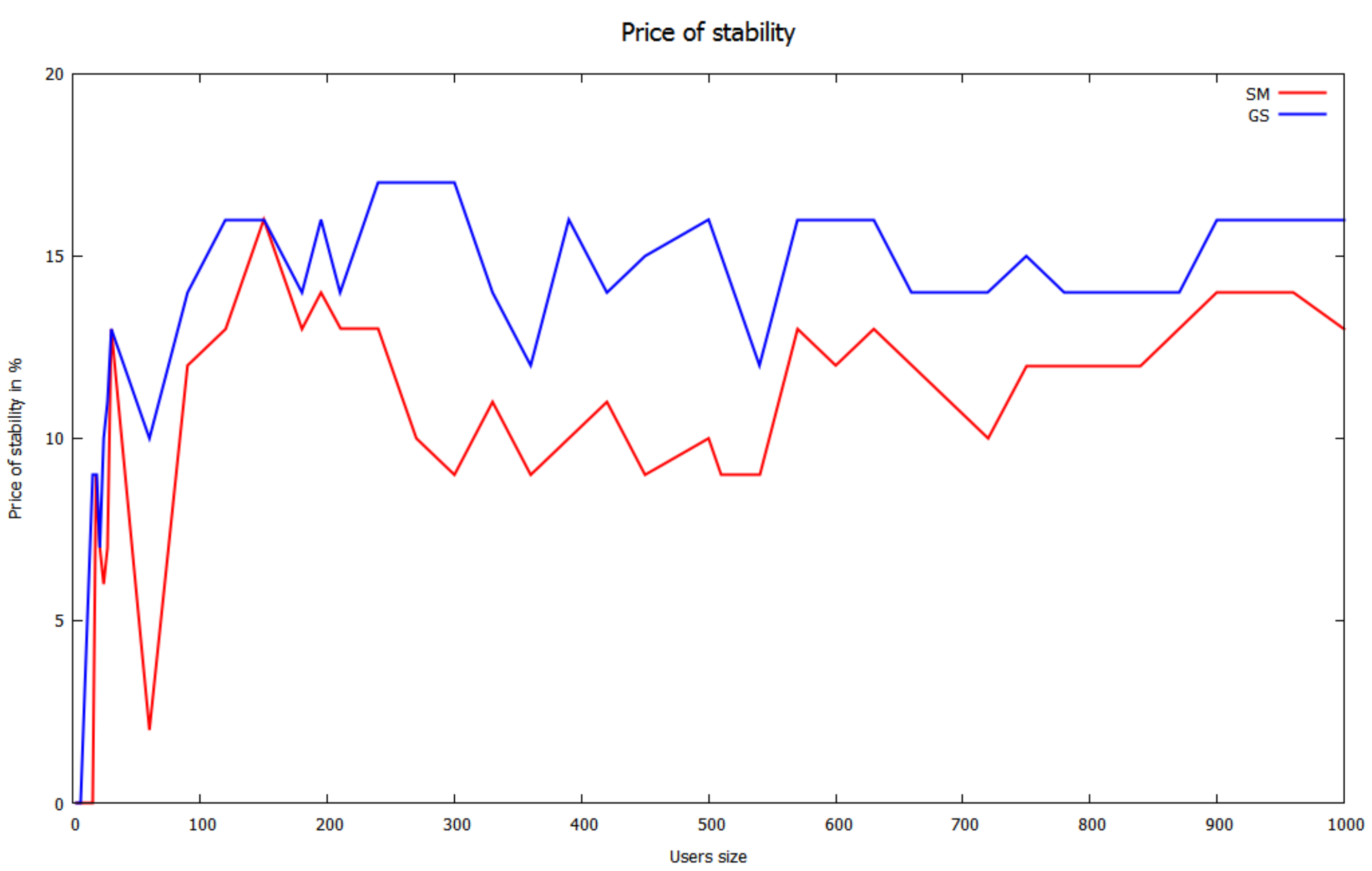}
      \caption{Price of stability within equal sets}
      \label{price}
\end{figure}
\begin{figure}
     \centering
     \includegraphics[scale=0.5]{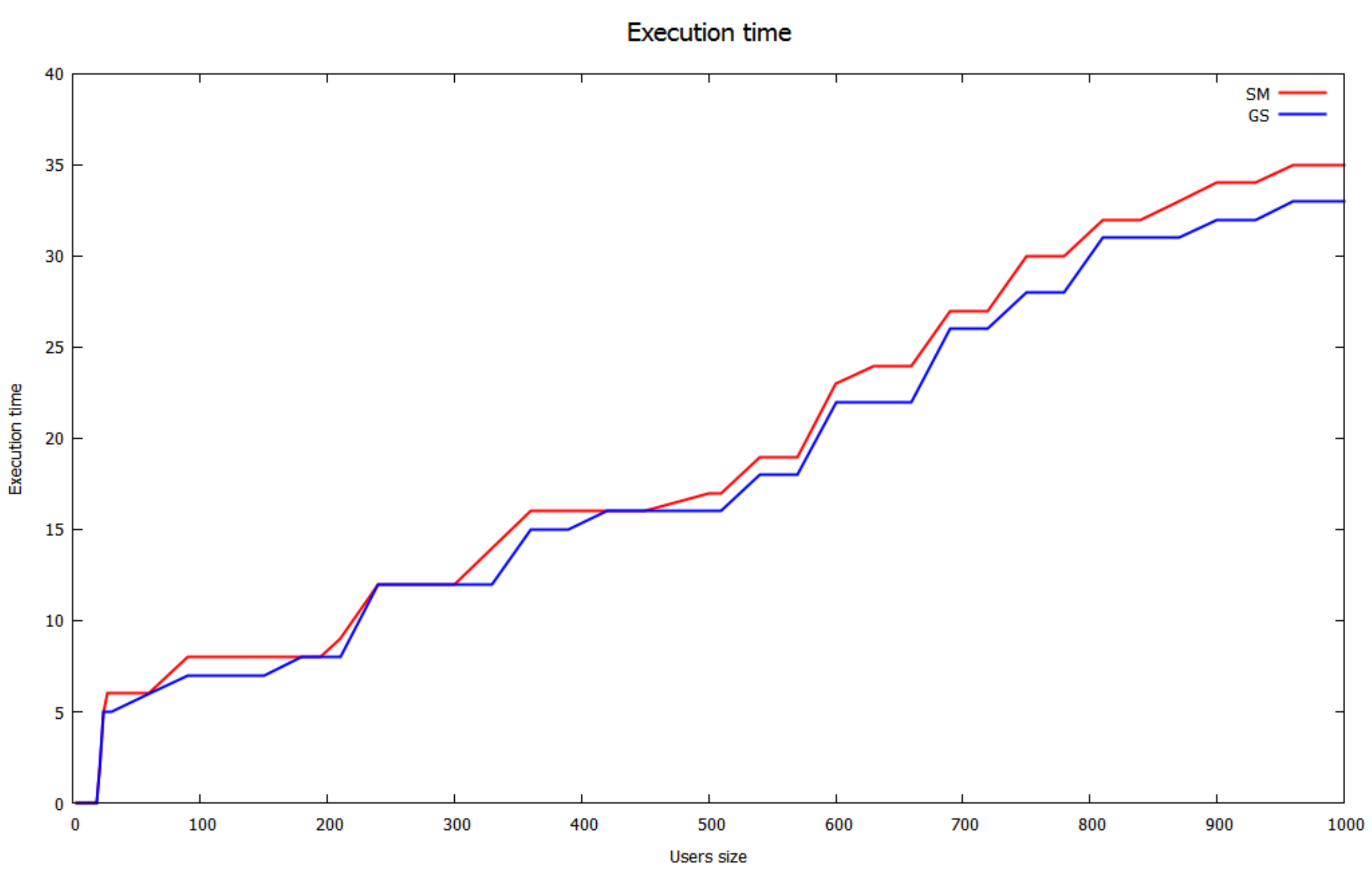}
      \caption{Execution time within equal sets}
      \label{time}
\end{figure}

\begin{figure}
    \centering
    \subfigure[Drivers=500]{\label{sub21} \includegraphics[scale=0.5]{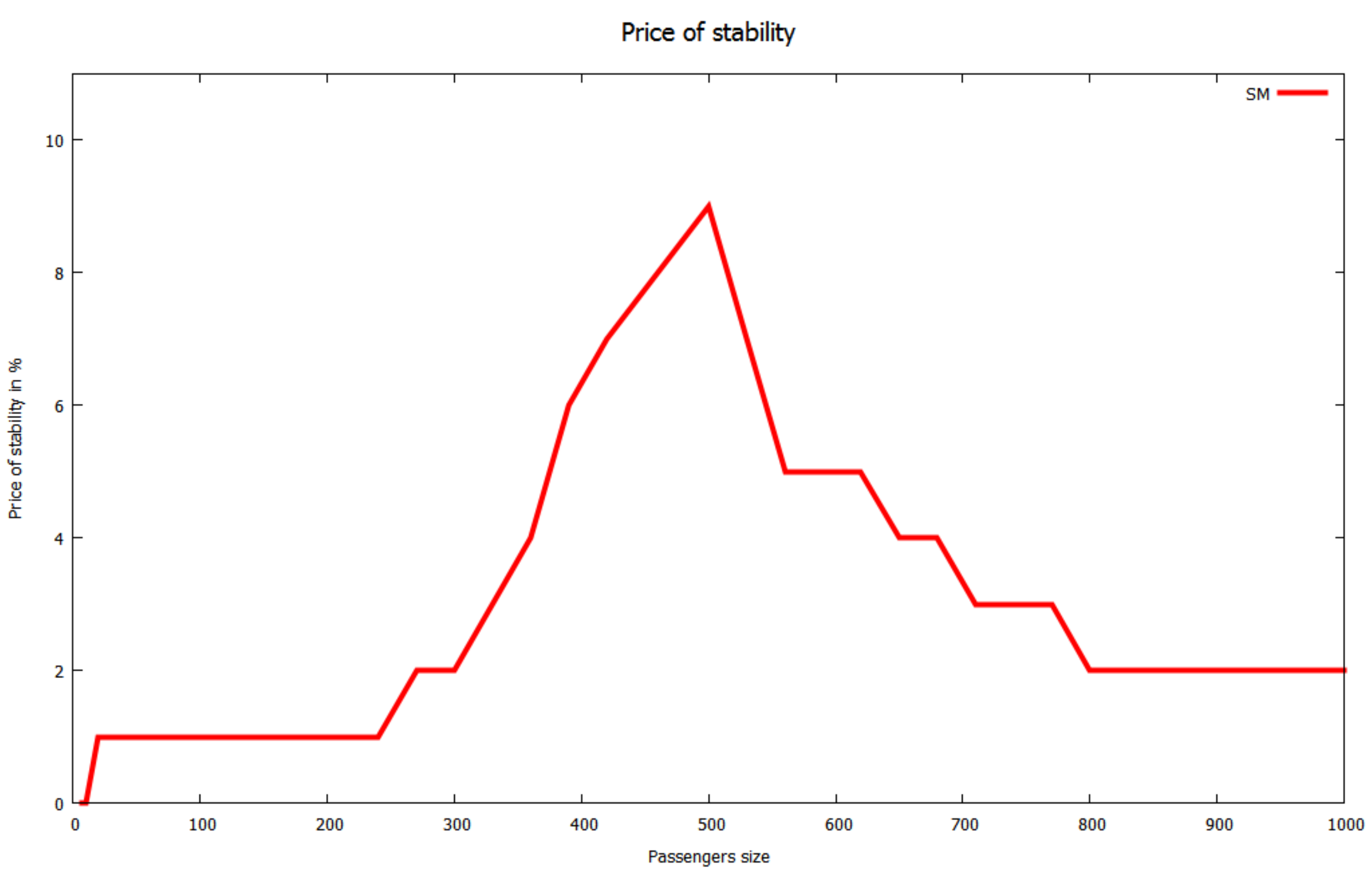}}
    \subfigure[Passengers=500]{\label{sub22} \includegraphics[scale=0.5]{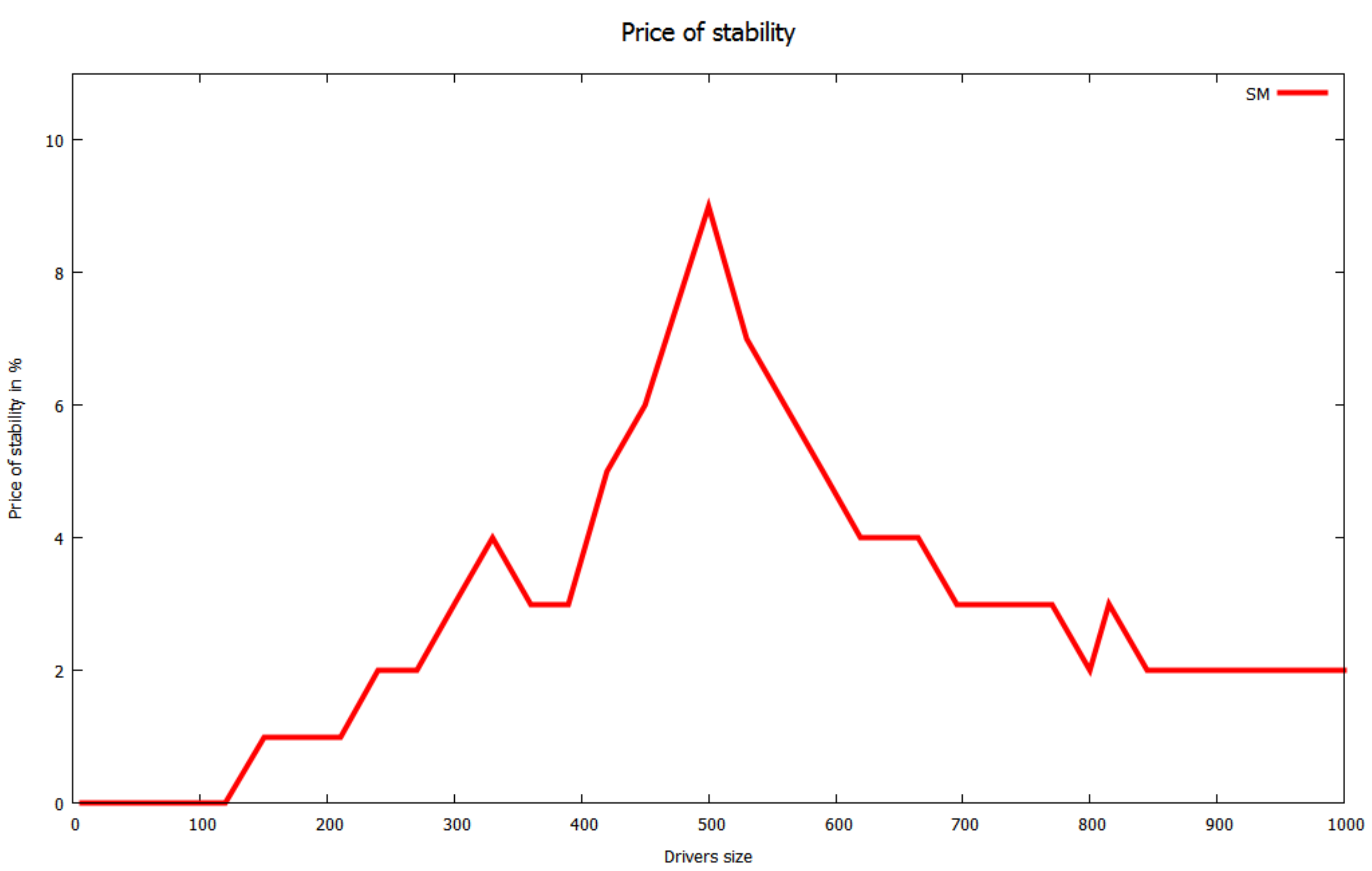}}
    \caption{Variation of the   price of stability within inequal sets}
    \label{price2}
\end{figure}

\subsubsection{Results}
 Fig.\ref{sub1}, Fig.\ref{sub2} and Fig.\ref{sub3}  show the evolution of the regret cost, the egalitarian cost and the sex equality cost of our algorithm compared to the GaleShapley algorithm w.r.t. the number of users in the system. 
We observe that the regret cost and the egalitarian cost of both algorithms increase as far as the number of users in the system increases. 
However, we note that the sex equality cost is independent of the users' number. This is explained by the fact that assigning equal importance to the preferences of passengers and drivers is not influenced by the number of users.\\

In addition, Fig.\ref{sub1}, Fig.\ref{sub2} and Fig.\ref{sub3} demonstrate that the two algorithms follow the same behavior within the three metrics. However, we note that $GS$ achieves in many cases far worse quality in all metrics. Indeed, $SM$ performs 5\% better than $GS$ in terms of regret cost and egalitarian cost. Consequently, in $SM$ algorithm the preferences of every individual are more considered to be equally important, i.e. it minimizes better the difference in happiness of all the passengers and the drivers. Similarly, we see in Fig.\ref{sub3} that $SM$ algorithm achieves a superior performance of 20\% in terms of sex equality cost, which indicates that, in our solution, the drivers are as pleased with the matching as the passengers.\\

Since we model a ridesharing provider that tries to achieve a right stable outcome, studying the effect of enforcing stability is worth of interest. Fig.\ref{price} investigates how large the price of stability might be when using the two different algorithms $SM$ and $GS$. As expected, we remark that the price of stability is independent of the users' number  and it is relatively significant and represents a 10\% approximate reduction of the objective value with $SM$ algorithm. Nevertheless, the costs of enforcing stability are significantly higher in $GS$ with a worse price of stability of 22\% compared to $SM$.\\

Studying the impact of the number of users on the  execution time of the different algorithms is of paramount importance. Indeed, Fig.\ref{time} shows that the execution time of the three algorithms varies substantially according to the users' number. $SM$ runs in approximately the same time as $GS$, whereas the latter slightly outperforms $SM$ in some cases. This is owing to the fact that within the $SM$ algorithm, additional tests are performed in order to check the  set sizes.\\

To assess the performance of our system in an unequal set case, we present the price of stability in different data sizes. In Fig.\ref{price2} \subref{sub21}, we vary the number of passengers from 5 to 1000 while fixing the number of drivers to 500. In Fig.\ref{price2} \subref{sub22}, we vary the number of drivers from 5 to 1000 while fixing the number of passengers to 500. 
The figures demonstrate that enforcing equal datasets comes at the cost of deteriorating the system-wide solution quality. Indeed, the price of stability is relatively small and represents a 3\% approximate reduction in the objective value, however as expected, it reaches a peak when the size of drivers is equal to that of passengers. This is explained by the originality and efficiency of the $SM$ algorithm, for unequal sets, which finds the optimal stable solution for the smaller set. In fact, $SM$ operates such that we obtain the smaller set's optimal solution. Thus, if there are fewer drivers than passengers, the drivers' optimal solution is reached. Nevertheless, if there are more drivers than passengers, the passengers' optimal solution is obtained. This stable solution is close to the unconstrained optimal solution.

\section{Conclusion}	
The solution we propose is particularly well suited to our purpose. It is in this context that the main lines of this chapter are written. Indeed, to make the theoretical performances coincide with the practical performances, we took good care to make the appropriate technical, technological and architectural choices and that would not interfere with our projects of optimization of the quality of service.
According to our strategic and methodological choices of analysis, design, modeling and specification of our system, we have brought into play the concepts of operational research, artificial intelligence and sustainable development.
An extensive experimental evaluation has shown that SMRM performs better than its competitor in terms of multi-criteria evaluation, stability quality and stability cost. The results have also demonstrated the efficiency of our solution, for unequal sets, which finds the optimal stable solution for the smaller set.

\chapter*{Conclusion}
\addcontentsline{toc}{chapter}{Conclusion}
The car occupies an important place in the individuals' life and remains always a source of very important and essential mobility. This does not preclude the fact that it has several disadvantages from several points of view (financial, environmental, societal, etc.).
\paragraph{}
In order to meet the needs of innovation while remaining in a respectful framework of fundamental environmental conditions for a healthy climate, ridesharing is presented as the key solution for a development without harmful consequences in addition to being of very advantageous economic order.
\paragraph{}
Based on an elaborate study of this concept, we have outlined the main lines of our proposal based on the limitations of existing works. We plan then through this master to remedy the absence of personal constraints and stability of matching aspects that make great weakness of the existing systems. Our interests, being among others of a practical nature, aim at making available to the general public, particular users or researchers an operational and friendly system. But also a system that promises the satisfaction of drivers and passengers' social preferences, considering the notion of stability for rideshare matches.
\paragraph{}
We focused our efforts on the implementation of an effective strategy of resolution taking advantage of a mixture of concepts; namely the MCDMs and the SMP. An alliance of these two fundamentally rich concepts was the main contribution of our SMRM system.
\paragraph{}
We broke SMRM into three main components: Preference Satisfier, Multi-Criteria Ranking, and Stable Matching. The Preference Satisfier component computes a judgment matrix for each user based on their preferences and weights as well as on the profile and received evaluations of each correspondent. Then, the Multi-Criteria Ranking component is based on TOPSIS method as correspondents ranking tool to evaluate the multiple constraints simultaneously. Finally, the stable matching component relies on the approach of stable marriage and returns the stable matching where there is no pair of participants that both prefer each other to the partners that they are matched to.
\paragraph{}
An extensive experimental evaluation has shown that SMRM selects the most appropriate candidates with respect to the passenger preferences. In addition, we manifest that the stable matching component performs better than its competitor in terms of stability quality and stability cost. The results have also demonstrated the efficiency of our solution, for unequal sets, which finds the optimal stable solution for the smaller set.
\paragraph{}
As future work, we aim to develop machine learning techniques that could create collective knowledge on user preferences, which will be exploited by SMRM in making the convenient recommendations. 
Indeed, with the prevalence of GPS-enabled devices, the exponentially growing popularity of social networking and the voluntary sharing of personal information online, we not only learn about the users' life experiences but also about their life modes. Ridesharing recommendations from online data can be seen as a kind of personal optimizing service, which may help to improve users experience on ridesharing. 
\paragraph{}
Additionally, in our approach, we considered the situation, in which drivers and passengers fulfill spatio-temporal constraints. This work could be extended on developing a spatio-temporal matching model using a skyline method. Thereby, the Skyline query is used to find a set of non-dominated correspondents by filtering the interesting matches of a potentially large dataset based on origin, destination, and time constraints.
With a growing number of users involving multi-constraints, skyline queries can be used to answer this problem accurately and efficiently.

\nocite{hamrouni2008succinctt}
\nocite{draheim2017generalized}
\nocite{draheim2017semantics}
\nocite{auer2009extending}
\nocite{draheim2010service}
\nocite{atkinson2010typed}
\nocite{mouakher2019efficient}
\nocite{mouakher2016qualitycover}
\nocite{houari2018nbf}
\nocite{houari2018new}
\nocite{ferjani2012formal}
\nocite{bouker2014mining}
\nocite{hamrouni2013looking}

\nocite{allani2016dpms}
\nocite{allani2016novel}
\nocite{allani2018scalable}

\backmatter

\addcontentsline{toc}{chapter}{\bibname}
\bibliographystyle{unsrt}
\bibliography{biblio.aux}

\begin{thebibliography}{10}

\bibitem{r35}
{\em Statista The portal for statistics}, accessed April, 2018).
\newblock
  \url{https://www.statista.com/statistics/281134/number-of-vehicles-in-use-worldwide/}.

\bibitem{rr8}
Huguenin-Richard Florence.
\newblock Mobilité urbaine : de l'automobilisme à l'éco-mobilité. un long
  chemin...
\newblock {\em Vincent Moriniaux. Mobilités, Armand Colin}, pages 109--137,
  2010.

\bibitem{rr9}
Jean-Christophe Ballet and Robert Clavel.
\newblock Le covoiturage en france et en europe : état des lieux et
  perspectives.
\newblock {\em Centre d'études sur les réseaux, les transports, l'urbanisme et
  les constructions publiques}, pages 1--88, 2009.

\bibitem{rr2}
{\em Energy, transport and environment indicators}.
\newblock 2017 edition.
\newblock
  \url{http://ec.europa.eu/eurostat/documents/3217494/8435375/KS-DK-17-001-EN-N.pdf}.

\bibitem{rr3}
Rocci Anaïs.
\newblock {\em Journée ATEC-ITS, regards croisés sur l'offre de véhicules en
  libre service}.
\newblock Journée ATEC-ITS, regards croisés sur l'offre de véhicules en libre
  service, 2008.

\bibitem{rr4}
Lucie TORTEL.
\newblock {\em La voiture, cet incontournable objet du désir: le rapport de
  l'individu à la voiture: approche psychologique, approche sémiologique,
  approche philosophique, approche sociologique}.
\newblock Certu, 2001.

\bibitem{rr6}
Alberto Bull.
\newblock {\em Traffic Congestion: The Problem and how to Deal with it}.
\newblock Economic Commission for Latin America and the Caribbean, Deutsche
  Gesellschaft für Technische Zusammenarbeit, United Nations, 2003.

\bibitem{rr10}
{\em Ademe, dossier de press : Mobilite durable}.
\newblock 2013 edition.
\newblock
  \url{http://www.ademe.fr/sites/default/_les/assets/documents/dossier-presse-mobilite-durable.pdf}.

\bibitem{c2r2}
{\em Careers at BlaBlaCar}, accessed April, 2018.
\newblock \url{https://www.cleverism.com/company/blablacar/}.

\bibitem{c2r3}
{\em Covoiturage: IDvroom, le BlaBlaCar de la SNCF}, accessed April, 2018).
\newblock
  \url{https://www.linformaticien.com/actualites/id/34040/covoiturage-idvroom-le-blablacar-de-la-sncf.aspx}.

\bibitem{c2r4}
Niels Agatz, Alan Erera, Martin Savelsbergh, and Xing Wang.
\newblock Optimization for dynamic ride-sharing: A review.
\newblock 223:295--303, 12 2012.

\bibitem{c2r5}
David Pentico.
\newblock Assignment problems: A golden anniversary survey.
\newblock 176:774--793, 01 2007.

\bibitem{c2r6}
Roberto Baldacci, Vittorio Maniezzo, and Aristide Mingozzi.
\newblock An exact method for the car pooling problem based on lagrangean
  column generation.
\newblock {\em Operations Research}, 52(3):422--439, 2004.

\bibitem{c2r7}
Jean-François Cordeau and Gilbert Laporte.
\newblock The dial-a-ride problem (darp): Models and algorithms.
\newblock 153:29--46, 06 2007.

\bibitem{c2r8}
Drews Florian and Luxen Dennis.
\newblock Multi-hop ride sharing.
\newblock 2013.

\bibitem{c2r9}
Wesam Herbawi and Michael Weber.
\newblock Evolutionary multiobjective route planning in dynamic multi-hop
  ridesharing.
\newblock 6622:84--95, 04 2011.

\bibitem{r1}
Wen He, Deyi Li, Tianlei Zhang, Lifeng An, Mu~Guo, and Guisheng Chen.
\newblock Mining regular routes from gps data for ridesharing recommendations.
\newblock In {\em Proceedings of the ACM SIGKDD International Workshop on Urban
  Computing}, UrbComp '12, pages 79--86, New York, NY, USA, 2012. ACM.

\bibitem{r2}
Angela Di~Febbraro, E~Gattorna, and Nicola Sacco.
\newblock Optimization of dynamic ridesharing systems.
\newblock {\em Transportation Research Record: Journal of the Transportation
  Research Board}, 2359:44--50, 2013.

\bibitem{r3}
Yeqian Lin, Wenquan Li, Feng Qiu, and He~Xu.
\newblock Research on optimization of vehicle routing problem for ride-sharing
  taxi.
\newblock {\em Procedia - Social and Behavioral Sciences}, 43:494 -- 502, 2012.

\bibitem{r4}
Dominik Pelzer, Jiajian Xiao, Daniel Zehe, Michael Lees, Alois C.~Knoll, and
  Heiko Aydt.
\newblock A partition-based match making algorithm for dynamic ridesharing.
\newblock 16:1--12, 2015.

\bibitem{r5}
Yan Huang, Favyen Bastani, Ruoming Jin, and Xiaoyang~Sean Wang.
\newblock Large scale real-time ridesharing with service guarantee on road
  networks.
\newblock {\em Proc. VLDB Endow.}, 7(14):2017--2028, 2014.

\bibitem{r6}
Maximilian Schreieck, Hazem Safetli, Sajjad~Ali Siddiqui, Christoph Pflügler,
  Manuel Wiesche, and Helmut Krcmar.
\newblock A matching algorithm for dynamic ridesharing.
\newblock {\em Transportation Research Procedia}, 19:272 -- 285, 2016.

\bibitem{r7}
Blerim Cici, Athina Markopoulou, and Nikolaos Laoutaris.
\newblock Sors: A scalable online ridesharing system.
\newblock In {\em Proceedings of the 9th ACM SIGSPATIAL International Workshop
  on Computational Transportation Science}, IWCTS '16, pages 13--18, New York,
  NY, USA, 2016. ACM.

\bibitem{r8}
Bin Cao, Louai Alarabi, Mohamed~F. Mokbel, and Anas Basalamah.
\newblock Sharek: A scalable dynamic ride sharing system.
\newblock In {\em Proceedings of the 2015 16th IEEE International Conference on
  Mobile Data Management - Volume 01}, MDM '15, pages 4--13, Washington, DC,
  USA, 2015. IEEE Computer Society.

\bibitem{r9}
G.~Dimitrakopoulos, P.~Demestichas, and V.~Koutra.
\newblock Intelligent management functionality for improving transportation
  efficiency by means of the car pooling concept.
\newblock {\em IEEE Transactions on Intelligent Transportation Systems},
  13(2):424--436, 2012.

\bibitem{r10}
R.~B. Khot and P.~K. Gokhale.
\newblock Pub-sub model based on preference matching for carpooling framework.
\newblock In {\em 2016 International Conference on Inventive Computation
  Technologies (ICICT)}, volume~3, pages 1--4, 2016.

\bibitem{r11}
Jamal Yousaf, Juanzi Li, Lu~Chen, Jie Tang, and Xiaowen Dai.
\newblock Generalized multipath planning model for ride-sharing systems.
\newblock {\em Frontiers of Computer Science}, 8(1):100--118, 2014.

\bibitem{r12}
Phathinan THAITHATKUL, Toru SEO, Takahiko KUSAKABE, and Yasuo ASAKURA.
\newblock A passengers matching problem in ridesharing systems by considering
  user preference.
\newblock {\em Journal of the Eastern Asia Society for Transportation Studies},
  11:1416--1432, 2015.

\bibitem{r13}
Maurizio Bruglieri, Diego Ciccarelli, Alberto Colorni, and Alessandro Luè.
\newblock Poliunipool: a carpooling system for universities.
\newblock {\em Procedia - Social and Behavioral Sciences}, 20:558 -- 567, 2011.

\bibitem{r14}
Gyozo Gidofalvi, Gergely Herenyi, and Torben~Bach Pedersen.
\newblock Instant social ride-sharing.
\newblock In {\em Proceedings of the Fifteenth World Congress on Intelligent
  Transport Systems}, page~8, New York, NY, USA, 2008. Intelligent
  Transportation Society of America.

\bibitem{r15}
Alexander Kleiner, Bernhard Nebel, and Vittorio~Amos Ziparo.
\newblock A mechanism for dynamic ride sharing based on parallel auctions.
\newblock In {\em Proceedings of the Twenty-Second International Joint
  Conference on Artificial Intelligence - Volume Volume One}, IJCAI'11, pages
  266--272. AAAI Press, 2011.

\bibitem{c2r11}
R~Marler and Jasbir Arora.
\newblock Survey of multi-objective optimization methods for engineering.
\newblock 26:369--395, 04 2004.

\bibitem{r16}
Evangelos Triantaphyllou.
\newblock {\em Multi-Criteria Decision Making Methods}, pages 5--21.
\newblock Springer US, Boston, MA, 2000.

\bibitem{hamrouni2008succinct}
Tarek Hamrouni, Sadok Ben~Yahia, and Engelbert Mephu~Nguifo.
\newblock Succinct minimal generators: Theoretical foundations and
  applications.
\newblock {\em International journal of foundations of computer science},
  19(02):271--296, 2008.

\bibitem{gasmi2007extraction}
Ghada Gasmi, Sadok Ben~Yahia, Engelbert~Mephu Nguifo, and Slim Bouker.
\newblock Extraction of association rules based on literalsets.
\newblock In {\em International Conference on Data Warehousing and Knowledge
  Discovery}, pages 293--302. Springer, 2007.

\bibitem{c3r1}
Bernard Roy and Denis Bouyssou.
\newblock Aide multicritère à la décision : méthodes et cas.
\newblock page 695. London School of Economics and Political Science, 1993.

\bibitem{bouker2012ranking}
Slim Bouker, Rabie Saidi, Sadok Ben~Yahia, and Engelbert~Mephu Nguifo.
\newblock Ranking and selecting association rules based on dominance
  relationship.
\newblock In {\em Tools with Artificial Intelligence (ICTAI)}, volume~1, pages
  658--665. IEEE, 2012.

\bibitem{c3r3}
Maurice LANDRY.
\newblock L'aide à la décision comme support à la construction du sens dans
  l'organisation.
\newblock {\em Systèmes d'Information et Management (French Journal of
  Management Information Systems)}, 3(1):5--39, 1998.

\bibitem{ferjani2012formal}
Fethi Ferjani, Samir Elloumi, Ali Jaoua, Sadok Ben~Yahia, Sahar Ismail, and
  Sheikha Ravan.
\newblock Formal context coverage based on isolated labels: An efficient
  solution for text feature extraction.
\newblock {\em Information Sciences}, 188:198--214, 2012.

\bibitem{hamdi2013trust}
Sana Hamdi, Amel Bouzeghoub, Alda~Lopes Gancarski, and Sadok Ben~Yahia.
\newblock Trust inference computation for online social networks.
\newblock In {\em Trust, Security and Privacy in Computing and Communications
  (TrustCom)}, pages 210--217. IEEE, 2013.

\bibitem{yahia2004revisiting}
Sadok Ben~Yahia and Engelbert~Mephu Nguifo.
\newblock Revisiting generic bases of association rules.
\newblock In {\em International Conference on Data Warehousing and Knowledge
  Discovery}, pages 58--67. Springer, 2004.

\bibitem{bouzouita2006garc}
Ines Bouzouita, Samir Elloumi, and Sadok Ben~Yahia.
\newblock Garc: A new associative classification approach.
\newblock In {\em International Conference on Data Warehousing and Knowledge
  Discovery}, pages 554--565. Springer, 2006.

\bibitem{hamrouni2010generalization}
Tarek Hamrouni, Sadok Ben~Yahia, and Engelbert~Mephu Nguifo.
\newblock Generalization of association rules through disjunction.
\newblock {\em Annals of Mathematics and Artificial Intelligence},
  59(2):201--222, 2010.

\bibitem{brahmi2012omc}
Hanen Brahmi, Imen Brahmi, and Sadok Ben~Yahia.
\newblock Omc-ids: at the cross-roads of olap mining and intrusion detection.
\newblock In {\em Pacific-Asia Conference on Knowledge Discovery and Data
  Mining {PAKDD}}, pages 13--24. Springer, 2012.

\bibitem{jelassi2014efficient}
M~Nidhal Jelassi, Christine Largeron, and Sadok Ben~Yahia.
\newblock Efficient unveiling of multi-members in a social network.
\newblock {\em Journal of Systems and Software}, 94:30--38, 2014.

\bibitem{ayouni2011extracting}
Sarra Ayouni, Sadok Ben~Yahia, and Anne Laurent.
\newblock Extracting compact and information lossless sets of fuzzy association
  rules.
\newblock {\em Fuzzy Sets and Systems}, 183(1):1--25, 2011.

\bibitem{c3r7}
{Vincke, Philippe}.
\newblock Une m\'ethode interactive en programmation lin\'eaire \`a plusieurs
  fonctions \'economiques.
\newblock {\em R.A.I.R.O. Recherche op\'erationnelle}, 10:5--20, 1976.

\bibitem{DBLP:conf/ictai/YahiaN04}
Sadok~Ben Yahia and Engelbert~Mephu Nguifo.
\newblock Emulating a cooperative behavior in a generic association rule
  visualization tool.
\newblock In {\em International Conference on Tools with Artificial
  Intelligence {ICTAI}}, volume 110. CEUR-WS.org, 2004.

\bibitem{r20}
Grzegorz Filcek and Jacek {\.{Z}}ak.
\newblock The multiple criteria optimization problem of joint matching
  carpoolers and common route planning.
\newblock In Jerzy {\'{S}}wi{\k{a}}tek, Zofia Wilimowska, Leszek Borzemski, and
  Adam Grzech, editors, {\em Information Systems Architecture and Technology:
  Proceedings of 37th International Conference on Information Systems
  Architecture and Technology -- ISAT 2016 -- Part III}, pages 225--236, Cham,
  2017. Springer International Publishing.

\bibitem{r21}
Antonella Petrillo, Pasquale Carotenuto, Ilaria Baffo, and Fabio De~Felice.
\newblock A web-based multiple criteria decision support system for evaluation
  analysis of carpooling.
\newblock {\em Environment, Development and Sustainability}, 2017.

\bibitem{r22}
Wenxiang Li, Ye~Li, Jing Fan, Haopeng Deng, and Mohamed Bakillah.
\newblock Siting of carsharing stations based on spatial multi-criteria
  evaluation: A case study of shanghai evcard.
\newblock volume~9, page 152. Sustainability, 2017.

\bibitem{r23}
Anjali Awasthi and Satyaveer~S. Chauhan.
\newblock Using ahp and dempster-shafer theory for evaluating sustainable
  transport solutions.
\newblock {\em Environ. Model. Softw.}, 26(6):787--796, 2011.

\bibitem{c4r1}
Alvin E.~Roth.
\newblock Stability and polarization of interests in job matching.
\newblock 52:47--57, 02 1984.

\bibitem{c4r2}
D.~Gale and L.~S. Shapley.
\newblock College admissions and the stability of marriage.
\newblock {\em The American Mathematical Monthly}, 120(5):386--391, 2013.

\bibitem{c4r3}
Nobel~Prize Committee.
\newblock Alvin e. roth and lloyd s. shapley: Stable matching: Theory,
  evidence, and practical design.
\newblock Nobel Prize in Economics documents 2012-2, Nobel Prize Committee,
  2012.

\bibitem{c4r5}
Alvin~E. Roth and John~H. Vande~Vate.
\newblock Incentives in two-sided matching with random stable mechanisms.
\newblock {\em Economic Theory}, 1(1):31--44, Mar 1991.

\bibitem{c4r6}
Alvin~E. Roth and Elliott Peranson.
\newblock {The Redesign of the Matching Market for American Physicians: Some
  Engineering Aspects of Economic Design}.
\newblock NBER Working Papers 6963, National Bureau of Economic Research, Inc,
  February 1999.

\bibitem{c4r7}
Alvin~E. Roth and Marilda Sotomayor.
\newblock The college admissions problem revisited.
\newblock {\em Econometrica}, 57(3):559--570, 1989.

\bibitem{c4r8}
Alvin~E. Roth, Tayfun Sönmez, and M.~Utku Ünver.
\newblock Kidney exchange.
\newblock {\em The Quarterly Journal of Economics}, 119(2):457--488, 2004.

\bibitem{c4r9}
{\em carms. Canadian resident matching service}, accessed July, 2018).
\newblock \url{http://www.carms.ca/}.

\bibitem{c4r10}
{\em jrmp. Japan residency matching program}, accessed July, 2018).
\newblock \url{http://www.jrmp.jp/}.

\bibitem{c4r11}
Xing Wang, Niels Agatz, and Alan Erera.
\newblock Stable matching for dynamic ride-sharing systems.
\newblock {\em ERIM Report Series Research in Management}, 2:34, 2014.

\bibitem{c4r12}
D.~G. McVitie and L.~B. Wilson.
\newblock Stable marriage assignment for unequal sets.
\newblock {\em BIT Numerical Mathematics}, 10(3):295--309, 1970.

\bibitem{r32}
Majid Behzadian, Sina Otaghsara, Morteza Yazdani, and Joshua Ignatius.
\newblock A state-of the-art survey of topsis applications.
\newblock 39:13051--13069, 2012.

\bibitem{r33}
Alvin~E. Roth, Uriel~G. Rothblum, and John~H. Vande~Vate.
\newblock Stable matchings, optimal assignments, and linear programming.
\newblock {\em Math. Oper. Res.}, 18(4):803--828, 1993.

\bibitem{r34}
{\em Geomatic ApS - Center for Geoinformatik}, accessed January, 2018.
\newblock \url{http://www.geomatic.dk}.

\bibitem{hamrouni2008succinctt}
Tarek Hamrouni, Sadok Ben~Yahia, and Engelbert~Mephu Nguifo.
\newblock Succinct system of minimal generators: A thorough study, limitations
  and new definitions.
\newblock In {\em Concept Lattices and Their Applications}, pages 80--95.
  Springer, 2008.

\bibitem{draheim2017generalized}
Dirk Draheim.
\newblock {\em Generalized Jeffrey Conditionalization - {A} Frequentist
  Semantics of Partial Conditionalization}.
\newblock Springer, 2017.

\bibitem{draheim2017semantics}
Dirk Draheim.
\newblock {\em Semantics of the Probabilistic Typed Lambda Calculus - Markov
  Chain Semantics, Termination Behavior, and Denotational Semantics}.
\newblock Springer, 2017.

\bibitem{auer2009extending}
Dagmar Auer, Verena Geist, and Dirk Draheim.
\newblock Extending {BPMN} with submit/response-style user interaction
  modeling.
\newblock In {\em IEEE Conference on Commerce and Enterprise Computing
  {(CEC)}}, pages 368--374. IEEE, 2009.

\bibitem{draheim2010service}
Dirk Draheim.
\newblock The service-oriented metaphor deciphered.
\newblock {\em Journal of Computing Science and Engineering {(JCSE)}},
  4(4):253--275, 2010.

\bibitem{atkinson2010typed}
Colin Atkinson, Dirk Draheim, and Verena Geist.
\newblock Typed business process specification.
\newblock In {\em 2010 14th IEEE International Enterprise Distributed Object
  Computing Conference {(EDOC)}}, pages 69--78. IEEE, 2010.

\bibitem{mouakher2019efficient}
Amira Mouakher and Sadok Ben~Yahia.
\newblock On the efficient stability computation for the selection of
  interesting formal concepts.
\newblock {\em Information Sciences}, 472:15--34, 2019.

\bibitem{mouakher2016qualitycover}
Amira Mouakher and Sadok Ben~Yahia.
\newblock Qualitycover: Efficient binary relation coverage guided by induced
  knowledge quality.
\newblock {\em Information Sciences}, 355-356:58--73, 2016.

\bibitem{houari2018nbf}
Amina Houari, Wassim Ayadi, and Sadok Ben~Yahia.
\newblock Nbf: An fca-based algorithm to identify negative correlation
  biclusters of {DNA} microarray data.
\newblock In {\em 2018 IEEE 32nd International Conference on Advanced
  Information Networking and Applications {(AINA)}}, pages 1003--1010. IEEE,
  2018.

\bibitem{houari2018new}
Amina Houari, Wassim Ayadi, and Sadok Ben~Yahia.
\newblock A new fca-based method for identifying biclusters in gene expression
  data.
\newblock {\em International Journal of Machine Learning and Cybernetics},
  pages 1--15, 2018.

\bibitem{bouker2014mining}
Slim Bouker, Rabie Saidi, Sadok Ben~Yahia, and Engelbert Mephu~Nguifo.
\newblock Mining undominated association rules through interestingness
  measures.
\newblock {\em International Journal on Artificial Intelligence Tools},
  23(04):1460011, 2014.

\bibitem{hamrouni2013looking}
Tarek Hamrouni, S~Ben~Yahia, and E~Mephu Nguifo.
\newblock Looking for a structural characterization of the sparseness measure
  of (frequent closed) itemset contexts.
\newblock {\em Information Sciences}, 222:343--361, 2013.

\bibitem{allani2016dpms}
Sabri Allani, Taoufik Yeferny, Richard Chbeir, and Sadok~Ben Yahia.
\newblock Dpms: A swift data dissemination protocol based on map splitting.
\newblock In {\em Computer Software and Applications Conference (COMPSAC), 2016
  IEEE 40th Annual}, volume~1, pages 817--822. IEEE, 2016.

\bibitem{allani2016novel}
Sabri Allani, Taoufik Yeferny, Richard Chbeir, and Sadok~Ben Yahia.
\newblock A novel vanet data dissemination approach based on geospatial data.
\newblock {\em Procedia Computer Science}, 98:572--577, 2016.

\bibitem{allani2018scalable}
Sabri Allani, Taoufik Yeferny, and Richard Chbeir.
\newblock A scalable data dissemination protocol based on vehicles trajectories
  analysis.
\newblock {\em Ad Hoc Networks}, 71:31--44, 2018.

\end{thebibliography}

\end{document}